\documentclass[fleqn,usenatbib]{mnras}

\usepackage[T1]{fontenc}
\usepackage{ae,aecompl}

\usepackage{graphicx}	
\usepackage{amsmath}	
\usepackage{amssymb}	
\usepackage{verbatim}
\usepackage[flushleft]{threeparttable}
\usepackage[dvipsnames]{xcolor}
\usepackage{soul}
\usepackage{newtxtext,newtxmath}

\newcommand{\qso}{J\,1652$+$2650}
\newcommand{\qsolong}{SDSS~J\,165252.67$+$265001.96}
\newcommand{\qsot}{Q\,2359$-$1241}

\newcommand{\HI}{H\,{\sc i}}

\newcommand{\SiIV}{Si\,{\sc iv}}

\newcommand{\CIV}{C\,{\sc iv}}
\newcommand{\NV}{N\,{\sc v}}
\newcommand{\OVI}{O\,{\sc vi}}
\newcommand{\NaI}{Na\,{\sc i}}
\newcommand{\CaI}{Ca\,{\sc i}}
\newcommand{\CaII}{Ca\,{\sc ii}}
\newcommand{\FeII}{Fe\,{\sc ii}}
\newcommand{\MnII}{Mn\,{\sc ii}}
\newcommand{\MnIIs}{Mn\,{\sc ii}$^\star$}
\newcommand{\MgI}{Mg\,{\sc i}}
\newcommand{\MgII}{Mg\,{\sc ii}}
\newcommand{\HeI}{He\,{\sc i}}
\newcommand{\HeIs}{He\,{\sc i}$^\star$}
\newcommand{\HeII}{He\,{\sc ii}}
\newcommand{\OII}{[\ion{O}{ii}]}

\newcommand{\Cloudy}{{\sc Cloudy}}

\definecolor{green}{rgb}{0,0.4,0}

\newcommand{\ioffe}{Ioffe Institute, {Polyteknicheskaya 26}, 194021 Saint-Petersburg, Russia}
\newcommand{\eso}{European Southern Observatory, Alonso de C\'ordova 3107, Vitacura, Casilla 19001, Santiago, Chile}
\newcommand{\fcla}{Franco-Chilean Laboratory for Astronomy, IRL\,3386, CNRS and U. de Chile, Casilla 36-D, Santiago, Chile}
\newcommand{\iap}{Institut d'Astrophysique de Paris, CNRS-SU, UMR\,7095, 98bis bd Arago, 75014 Paris, France}
\newcommand{\lyon}{Centre de Recherche Astrophysique de Lyon, UMR\,5574, 9 avenue Charles Andr\'e, 69230 Saint-Genis-Laval, France}
\newcommand{\das}{Departamento de Astronom\'\i a, Universidad de Chile, Casilla 36-D, Santiago, Chile}

\title[FeLoBAL towards \qso]{Low-ionization iron-rich Broad Absorption-Line Quasar SDSS~\qso:
Physical conditions in the ejected gas from excited \FeII\ and metastable \HeI\thanks{Based on
observations carried out in visitor mode at the European Southern Observatory under ESO program
ID 0103.A-0529(A) (PI: Ledoux).}}

\author[]{
S.~A.~Balashev,$^1$\thanks{E-mail: s.balashev@gmail.com},
C.~Ledoux,$^2$
P.~Noterdaeme,$^{3,4}$
P.~Boiss\'e,$^4$
J.-K.~Krogager,$^5$\newauthor
S.~L\'opez$^6$ and
K.~N.~Telikova$^{1}$\\
$^1$ \ioffe \\
$^2$ \eso \\
$^3$ \fcla \\
$^4$ \iap \\
$^5$ \lyon \\
$^6$ \das
}

\date{Accepted 2023. Received 2023; in original form 2023}

\pubyear{2023}

\begin{document}
\label{firstpage}
\pagerange{\pageref{firstpage}--\pageref{lastpage}}
\maketitle

\begin{abstract}

We present high-resolution VLT/UVES spectroscopy and a detailed analysis of the unique Broad Absorption-Line system towards the
quasar \qsolong. This system exhibits low-ionization metal absorption lines from the ground states and excited energy levels of \FeII\ and
\MnII, and the meta-stable $2\,^3S$ excited state of \HeI. The extended kinematics of the absorber encompasses three main clumps
with velocity offsets of $-5680$, $-4550$, and $-1770$~km~s$^{-1}$ from the quasar emission redshift, $z=0.3509\pm 0.0003$, derived from
\OII\ emission. Each clump shows moderate partial covering of the background continuum source, $C_f\approx [0.53; 0.24; 0.81]$. 
We discuss the excitation mechanisms at play in the gas, which we use to constrain the distance of the clouds from the Active Galactic Nucleus (AGN) as well as the density, temperature, and typical sizes of the clouds. The number density is found to be $n_{\rm H}\sim 10^4$~cm$^{-3}$ and the temperature $T_{\rm e}\sim 10^{4}$~K, with longitudinal cloudlet sizes of $\gtrsim 0.01$~pc. \Cloudy\ photo-ionization modelling of \HeIs, which is also produced at the interface between the neutral and ionized phases, assuming the number densities derived from \FeII, constrains the ionization parameter to be $\log U\sim -3$. This corresponds to distances of a few 100~pc from the AGN. We discuss these results in the more general context of associated absorption-line systems and propose a connection between FeLoBALs and the recently-identified molecular-rich intrinsic absorbers. Studies of significant samples of FeLoBALs, even though rare per se, will soon be
possible thanks to large dedicated surveys paired with high-resolution spectroscopic follow-ups.

\end{abstract}



\section{Introduction}
\label{sect:introduction}

A key characteristic of around 20\% of optically-selected quasars is the occurrence of broad absorption-line
(BAL) systems along the line-of-sight to the quasar (\citealt{Tolea2002}; \citealt{Hewett2003}; \citealt{Reichard2003}; \citealt{Knigge2008};
\citealt{Gibson2009}). BAL systems are typically associated with highly-ionized metals, e.g., \CIV\ and \OVI, and their wide kinematic
spreads, velocity offsets, and partial covering factors all indicate that they are produced by out-flowing gas. Observations of such
outflows provide a direct test of quasar feedback models.

One-tenth of BAL systems show associated wide \MgII\ absorption \citep{Trump2006} and are called low-ionization BALs (hereafter {LoBALs}). An
even smaller fraction, totalling only $\sim 0.3\%$ of the global quasar population, in addition, exhibits \FeII\ absorption and is hence called
FeLoBALs. A qualifying feature of FeLoBALs is the detection of \FeII\ in its various fine-structure energy levels of the lowest
electronic states. These levels may be excited by collisions or UV pumping, and their relative abundance can provide robust estimates of
critical physical parameters. Interestingly, the modelling of FeLoBALs indicates they contain some neutral gas and likely occur at the interface
between the ionized and neutral media \citep{Korista2008}. Another feature of FeLoBALs, which was gradually recognised, is the presence of
absorption lines corresponding to transitions from the first excited level of neutral helium, \HeI$^*$ \citep{Arav2001, Aoki2011, Leighly2011}.
This is observed in FeLoBALs but also more generally in LoBALs \citep{Liu2015}. These lines have also been detected in the host galaxies
of a few GRBs \citep{Fynbo2014}. \HeI$^*$ is predominately populated by recombination of \HeII\ and the measured column densities of \HeI$^*$ provide
a measure of the total column density of the ionized medium. This can constrain the physical conditions in the outflowing gas
and determine the total mass budget to draw a more complete physical picture of quasar activity. For example, rapid cooling followed by the phase transition and
subsequent condensation in an outflowing medium can result in the escape of small chunks of the medium from the outflowing gas. Such cloudlets can precipitate back
onto the central engine and sustain the formation of the broad-line region around the central powering source \citep{Elvis2017}.

Because the incidence rate of FeLoBALs in quasars is low, only a small sample of such systems was found in the Sloan Digital Sky
Survey (SDSS) database \cite[e.g.,][]{Trump2006,Farrah2012,Choi2022}. Most importantly, only about a dozen such systems was studied so far by
means of high-resolution near-UV and visual spectroscopy, i.e.: Q\,0059$-$2735 {\citep{Hazard1987,Wampler1995,Xu2021}}, Q\,2359$-$1241 \citep{Arav2001,Arav2008},
FIRST\,104459.6$+$365605 \citep{Becker2000,deKool2001}, FBQS\,0840$+$3633 \citep{Becker1997,deKool2002a}, FIRST~J\,121442.3$+$280329
\citep{Becker2000,deKool2002b}, SDSS~J\,030000.56$+$004828.0 \citep{Hall2003}, SDSS~J\,0318$-$0600 \citep{Dunn2010,Bautista2010}, AKARI~J\,1757$+$5907
\citep{Aoki2011}, PG\,1411$+$442 \citep{Hamann2019}, SDSS~J\,2357$-$0048 \citep{Byun2022a}, SDSS~J\,1439$-$0106 \citep{Byun2022b}, {SDSS J\,0242$+$0049 \citep{Byun2022c}, SDSS J\,1130$+$0411 \citep{Walker2022},} Mrk\,231 \citep{Boroson1991,Smith1995,Veilleux2016}, and NGC\,4151 \citep{Crenshaw2000,Kraemer2001}. Among these, only a few systems exhibit mild
line saturation and overlapping, which allow one to resolve the fine-structure lines and therefore derive robust constraints on the gas physical
conditions \citep{Arav2008}. Moreover, each previously-studied system appears to be fairly specific, i.e., FeLoBALs show a broad range of properties,
which means any new observation and detailed analysis potentially bring new valuable clues to understanding the physics and environmental properties of AGN outflows.

In this paper, we report the serendipitous discovery of a multi-clump FeLoBAL towards \qsolong, which we refer to in the following as \qso. We present high-quality VLT/UVES data of this quasar and
the spectroscopic analysis of the absorption system and discuss the excitation mechanisms at play in the gas. Our goal is to infer
the physical properties of FeLoBAL clouds and estimate their distance from the central engine.

\section{Observations and data reduction}
\label{sect:data}

We selected the bright quasar \qso\ \citep[$B=18.2$; $V=17.7$; $z_{\rm em}=0.35$;][]{Veron2010} with the primary goal to search
for CN, CH, and CH$^+$ molecules in absorption based on the detection of strong associated \NaI\ lines at $z\approx 0.33$ in
the SDSS spectrum \citep{Paris2018,Negrete2018}{, which is shown in Fig.~\ref{fig:SDSS}}. We observed the target in visitor mode on the night of July 27, 2019, with UVES, the
Ultraviolet and Visual Echelle Spectrograph \citep{Dekker2000} installed at the Nasmyth-B focus of the ESO Very Large
Telescope Unit-2, Kueyen. The total on-source integration time was four hours, subdivided evenly into three exposures taken in a row.
The instrumental setup used Dichroic beam splitter \#1 and positioning of the cross-dispersers at central wavelengths of 390~nm
and 590~nm in the Blue and Red spectroscopic arms, respectively. In each arm, the slit widths were fixed to $1\arcsec$ and CCD pixels were
binned $2\times 2$. While observing, the weather conditions were excellent with clear sky transparency and a measured Differential Image
Motion Monitor \citep{Sarazin1990} seeing of $0\farcs 6$. Despite a relatively high airmass (1.63--1.97), the source was recorded on the
detectors with a spatial PSF trace of only $1\farcs 1$ FWHM in the Blue ($1\farcs 0$ FWHM in the Red).

The raw data from the telescope was reduced offline applying the recipes of the UVES pipeline v5.10.4 running on the ESO Reflex
platform. During this process, the spectral format of the data was compared to a physical model of the instrument, to which a slight CCD
rotation was applied ($-0.05\degr$ in the Blue; $+0.05\degr$ in the Red). ThAr reference frames acquired in the morning following
the observations were used to derive wavelength-calibration solutions, which showed residuals of 1.53~m\AA\ RMS in the Blue (4.25~m\AA\ RMS
in the Red). The object and sky spectra were extracted simultaneously and optimally, and cosmic-ray hits were removed efficiently using a
$\kappa$-$\sigma$ clipping factor of five. The wavelength scale was converted to the helio-vacuum rest frame.
Individual 1D exposures were then scaled and combined together by weighing each pixel by its S/N. The S/N of the final science product is
$\sim 15$ per pixel at $325<\lambda_{\rm obs}<455$~nm and $\sim 32$ per pixel at $490<\lambda_{\rm obs}<690$~nm. With a delivered
resolving power of 50\,000, the instrumental line-spread function is 6~km~s$^{-1}$ FWHM.

\section{Data analysis}
\label{sect:analysis}

\subsection{Quasar spectrum and systemic redshift}
\label{sect:redshift}

\begin{figure*}
\centering
\includegraphics[trim={0.0cm 0.0cm 0.0cm 0.0cm},clip,width=\textwidth]{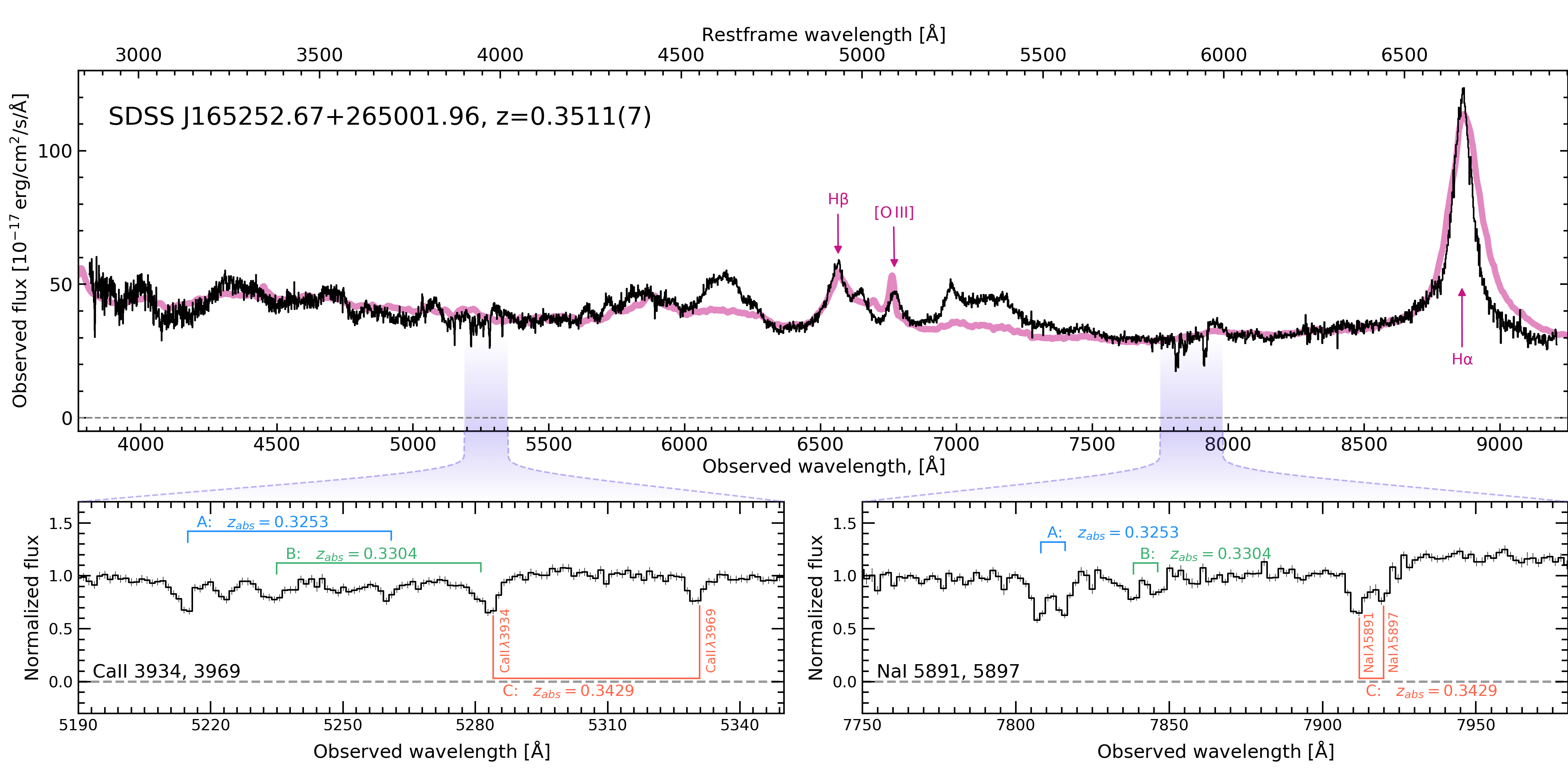}\\
\caption{SDSS spectrum of \qso. The top panel shows the full range of SDSS spectrum plotted by black line. The quasar emission lines are marked by pink arrows, while pink line represents the composite QSO spectrum reddened by dust (see Sect.~\ref{sect:redshift}). The bottom panels show a zoom view of \CaII\ and \NaI\ absorption line regions (these lines were used to preselect \qso\ from SDSS database, see Sect.~\ref{sect:data}). The red, green and blue lines indicate a absorption line complexes, that were resolved using UVES spectrum, see Sect.~\ref{sect:overview}.
\label{fig:SDSS}}
\end{figure*}

\qso\ exhibits moderate reddening. Based on the SDSS spectrum, {shown in Fig.~\ref{fig:SDSS}}, and using the Type~I quasar template from \cite{Selsing2016}, we followed a procedure similar to that employed by \citet{Balashev2017,Balashev2019} and derived that $A_V\approx 1.2${, assuming standard galactic extinction law \citep{Fitzpatrick2007}.} This is quite large
compared to intervening quasar absorbers, e.g., DLAs \citep{Murphy2016}. We also note that this quasar shows iron-emission line
complexes in the spectral regions around 4600~\AA\ and 5300~\AA\ in the quasar rest frame, that are enhanced by a factor of $\sim 4$
relative to the fiducial quasar template {(see Fig.~\ref{fig:SDSS})}.

To determine the quasar emission redshift accurately, we followed the recommendations of \citet{Shen2016} that \CaII\ and \OII\ should be
considered the most reliable systemic-redshift indicators. In the case of \qso, the blue side of the \CaII\ profile is affected by strong
self-absorption, so we are left with \OII, which according to \citet{Shen2016} is not significantly shifted relative to \CaII. Based on
a single-component Gaussian fit, we measured $z_{\rm em}=0.3511(7)$ when considering the \OII,$\lambda$3727.092 transition line alone,
and $z_{\rm em}=0.3506(7)$ when using the mean wavelength of the \OII,$\lambda\lambda$3727.092,3729.875 doublet. This translates
into $z_{\rm em}=0.3509\pm 0.0003$, which we consider as our most-accurate determination of the quasar systemic redshift.
The H$\beta$ emission line is observed at a redshift of $z\approx 0.3494$, implying a velocity blue-shift
of $\Delta V\sim -330$~km\,s$^{-1}$ relative to \OII. This is consistent with the findings of \citet{Shen2016} using their own sample.

\subsection{Absorption-line system overview}
\label{sect:overview}

\begin{figure*}
\centering
\includegraphics[trim={0.0cm 0.0cm 0.0cm 0.0cm},clip,width=\textwidth]{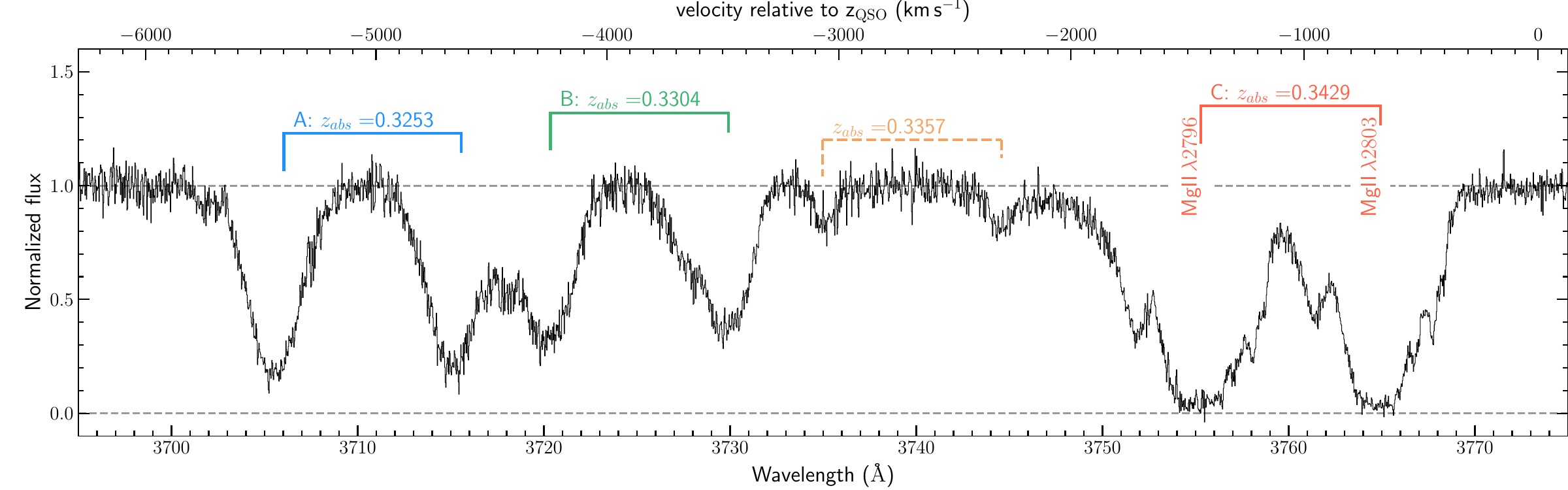}\\
\vspace{0.4cm}
\includegraphics[trim={0.0cm 0.0cm 0.0cm 0.0cm},clip,width=\textwidth]{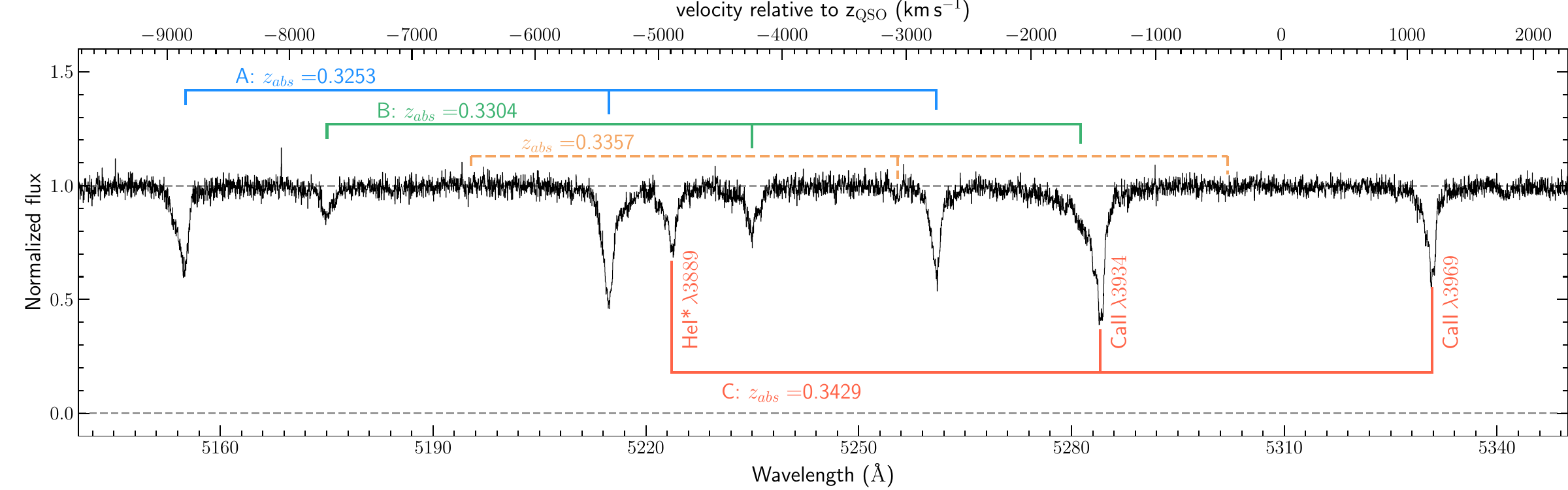}\caption{Portions of the normalized UVES spectrum showing the kinematics of \MgII\ (upper panel), \CaII\ H and K, and \HeIs\,$\lambda3889$
(lower panel) in the FeLoBAL towards \qso. The \MgII\ absorption-line complex at $z\approx 0.3357$ is much weaker than complexes $A$, $B$, and $C$,
and therefore is not included in the following Voigt-profile fitting analysis. In both panels, the top axis shows the velocity of the strongest
transition, i.e., \MgII$\lambda2796$ (upper panel) or \CaII$\lambda3934$ (lower panel), relative to the quasar systemic redshift.
\label{fig:overview_MgII_and_CaII}}
\end{figure*}

The FeLoBAL\footnote{Based on \MgII\ lines this system do not satisfy standard BAL definition \cite{Weymann1981}, and should be attributed to the mini-BAL. However, \CIV\ lines (that are typically used in BAL definition and usually indicate much wider profiles than \MgII\ lines) out of the spectral range. Therefore, we will keep denotation of this system as FeLoBAL, which is supported by example of \qsot \citep{Arav2001}, for which \MgII\ lines indicate a similar width as in \qso, but HST observations confirm large width of \CIV\ lines.} on the line-of-sight to \qso\ consists of multiple prominent absorption lines from \MgII, \CaII, \HeIs\ (i.e., the meta-stable excited
state $2^3\rm S$), \MgI, \FeII, and \MnII, all covered by the UVES spectrum. The system is composed of three
main, kinematically-detached absorption-line complexes,\footnote{When selecting this quasar to search for intervening molecular absorption, we assumed
the line-of-sight could intersect three galaxies (possibly located in a cluster hosting the quasar). It turns out that the gas is associated with the
quasar active nucleus itself in spite of its low ionization.} i.e., at $z_{\rm abs}=0.32531$ ($\Delta V\approx -5680$~km\,s$^{-1}$), 0.33043
($\Delta V\approx -4550$~km\,s$^{-1}$), and 0.34292 ($\Delta V\approx -1770$~km\,s$^{-1}$), where the reddest and bluest clumps exhibit the
strongest \MgII\ and \CaII\ absorption overall (see Fig.~\ref{fig:overview_MgII_and_CaII}). In the following, we refer to these three complexes as $A$,
$B$, and $C$, in order of increasing redshift. Each complex has at least a few velocity components resolved by eye within its own profile.
Weak absorption is also visible in \MgII\ at $z_{\rm abs}=0.3357$ ($\Delta V\approx -3390$~km\,s$^{-1}$) and, tentatively, also in \CaII\,$\lambda$3934.

\FeII,$\lambda\lambda$2586,2600 ground-state absorption lines as well as lines from the fine-structure energy levels of the two LS states
$\rm 3d^{6}4s\,a^6D$ (ground) and $\rm 3d^{6}4s\,a^4D$ (second excited state, which is encompassing the \FeII$^{9*}$--\FeII$^{12*}$ levels) are detected
in this system (see Figs.~\ref{fig:fit_overview}, \ref{fig:FeII_low_1}, \ref{fig:FeII_low_2}, and \ref{fig:FeII_high}). Transition lines from the first excited LS state
$\rm 3d^7\,a^4F$ (i.e., 5$^{\rm th}$ to 8$^{\rm th}$ excited levels above the ground state) are not covered by our spectrum, being located bluewards of the
observed wavelength range. The \MnII\,$\lambda\lambda\lambda$2576,2594,2606 triplet and \MnII$^\star$ transition
lines (i.e., $\lambda\lambda\lambda$2933,2940,2950) from the first excited level of \MnII, with an excitation energy of 9473~cm$^{-1}$, are detected most
clearly in the $C$ (reddest) and $A$ (bluest) clumps (see Fig.~\ref{fig:MnII}). Such transitions were detected before only in a few FeLoBALs
\citep[e.g., FBQS~J\,1151$+$3822;][]{Lucy2014}.

\CaI\,$\lambda$4227 and CH$^+$\,$\lambda\lambda$3958,4233 absorptions are not detected. Using the oscillator strengths of CH$^+$ lines from
 \citet{Weselak2009}, we derive a 2$\sigma$ upper limit on the column density of {$N(\rm CH^+)=10^{13}\rm\,cm^{-2}$} for each of the three complexes. 
We detect possible \NaI\ emission lines in both the UVES and SDSS spectra. All absorption lines in the UVES spectrum were identified, except a weak line
at $\lambda_{\rm obs}=5341$~\AA, which has a similarly wide profile as the FeLoBAL lines. Searching an identification in the NIST database at the
different BAL sub-redshifts did not provide any satisfactory solution{, therefore it has likely spurious nature.} 

\subsection{Evidence of partial flux covering from \MgII}
\label{sect:mgii}

\begin{figure}
\centering
\includegraphics[trim={0.0cm 0.0cm 0.0cm 0.0cm},clip,width=\columnwidth]{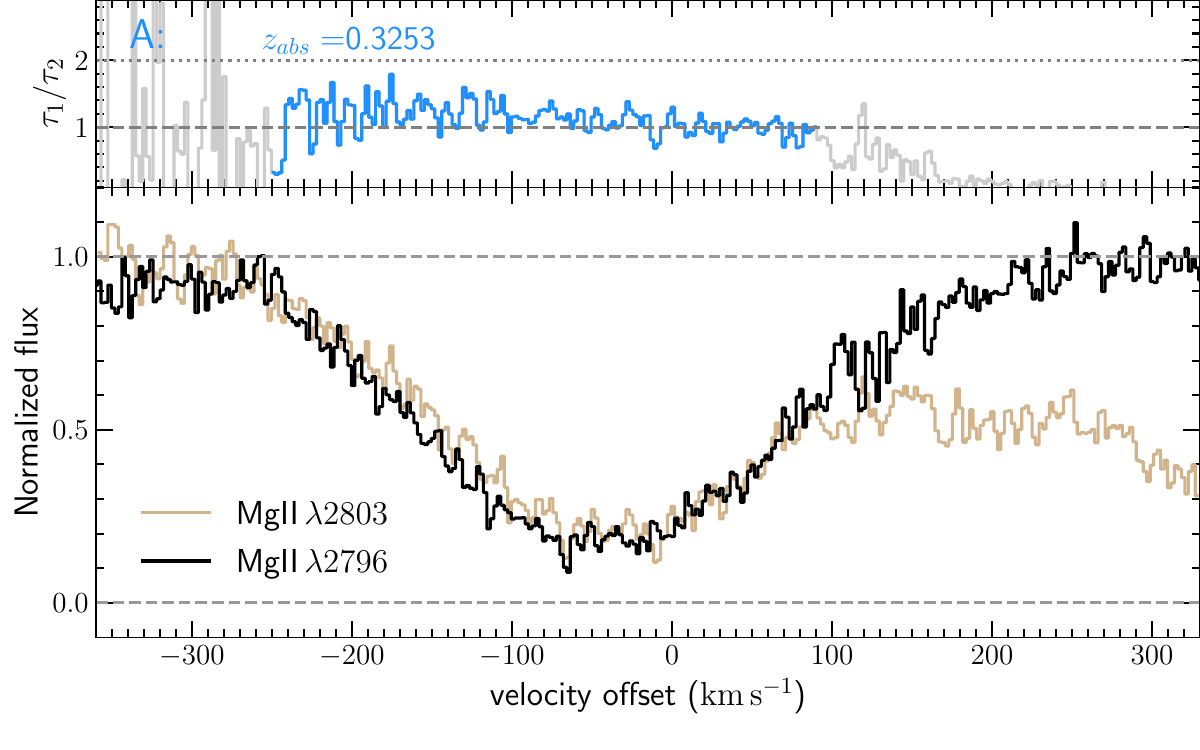}
\includegraphics[trim={0.0cm 0.0cm 0.0cm 0.0cm},clip,width=\columnwidth]{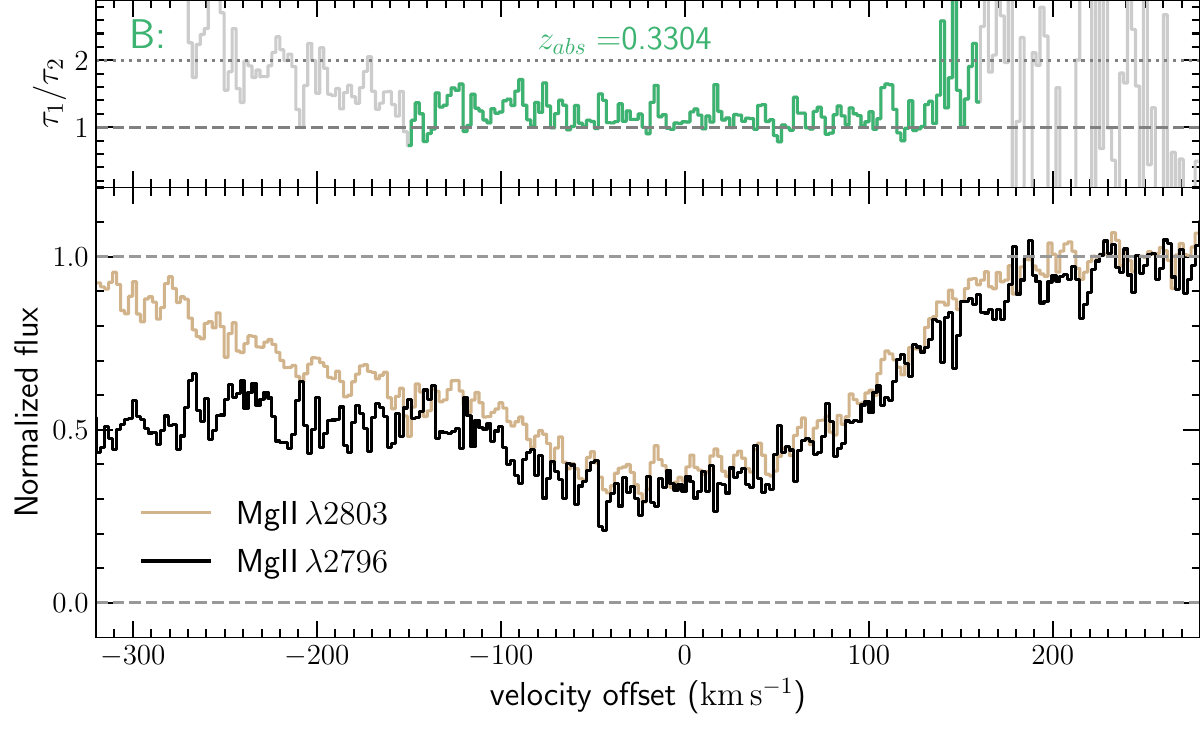}
\includegraphics[trim={0.0cm 0.6cm 0.0cm 0.0cm},clip,width=\columnwidth]{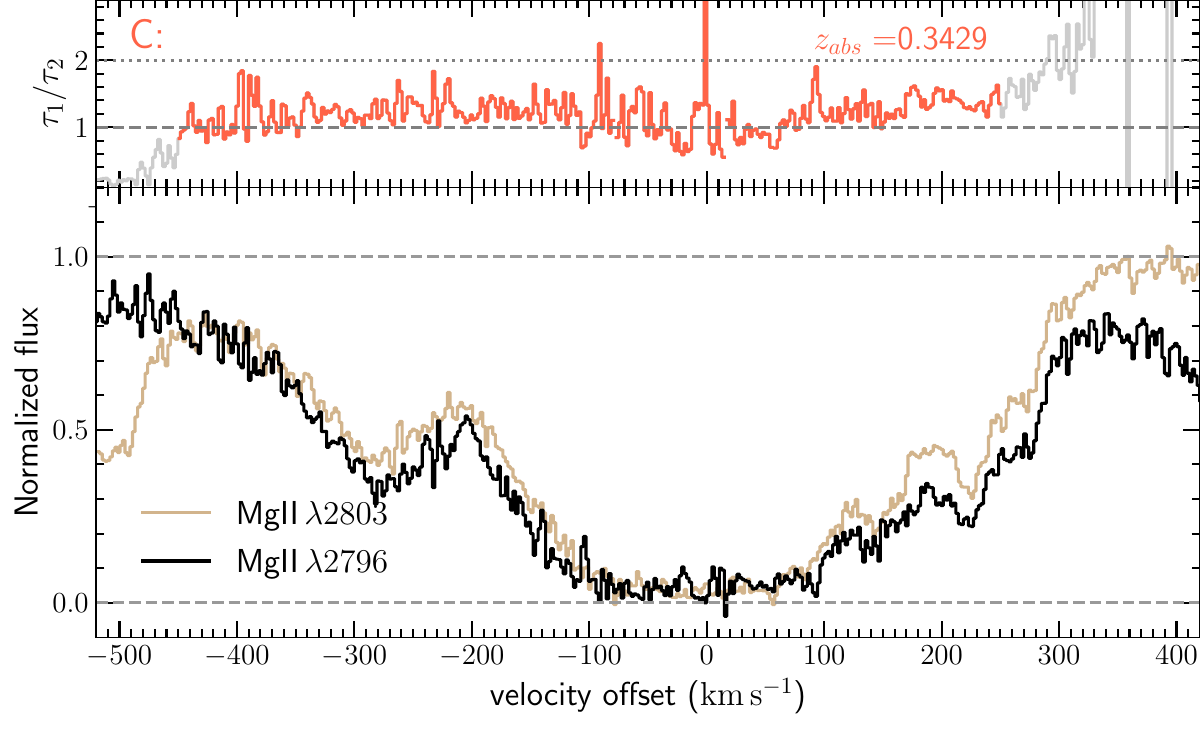}
\caption{{\sl Top to Bottom}: \MgII\ in absorption-line complexes $A$, $B$, and $C$, at $z_{\rm abs}=0.32531$, 0.33043, and 0.34292,
respectively. {\sl Upper insets in each panel:} Apparent optical-depth ratio of the \MgII\ doublet. The regions where the lines are not
apparently blended are highlighted in colour (blue, green, or red). The dashed and dotted horizontal lines correspond to the ratio expected
in the cases of complete line saturation ($\tau_1/\tau_2 =1$) and optically-thin lines ($\tau_1/\tau_2 =2$), respectively.
\label{fig:aod_MgII}}
\end{figure}

Singly-ionized magnesium exhibits an overall similar velocity structure as \CaII, but much stronger lines, hence additional components are detected in
the \MgII\ profiles. Since the \MgII\ lines are saturated (see Fig.~\ref{fig:overview_MgII_and_CaII}), it would not be reasonable to fit such
complex profiles as the velocity decomposition would be unreliable and solutions degenerate. However, we can obtain qualitative results by comparing
the apparent optical-depth ratios of the \MgII\ doublet lines, $\tau_1/\tau_2$ (where $\tau_1$ and $\tau_2$ are the apparent optical depths
in \MgII\,$\lambda2796$ and \MgII\,$\lambda2803$, respectively). This is shown in Fig.~\ref{fig:aod_MgII} for all three absorption-lines complexes, $A$,
$B$, and $C$. {When the absorber fully covers the
continuum source, the apparent optical-depth ratio of the \MgII\ doublet is expected to be $\tau_1/\tau_2 = f_1 \lambda_1 / f_2 \lambda_2 \approx 2$ (where $f_i$ and $\lambda_i$
are the line oscillator strengths and wavelengths, respectively), unless the lines are not fully saturated.} One can see in Fig.~\ref{fig:aod_MgII} that in our case $\tau_1/\tau_2$ is close to unity along the entire profiles,
which, together with {\sl seemingly} saturated profiles, is evidence for partial flux covering. In the case of fully-saturated line profiles,
partial covering factors ($C_f$) can be roughly determined as $C_f \approx 1 - e^{-\tau_1}$. Therefore, a value of $\tau_1 /\tau_2$ close to unity even
in the line wings (where the optical depths $\tau_{1,2} < 1$) indicates that partial covering is likely changing through the profile, which may additionally complicate line-profile fitting. Using the flux residuals observed at the bottom of the profiles (where $\tau_1/\tau_2 \approx 1$), we derive upper limits on $C_f$ of
$\sim 0.82$, 0.68, and 0.96, in complexes $A$, $B$, and $C$, respectively. We also note that the \MgII\,$\lambda\lambda2796,2803$ lines are not blended
with each other, with the exception of the far wings of \MgII\,$\lambda2803$ and \MgII\,$\lambda2796$ in complexes $A$ and $B$, respectively. In addition,
the velocity differences between complexes $A$, $B$, and $C$, and the weaker complex at $z\approx 0.3357$, do not correspond to any of the strong high-ionization line-doublet splitting, i.e., \SiIV, \CIV, \NV, nor \OVI. Therefore, line locking \citep[see, e.g.,][]{Bowler2014} is
not clearly present in this system.

\subsection{Voigt-profile fitting of \CaII, \MgI, \HeIs, \FeII, and \MnII}
\label{sect:fits}

We performed simultaneous fits to \CaII, \MgI, \HeIs, \FeII\ and \MnII\ absorption lines using multiple-component Voigt profiles. While even
\CaII\,$\lambda\lambda3934,3969$, \MgI\,$\lambda2852$, and \HeIs\ ($\lambda\lambda\lambda2945,3188,3889$) lines are located in spectral regions of
high S/N, the weakness of some of the velocity components prevents us from fitting the lines individually. Additionally, \FeII\ and \MnII\ lines are
significantly blended with each other, not only between components of a given complex ($A$, $B$, or $C$) but also between components pertaining
to different complexes. Therefore, to obtain internally consistent fits, we tied the Doppler parameters in each component assuming them to be equal
for each species. This implicitly assumes that turbulent broadening dominates over thermal broadening (micro-turbulence assumption), which is
reasonable for the wide (FWHM~$>15$~km\,s$^{-1}$) profiles of this system.

As we mentioned, absorption lines from \FeII\ and \MnII\ in the UVES spectrum display a high degree of mutual blending and complexity, therefore,
in order to remove possible degeneracies, we assumed that the \FeII\ levels are populated by collision with electrons (as argued for the majority
of previously-studied FeLoBALs \citep{Korista2008,Dunn2010,Bautista2010,Byun2022b}. This assumption also minimizes the number of independent
variables in the analysis. Thus, for each velocity component, the column densities of \FeII\ levels are set by the total \FeII\ column density, the
electron density, and the temperature. The data for the strengths of collisions with electrons were taken from the {CHIANTI} 9.0.1
database \citep{Dere2019} and the atomic data from the NIST database. 
We did not find any data for the collisional excitation of \MnII\ levels. Therefore, we could not consider its excitation together with \FeII\ and
hence we derived the column densities of the \MnII\ and \MnIIs\ levels, independently from \FeII. The atomic data for \HeIs\ and \CaII\ were
taken from \citet{Drake07} and \citet{Safronova2011}, respectively. For lines from excited levels of \FeII\ and \MnII, we used the data from
\citet{Nave2013}, \citet{Schnabel2004}, and \citet{Kling2000}, respectively. For other species, we used the atomic data compiled by \citet{Morton2003}.

Similar to the analysis of \MgII\ lines in Sect.~\ref{sect:mgii}, the apparent optical depth of the \CaII\ (as well as \FeII) lines indicates
partial covering.
Therefore, we need to include partial covering in the line-profile fitting procedure and for this we used the simple model proposed
by \citeauthor{Barlow1997} (\citeyear{Barlow1997}; see also \citealt{Balashev2011}). Within this model, it is assumed that the velocity
components with non-unity covering factors spatially overlap. If it were not the case, this would introduce additional covering factors to describe mutual overlapping (see, e.g., the discussion in \citealt{Ishita2021}). When many components are intertwined in wavelength space, this requires a significant increase in the number of independent variables for the analysis (up to the factorial of $n$, where $n$ is the total number of components). This complicates the analysis and makes the derived results ambiguous. Therefore, we made no attempt here to include mutual covering in the fitting procedure but rather tied the covering factors of all the components within the same complex (i.e., $A$, $B$, or $C$) to be the same. The model employed here therefore can
only provide coarse estimates of the covering factors.

The likelihood function was constructed using visually-identified regions of the spectrum that are associated with the lines to be fitted assuming
a normal distribution of the pixel uncertainties. To obtain the posterior probability functions on the fit parameters (i.e., column densities, redshifts,
Doppler parameters, and covering factors), we used a Bayesian approach with affine-invariant sampler \citep{Goodman2010}.
We used flat priors on redshifts, Doppler parameters, covering factors, and logarithms of column densities. For the electron temperature (which
is relevant for lines from the excited \FeII\ levels), we used a Gaussian prior of $\log T_e=4.2\pm 0.5$, corresponding to the typical electron
temperatures of a fully-ionized medium, where excited \FeII\ levels are highly populated \citep[e.g.,][]{Korista2008}. 
The sampling was performed on the cluster running $\approx100$ processes in parallel (using several hundred walkers) which typically took a few days until convergence.
While this approach allows us to constrain the full shape of the posterior distribution function for each parameter, in the following we report the
fit results in a standard way. The point and interval estimates correspond to the maximum posterior probability and the 0.683 credible intervals
obtained from 1D marginalized posterior distribution functions.

\begin{figure*}
\centering
\includegraphics[trim={0.0cm 0.0cm 0.0cm 0.0cm},clip,width=0.32\textwidth]{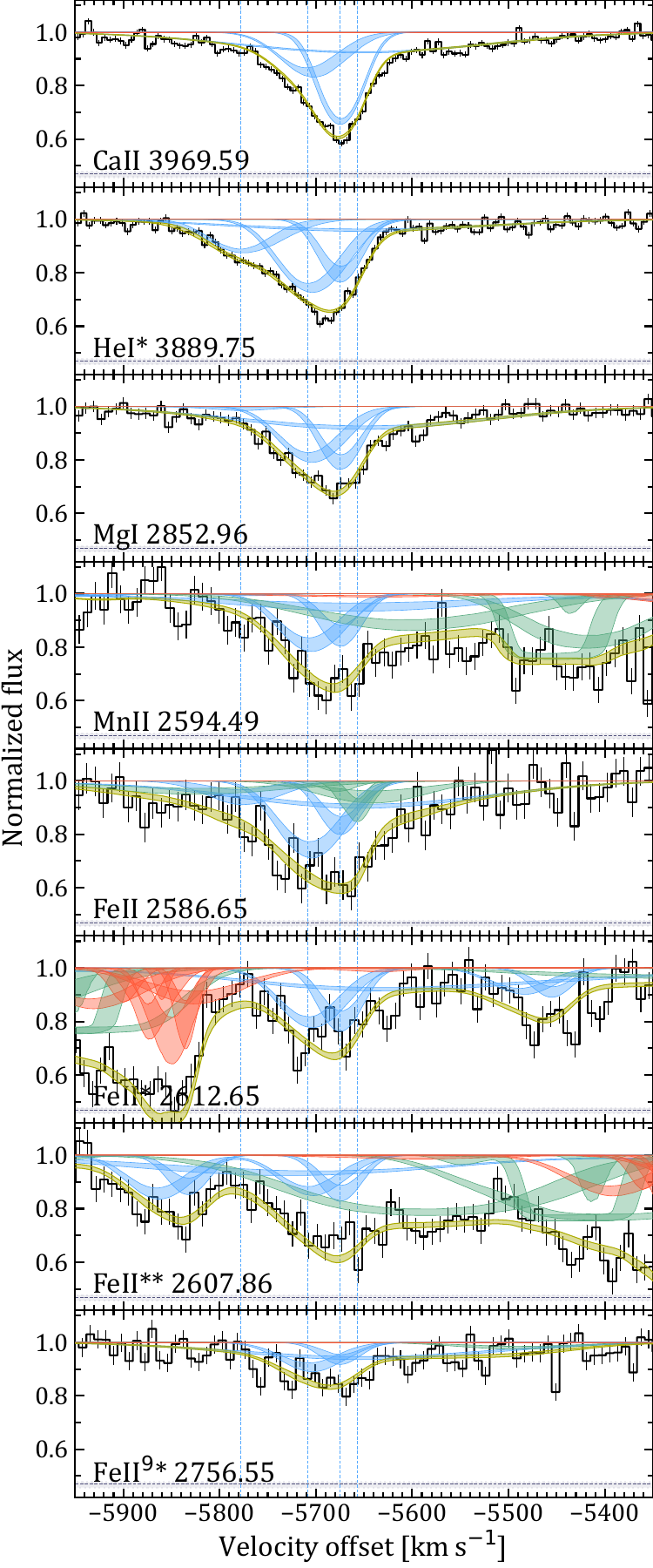}
\includegraphics[trim={0.0cm 0.0cm 0.0cm 0.0cm},clip,width=0.32\textwidth]{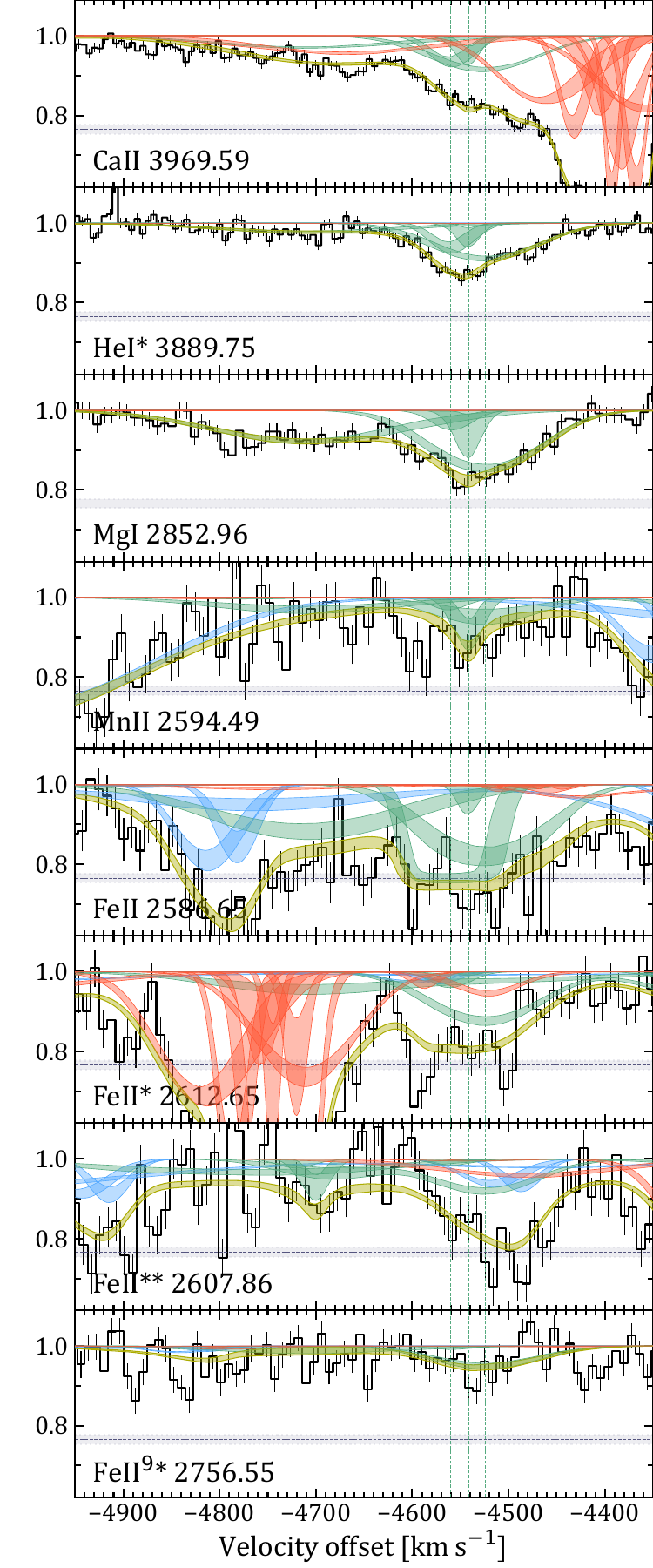}
\includegraphics[trim={0.0cm 0.0cm 0.0cm 0.0cm},clip,width=0.32\textwidth]{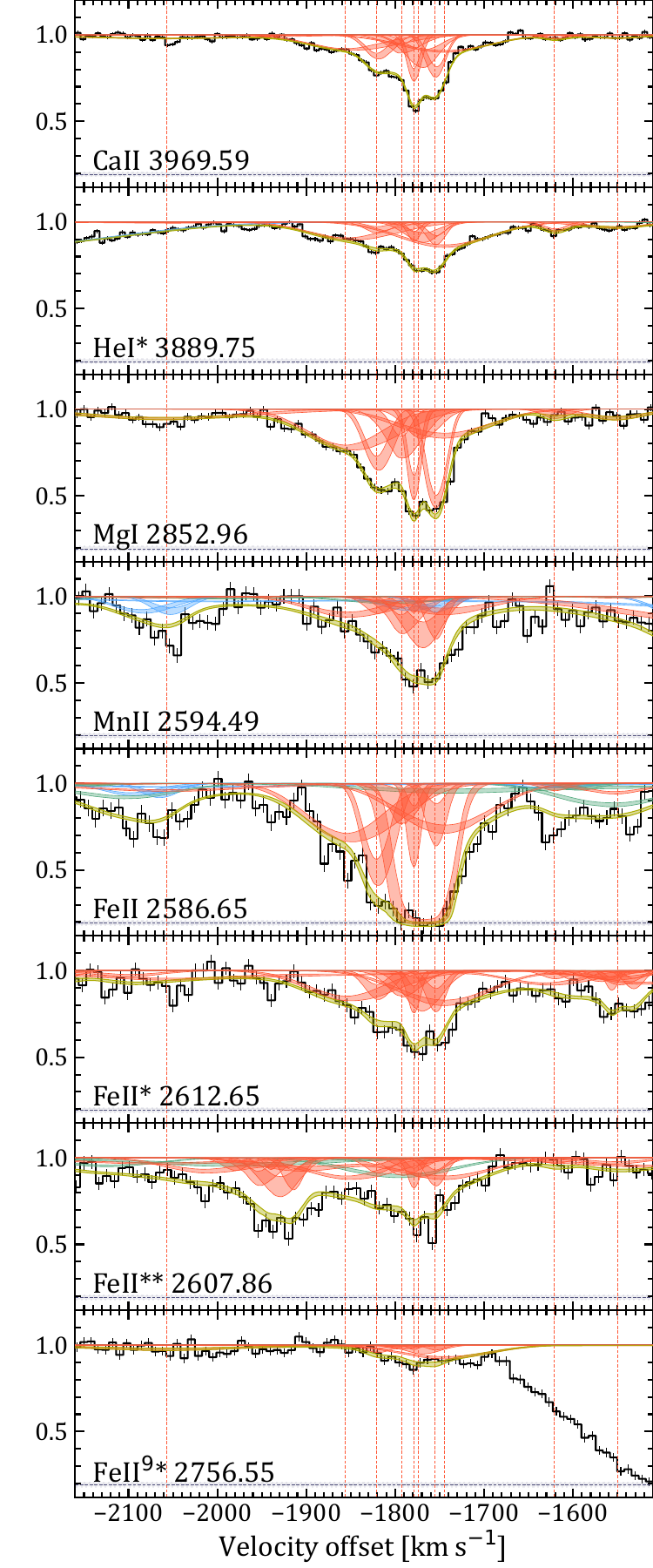}
\caption{{\sl Left to Right}: Voigt-profile fits to selected \CaII, \HeIs, \MgI, \MnII, and \FeII\ absorption lines in complexes $A$ (Left), $B$ (Middle), and $C$ (Right),
at $z_{\rm abs}=0.3253$, 0.3304, and 0.3429, respectively, towards \qso. The coloured stripes show a 0.683 credible interval of the line profiles sampled from posterior probability distributions of fitting parameters. The yellow represents the total line profile, while the blue, green, and red lines indicate individual components from complexes $A$, $B$, and $C$, respectively. The vertical lines show the positions of each component. Horizontal dashed lines and their surrounding grey areas indicate the extent of partial covering determined by fitting each clump independently with its own covering factor. The spectrum was rebinned to 0.1~\AA\ scale for presentation purposes. Note the different y-axis scaling in each column. The original spectrum and all absorption-line profiles are displayed in Figs.~\ref{fig:CaII_MgI_HeI}, \ref{fig:FeII_low_1}, \ref{fig:FeII_low_2}, \ref{fig:FeII_high}, and \ref{fig:MnII}.
\label{fig:fit_overview}
}
\end{figure*}

The results of Voigt-profile fitting are given in Table~\ref{tab:fit_results} and the modeled line profiles are briefly shown in Fig.~\ref{fig:fit_overview}, and fully displayed in Figs.~\ref{fig:CaII_MgI_HeI}, \ref{fig:FeII_low_1}, \ref{fig:FeII_low_2},
\ref{fig:FeII_high}, and \ref{fig:MnII}. The derived Doppler parameters span a wide range, from several up to two hundred km\,s$^{-1}$. The largest Doppler parameters found here should however be considered as upper limits only due to our inability to unambiguously resolve the line
profiles in individual components owing to insufficient spectrum quality and few available transitions. While the column-density ratios between components are not drastically varied some trends likely appear.
For example, the components in complex $A$ have a much larger \CaII-to-\MgI\ column-density ratio than the components in complexes $B$ or $C$. This likely indicates that the physical conditions vary from one complex to the other. Alternatively, this may indicate that the species under study are not co-spatial, which would weaken the assumption of identical Doppler parameters and location in velocity space for the considered species. Therefore, the derived uncertainties on the column densities should be considered with caution as they only describe uncertainties in a statistical sense. In complexes $A$, $B$, and $C$, the covering factors, $C_f$, are measured to be $0.53^{+0.01}_{-0.01}$, $0.24^{+0.01}_{-0.01}$, and $0.81^{+0.01}_{-0.01}$, respectively, where again the quoted uncertainties should be considered with caution given the assumptions discussed above.
The covering factors are mainly constrained by the \CaII\ and \FeII\ lines since the \HeIs\ and \MnII\ lines are weak (and hence are less sensitive to partial covering) and \MgI\ exhibits a single line. Therefore, the \MgI\ column densities reported in Table~\ref{tab:fit_results} are only reliable if the \MgI-bearing gas is co-spatial with \CaII.
One should also note that the covering factors derived here are smaller than those found for \MgII\ in Sect.~\ref{sect:mgii}. This indicates that the spatial extent of the \MgII-bearing gas is larger than that of \CaII\ and \FeII.

\begin{figure}
\centering
\includegraphics[trim={0.0cm 0.4cm 0.0cm 0.0cm},clip,width=\columnwidth]{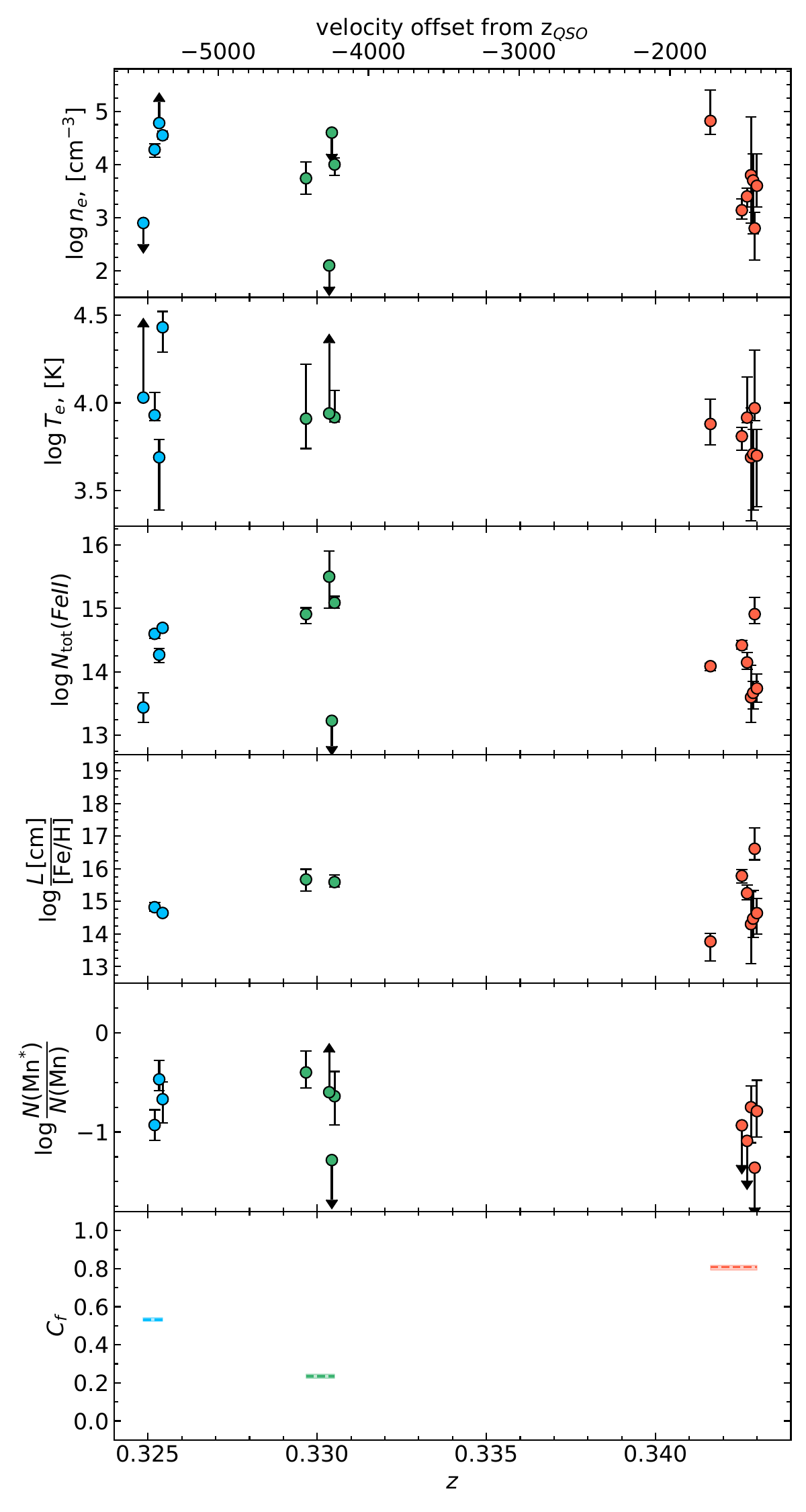}
\caption{Comparison of physical parameters in velocity complexes $A$, $B$, and $C$ (in blue, green, and red, respectively) towards \qso. The values are taken from Table~\ref{tab:fit_results}, except $\rm [Fe/H]$, which is the Fe gas-phase abundance relative to the solar. For presentation purposes, we only show the values that are reasonably well-constrained. In the lower panel, larger values of $C_f$ correspond to larger covering factors.
\label{fig:res_J1652}}
\end{figure}

In Fig.~\ref{fig:res_J1652}, we plot the physical parameters derived using the lines from excited levels of \FeII\ for velocity complexes $A$, $B$, and $C$. We found that the electron densities in different components lie in the range $n_e\approx [10^2; 10^5]$~cm$^{-3}$, which is expected given the detection of
highly-excited levels (with energies $\lesssim 10^4$~cm$^{-1}$). Interestingly, the electron densities in complex $A$ are found to be systematically higher than in $B$ or $C$. This suggests that complex $A$ is located closer to the central engine in a harsher environment, which in turn is in agreement with
the assumption that these complexes are produced in a decelerated-wind medium. {We however note, that the line profiles are quite complex and therefore exact velocity decomposition is quite complicated in this system and hence our solution is not necessary to be unique.}

The electron temperature is found to be in the wide range $T_{\rm e}\approx 10^{3.5}-10^{4.5}$~K, close to the observationally motivated chosen priors. 
This is not surprising since collisional population is less sensitive to the temperature than to the number density itself. Using the inferred electron densities (representing the number density, since the excited \FeII\ originates from ionized gas \citep[see, e.g.,][]{Korista2008}, and the total column density of \FeII\, one can derive the longitudinal extent of the absorbing clouds associated with each component, which we found to also span a wide range. When inferring this value, one should take into account the Fe gas abundances. If we assume a solar \FeII\ abundance, we get characteristic 
values of about $10^{15}$ and $10^{15.5}\,\rm cm$ for complexes $A$ and $B$, respectively, and a wide range of values for complex $C$. These values should be considered as lower limits only, since neither the metallicity nor the Fe depletion, nor the \FeII\ ionization correction, are known. Indeed, the depletion of Fe is much less than $<1$ and a very sensitive function of metallicity, and \FeII\ can be a subdominant form of Fe, even where excited \FeII\ levels are populated \citep[see, e.g.,][]{Korista2008}.  
The ratio of \MnIIs\ to \MnII\ column densities was found to be around $0.1-0.3$, similar to \FeII, where complex $A$ exhibits slightly higher
excitation than complex $C$ (\MnIIs\ is very unconstrained in complex $B$). If fitted individually, the covering factors derived here for each complex were found to be consistent between \CaII\ and \FeII, which indicates that \FeII\ and \CaII-bearing clouds likely have similar spatial extents.

\setlength{\tabcolsep}{1pt}
\renewcommand{\arraystretch}{1.3}
\begin{table*}
\caption{Results of simultaneous fits (shown in Figs.~\ref{fig:fit_overview}, \ref{fig:CaII_MgI_HeI}, \ref{fig:FeII_low_1}, \ref{fig:FeII_low_2}, \ref{fig:FeII_high}, and \ref{fig:MnII}) to absorption lines from \FeII, \CaII, \HeIs, \MgI, and \MnII\ in the FeLoBAL towards \qso. 
}\label{tab:fit_results}
\begin{tabular}{ccccccccccccc}
\hline
\hline
Comp. & z & $\Delta$v$^{\dagger}$ & b & $\log n$ & $\log T$ & $\log N_{\rm tot}$(\FeII) & $\log N$(CaII) & $\log N$(HeI$^{*}$) & $\log N$(MgI) & $\log N$(MnII) & $\log N$(MnII$^{*}$) & $C_{f}$$^{\rm \ddagger}$ \\
      &               & [km~s$^{-1}$]      & [km~s$^{-1}$] & [cm$^{-3}$] & [K]        &                           &             &                  &     \\
\hline
$A, 1$ & $0.324862(^{+23}_{-15})$ & -5778 & $46^{+7}_{-5}$ & $<2.9$ & $>4.0$ & $13.44^{+0.23}_{-0.24}$ & $<10.6$ & $13.48^{+0.08}_{-0.04}$ & $<11.1$ & $<11.3$ & $<11.9$ & $0.531^{+0.012}_{-0.008}$ \\
$A, 2$ & $0.325195(^{+14}_{-20})$ & -5704 & $44.3^{+1.8}_{-2.4}$ & $4.28^{+0.11}_{-0.14}$ & $3.9^{+0.1}_{-0.1}$ & $14.60^{+0.05}_{-0.07}$ & $12.90^{+0.05}_{-0.08}$ & $13.90^{+0.05}_{-0.05}$ & $12.39^{+0.07}_{-0.05}$ & $13.26^{+0.07}_{-0.08}$ & $12.33^{+0.13}_{-0.14}$ &  "  \\
$A, 3$ & $0.3253310(^{+30}_{-30})$ & -5674 & $29.0^{+0.8}_{-1.2}$ & $>4.8$ & $3.7^{+0.1}_{-0.3}$ & $14.27^{+0.10}_{-0.12}$ & $13.18^{+0.04}_{-0.04}$ & $13.59^{+0.10}_{-0.09}$ & $12.29^{+0.06}_{-0.10}$ & $13.04^{+0.10}_{-0.18}$ & $12.57^{+0.06}_{-0.06}$ &  "  \\
$A, 4$ & $0.325432(^{+17}_{-17})$ & -5651 & $181^{+3}_{-7}$ & $4.55^{+0.08}_{-0.07}$ & $4.4^{+0.1}_{-0.2}$ & $14.69^{+0.05}_{-0.03}$ & $13.154^{+0.019}_{-0.011}$ & $13.59^{+0.04}_{-0.04}$ & $12.57^{+0.04}_{-0.04}$ & $13.23^{+0.12}_{-0.14}$ & $12.56^{+0.10}_{-0.21}$ &  "  \\
\hline
$B, 5$ & $0.329670(^{+30}_{-40})$ & -4710 & $125^{+14}_{-9}$ & $3.7^{+0.3}_{-0.3}$ & $3.9^{+0.3}_{-0.2}$ & $14.91^{+0.10}_{-0.15}$ & $12.97^{+0.04}_{-0.06}$ & $13.55^{+0.07}_{-0.07}$ & $12.83^{+0.05}_{-0.06}$ & $13.29^{+0.10}_{-0.19}$ & $12.89^{+0.10}_{-0.12}$ & $0.235^{+0.011}_{-0.012}$ \\
$B, 6$ & $0.330355(^{+18}_{-21})$ & -4559 & $27^{+4}_{-4}$ & $<2.1$ & $>3.9$ & $15.5^{+0.4}_{-0.5}$ & $12.72^{+0.11}_{-0.15}$ & $13.41^{+0.11}_{-0.16}$ & $<12.1$ & $<12.9$ & $12.36^{+0.17}_{-0.23}$ &  "  \\
$B, 7$ & $0.330433(^{+12}_{-13})$ & -4542 & $15^{+4}_{-4}$ & $<4.6$ & $4.1^{+0.6}_{-0.3}$ & $<13.2$ & $12.39^{+0.18}_{-0.26}$ & $12.9^{+0.3}_{-0.3}$ & $<12.4$ & $12.97^{+0.24}_{-0.32}$ & $<11.6$ &  "  \\
$B, 8$ & $0.330513(^{+13}_{-17})$ & -4524 & $72^{+4}_{-3}$ & $4.0^{+0.1}_{-0.2}$ & $3.92^{+0.15}_{-0.03}$ & $15.09^{+0.10}_{-0.09}$ & $13.23^{+0.04}_{-0.05}$ & $13.98^{+0.03}_{-0.07}$ & $12.97^{+0.03}_{-0.08}$ & $13.27^{+0.15}_{-0.20}$ & $12.63^{+0.15}_{-0.25}$ &  "  \\
\hline
$C, 9$ & $0.341623(^{+30}_{-30})$ & -2059 & $140^{+17}_{-29}$ & $4.8^{+0.6}_{-0.3}$ & $3.9^{+0.2}_{-0.1}$ & $14.09^{+0.05}_{-0.07}$ & $12.30^{+0.05}_{-0.08}$ & $<12.7$ & $12.07^{+0.08}_{-0.06}$ & $<12.5$ & $12.35^{+0.09}_{-0.13}$ & $0.807^{+0.010}_{-0.014}$ \\
$C, 10$ & $0.342551(^{+32}_{-28})$ & -1853 & $62^{+6}_{-7}$ & $3.14^{+0.21}_{-0.17}$ & $3.8^{+0.1}_{-0.1}$ & $14.42^{+0.08}_{-0.06}$ & $12.56^{+0.06}_{-0.06}$ & $13.35^{+0.05}_{-0.07}$ & $12.42^{+0.05}_{-0.08}$ & $12.94^{+0.07}_{-0.10}$ & $<12$ &  "  \\
$C, 11$ & $0.342705(^{+6}_{-15})$ & -1819 & $20.2^{+1.5}_{-3.2}$ & $3.40^{+0.16}_{-0.20}$ & $3.9^{+0.2}_{-0.1}$ & $14.15^{+0.16}_{-0.11}$ & $12.29^{+0.05}_{-0.14}$ & $12.52^{+0.20}_{-0.18}$ & $12.15^{+0.06}_{-0.14}$ & $12.27^{+0.24}_{-0.38}$ & $<11$ &  "  \\
$C, 12$ & $0.342821(^{+25}_{-15})$ & -1793 & $14^{+11}_{-3}$ & $3.8^{+1.1}_{-0.9}$ & $3.7^{+0.3}_{-0.3}$ & $13.6^{+0.5}_{-0.4}$ & $12.05^{+0.22}_{-0.21}$ & $<12.3$ & $11.93^{+0.25}_{-0.21}$ & $12.59^{+0.28}_{-0.16}$ & $11.84^{+0.14}_{-0.22}$ &  " \\
$C, 13$ & $0.3428840(^{+30}_{-30})$ & -1779 & $8.9^{+1.3}_{-1.2}$ & $3.7^{+0.5}_{-1.0}$ & $3.7^{+0.2}_{-0.3}$ & $13.67^{+0.18}_{-0.25}$ & $12.25^{+0.11}_{-0.12}$ & $12.42^{+0.22}_{-0.12}$ & $12.06^{+0.09}_{-0.13}$ & $<12.1$ & $<11.1$ &  "  \\
$C, 14$ & $0.342930(^{+14}_{-14})$ & -1769 & $25^{+4}_{-4}$ & $2.8^{+0.3}_{-0.6}$ & $4.0^{+0.3}_{-0.1}$ & $14.91^{+0.26}_{-0.15}$ & $12.48^{+0.10}_{-0.19}$ & $12.99^{+0.09}_{-0.16}$ & $<12.1$ & $13.03^{+0.11}_{-0.27}$ & $<11.6$ &  "  \\
$C, 15$ & $0.342996(^{+4}_{-5})$ & -1754 & $14.2^{+1.1}_{-1.0}$ & $3.6^{+0.6}_{-0.4}$ & $3.7^{+0.2}_{-0.3}$ & $13.74^{+0.23}_{-0.22}$ & $12.39^{+0.09}_{-0.08}$ & $12.75^{+0.12}_{-0.12}$ & $12.34^{+0.05}_{-0.08}$ & $12.60^{+0.18}_{-0.27}$ & $11.81^{+0.16}_{-0.19}$ &  "  \\
$C, 16$ & $0.3430353(^{+59}_{-29})$ & -1745 & $68.6^{+3.2}_{-2.4}$ & $4.18^{+0.05}_{-0.07}$ & $4.0^{+0.1}_{-0.1}$ & $14.54^{+0.04}_{-0.03}$ & $12.67^{+0.02}_{-0.02}$ & $13.55^{+0.03}_{-0.02}$ & $12.27^{+0.05}_{-0.04}$ & $12.88^{+0.09}_{-0.11}$ & $12.15^{+0.14}_{-0.21}$ &  "  \\
$C, 17$ & $0.343601(^{+10}_{-7})$ & -1620 & $19^{+4}_{-3}$ & $5.2^{+1.3}_{-1.0}$ & $4.2^{+0.5}_{-0.3}$ & $13.16^{+0.24}_{-0.22}$ & $11.48^{+0.08}_{-0.10}$ & $12.53^{+0.09}_{-0.10}$ & $<11.4$ & $<11.9$ & $<11.9$ &  "  \\
$C, 18$ & $0.343902(^{+16}_{-20})$ & -1553 & $47^{+8}_{-6}$ & $4.5^{+0.5}_{-0.4}$ & $4.2^{+0.3}_{-0.5}$ & $13.71^{+0.07}_{-0.13}$ & $11.79^{+0.08}_{-0.04}$ & $12.75^{+0.06}_{-0.09}$ & $11.63^{+0.09}_{-0.09}$ & $<11.9$ & $<11.9$ &  "  \\
\hline
\end{tabular}
\begin{tablenotes}
\item $^{\rm \dagger}$ relative to $z_{\rm sys}=0.3509$ (as determined in Sect.~\ref{sect:redshift}).
\item $^{\rm \ddagger}$ covering factor
\end{tablenotes}
\end{table*}

\subsubsection{On the possibility of UV pumping}
\label{sec:pumping}
We tried to model the excitation of the observed \FeII\ levels by UV pumping instead of collisions with electrons since the UV flux can be
very high for gas in the vicinity of the central engine. To do this, we used the data of transition probabilities from the NIST database to
calculate the excitation through UV pumping. We note that UV pumping can easily be incorporated into the fit only in
the optically-thin regime (corresponding to $\log N(\text{Fe\,{\sc ii}})\lesssim 13$), which is not the case for most components.
Therefore, a complex multiple-zone excitation model should be implemented that takes into account radiative transfer fully, since
the UV excitation at some position depends on the line profiles, and therefore on the excitation balance at the regions closer to the
radiation source. This implementation however is impractical for such a complex line-profile fitting as we have towards \qso. However, we
can draw qualitative conclusions from the following two limiting cases: the optically-thin limit, or assuming constant dilution of the
excitation using typically-observed column densities.

\begin{figure*}
\centering
\includegraphics[trim={0.0cm 0.0cm 0.0cm 0.0cm},clip,width=\textwidth]{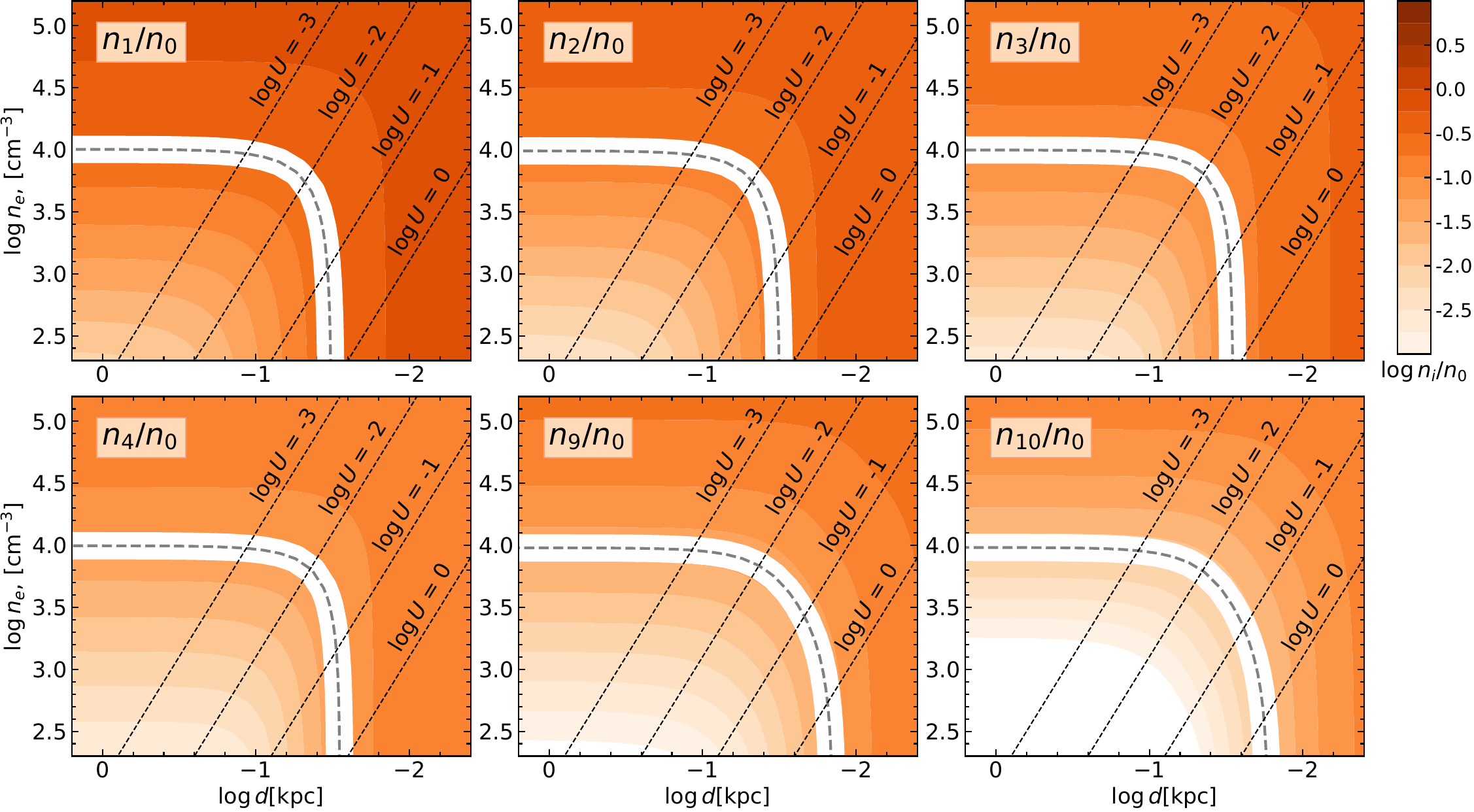}
\caption{Excitation of \FeII\ levels as a function of electron density, $n_e$, and distance to the central engine, $d$. The latter is calculated
from the measured photometric flux of \qso\ assuming a typical quasar spectral shape. Calculations were performed in the optically-thin limit,
hence any constraint on the distance that can be obtained if UV pumping dominates the \FeII\ excitation must be considered as an upper limit. Additionally, we plot lines of constant ionizing parameter, $U$, calculated by scaling the UV flux and assuming bluewards of the Lyman limit an AGN spectrum with power-law shape with index $-1.2$. One can see that, for UV pumping to dominate the excitation of \FeII\ levels, the ionizing
parameter must be larger than 0.01.
\label{fig:FeII_rad}}
\end{figure*}

In the optically-thin case, we found that UV pumping does not provide satisfactory fits, since it cannot reproduce the observed
excitation of the \FeII\ levels as well 
as collisional excitation can do. To qualitatively illustrate this, we plot in
Fig.~\ref{fig:FeII_rad} the excitation of \FeII\ levels as a function of electron density and UV field. The UV field is expressed in terms
of distance to the central engine estimated from the observed $r$-band \qso\ magnitude of $17.0$ assuming a typical quasar spectral
shape. 
One can see that the \FeII\ excitation described by typically-estimated electron densities (for example, at
$n_e\approx 10^4\,\rm cm^{-3}$, which are shown by a dashed line in each panel of Fig.~\ref{fig:FeII_rad}), corresponds to roughly
a factor of two difference in distance (hence a factor of four difference in UV flux) that is needed to describe the fine-structure
levels of the ground term ($\rm3d^{6}4s\,^6D$, representing excited levels from 1$^{\rm th}$ to 4$^{\rm th}$) and the second excited ($\rm3d^{6}4s\,^4D$, representing excited levels from 9$^{\rm th}$ to 12$^{\rm th}$) \FeII\ term. 
From this diagram, one can see that if the excitation is described by UV pumping the absorbing gas must be located a few tens of parsec 
away from the central engine. However, if UV pumping dominates the excitation of the \FeII\ levels, this results in very large values of the ionization parameter, $\log U\gtrsim -1$, that is difficult to reconcile with the survival of \FeII\ and other associated low-ionization species (e.g. \CaII\ and \NaI). 
Additionally, the observed column density of \HeIs\ indicates $\log U\approx -3$ for dense gas, which is discussed later in Sect.~\ref{sect:modeling}.
All this indicates that UV pumping is unlikely to be the dominant excitation process at play in the gas. Using calculations in the optically-thick regime and assuming $\log N(\text{Fe\,{\sc ii}})=14$ we found less disagreement in the UV fluxes required to populate the low and high \FeII\ levels. However, the optically-thick case implies smaller distances to the central engine and hence even higher
ionization parameters, in comparison to the optically-thin case.

\section{Photo-ionization model}
\label{sect:modeling}

We modelled the abundance of \HeIs\ to estimate the physical conditions of the gas associated with this small-detached narrow/low-ionization
BAL towards \qso. As discussed by, e.g., \citet{Arav2001} and \citet{Korista2008}, the meta-stable $2\,^3S$ level of \HeIs\ is
mostly populated through \HeII\ recombination and depopulated by radiative transition and collisional de-excitation. Therefore,
\HeIs\ predominantly originates from a layer of ionized gas where helium is in the form of \HeII\ and $n_e\approx n_{\rm H}$, and the \HeIs\
column density is sensitive to the number density and ionizing flux. In that sense, \HeIs\ is an exceptional diagnostic of the physical
conditions, almost independent of metallicity and depletion, unlike other metals. This is particularly relevant for the FeLoBAL under
study since we can measure neither the abundance of \HI\ nor the total abundance of any metal (i.e., only the singly-ionized state of each
species is constrained), hence we have neither a measurement of metallicity nor metal depletion in this system. This limitation is also an issue for most of the previously studied FeLoBAL systems, and the assumption of a particular metallicity value can significantly affect the physical conditions derived from the photo-ionization modelling \citep[e.g.,][]{Byun2022a}.

We used the latest public version of the \Cloudy\ software package C17.02 \citep{Ferland2017} to model a slab of gas in the vicinity of
the AGN. Our basic setup is a cloud of constant density that is illuminated on one side by a strong UV field with a typical AGN spectrum. We assumed a metallicity of $0.3$ solar\footnote{We took solar relative abundances of the metals, i.e. we did not use any depletion factor. While in the most known FeLoBALs the metallicity is found to be around solar value \citep[e.g.][]{Arav2001,Aoki2011,Byun2022a}, our chosen value 0.3 mimics possible Fe depletion, which typically large (up to 2 dex) at solar metallicity.}, a characteristic value for such clouds, but we also checked that the exact metallicity value has little impact on the derived \HeIs\ column densities. Temperature balance was calculated self-consistently. As a stopping criterion, we used a total \FeII\ column density of $10^{15}$\,cm$^{-2}$ corresponding to the higher end of \FeII\ column densities observed within the FeLoBAL components.

\begin{figure}
\centering
\includegraphics[trim={0.0cm 0.0cm 0.0cm 0.0cm},clip,width=0.5\textwidth]{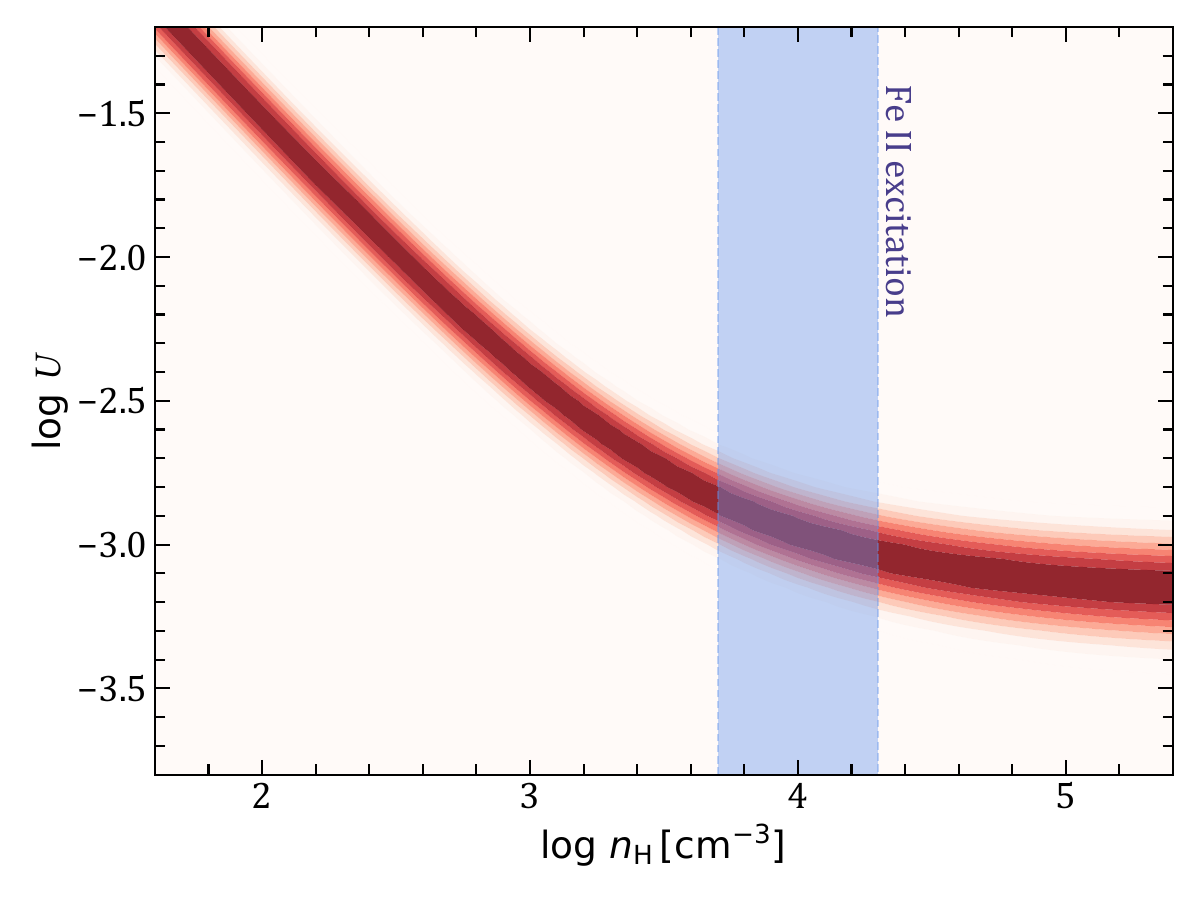}
\caption{Constraints on the ionization parameter and number density from \Cloudy\ photo-ionization modeling of \HeIs\ in the
FeLoBAL towards \qso. The region displayed in brown shows the estimated probability density function constructed by comparing
the calculated \HeIs\ column densities with the observed typical value of $\log N=13.7\pm 0.1$ (see Table~\ref{tab:fit_results}). The region displayed in
blue shows the range of $n_{\rm H}$ values favoured by the measured \FeII\ excitation.
\label{fig:cloudy}}
\end{figure}

We ran a grid of photo-ionization models by varying two main parameters: the number density and the ionization parameter, within the
ranges of $\log [1; 6]$ and $\log [-4; 0]$, respectively. Fig.~\ref{fig:cloudy} shows the constraints on each parameter derived from the comparison
of the modelled \HeIs\ column density with the fiducial value of $13.7\pm 0.1$, typical of high column-density components. One can see
that the modelling provides estimates on the number of ionizing photons of $\log (U n_{\rm H})\approx 0.5$
for $\log n_{\rm H}\lesssim 4$, and $\log U\approx -3$ for $\log n_{\rm H} \gtrsim 4$. The latter solution to preferred value by the excitation
of \FeII, which provides an independent constraint on $\log n_{\rm H}\approx 4$. Since the excited \FeII\ levels predominately arise from
the ionized medium (as they are excited by collision with electrons; see also \citealt{Korista2008}), this suggests that
$n_{\rm e}\approx n_{\rm H}$ and hence likely, the \HeIs\ abundance provides an estimate of $\log U \sim -3$. While the exact value of
$\log U$ for each component depends on the observed \HeIs\ column density, we refrain from using the latter because with such a modelling we cannot be confident regarding constrained \FeII\ column densities, as mutual covering may impact the derived column densities. We also checked that the \Cloudy\ modelling roughly reproduces the \CaII, \MgI, and \MnII\ column densities. However, as we mentioned above, using the abundance of these species is limited due to unconstrained total metallicities and depletion patterns.

\section{Discussion}
\label{sect:discussion}

\subsection{Case of FeLoBAL towards \qsot}
\label{sect:q2359}

One of the most comprehensive studies so far of a FeLoBAL by means of high-resolution spectroscopy concerns \qsot\ \citep{Arav2001}.
A broad and deep VLT/UVES spectrum of this quasar allowed \citet{Arav2008} to detect \FeII\ lines up to the $8^{\rm th}$ excited level (excitation energy of 7955\,cm$^{-1}$)
above the ground state, and to constrain the physical conditions in the associated medium \citep{Korista2008}.
A sophisticated fitting model was used, describing partial covering of the source by the absorbing clouds based on a power-law distribution
\citep[see][]{Arav2008}. In our present study, we used a model of uniform partial covering instead, which is dictated by the observation of complex
mutually-blended absorption-line profiles. In the case of \qsot, the line velocity structure is simpler, with only a few visually-distinct velocity
components, and the lines are not significantly saturated. This allows one to independently constrain the column density in each \FeII\ level and for
each component, and then describe the population of \FeII\ levels to constrain the excitation mechanisms. Additionally, due to its higher redshift,
the FeLoBAL towards \qsot\ allows one to constrain the column density of intermediate \FeII\ levels with energies between 1872$-$3117\,cm$^{-1}$
(corresponding to the $\rm a^4F$ term of the 3d$^7$ configuration). We found that these levels are important to disentangle between radiative pumping and
collisions with electrons. Certainly, such an independent determination of \FeII\ column densities is more robust than tightening them assuming a dominant
excitation mechanism, as we did for \qso. Therefore, we endeavoured to test our procedure by also fitting the FeLoBAL towards \qsot\ using the spectrum taken from the SQUAD database \citep{Murphy2019}.

To fit the \FeII\ lines, we used an eight-component fit, out of which four components exhibit higher excitation and are mutually blended, and
four components show a low level of excitation with only the first few excited levels above the ground state detected \cite{Bautista2010}. We used the same Doppler parameters
for all of the \FeII\ levels in a given component. We added an independent covering factor to each component, yet tying the covering factor to be equal
in two closely-associated weak components at $z=0.8611$. The quasar continuum was re-constructed locally by interpolating the spectrum free from
absorption features. We note that the continuum placement may be important for weak lines since the line profiles are fairly broad. The line-fitting
procedure we used is the same as described for \qso\ (see Sects.~\ref{sect:fits}). The fitting results are listed in
Table~\ref{tab:2359_res} and the modelled line profiles are shown in the Appendix, in Figs.~\ref{fig:J2359_FeII_low_1} to \ref{fig:J2359_FeII_high_2}.
In comparison to the study of \citet{Arav2008} and \citet{Korista2008}, we were able to identify a larger number of \FeII\ levels, up to the
$12^{\rm th}$ excited level of \FeII\ (excitation energy of $\sim$8850\,cm$^{-1}$) above the ground state. 

\setlength{\tabcolsep}{2pt}
\begin{table*}
\caption{Results of simultaneous fits to absorption lines from \FeII\ ground state and excited levels in the FeLoBAL towards \qsot.}\label{tab:2359_res}
\begin{tabular}{lcccccccc}
\hline
\hline
Comp. & \#1 & \#2 & \#3 & \#4 & \#5 & \#6 & \#7 & \#8 \\
\hline
$z_{\rm abs}$ & $0.859761(^{+10}_{-16})$ & $0.859922(^{+7}_{-7})$ & $0.8599318(^{+14}_{-9})$ & $0.860189(^{+6}_{-5})$ & $0.861056(^{+11}_{-6})$ & $0.861127(^{+26}_{-15})$ & $0.8618265(^{+17}_{-7})$ & $0.8626427(^{+13}_{-10})$ \\
$\Delta v^{\rm \dagger}$ [km~s$^{-1}$] & -1325 & -1299 & -1298 & -1256 & -1117 & -1105 & -992 & -861 \\
$b$                [km~s$^{-1}$] & $85.3^{+3.4}_{-1.9}$ & $47.5^{+2.0}_{-1.8}$ & $13.5^{+0.3}_{-0.3}$ & $15.9^{+1.0}_{-1.2}$ & $7.5^{+3.2}_{-1.5}$ & $20.1^{+2.5}_{-3.1}$ & $5.7^{+0.4}_{-0.4}$ & $7.5^{+0.3}_{-0.3}$ \\
\hline
$\log N$(\FeII,g.s.) & $15.39^{+0.12}_{-0.06}$ & $14.39^{+0.07}_{-0.04}$ & $13.63^{+0.03}_{-0.03}$ & $13.40^{+0.09}_{-0.17}$ & $13.45^{+0.18}_{-0.18}$ & $13.58^{+0.13}_{-0.17}$ & $13.01^{+0.07}_{-0.10}$ & $13.24^{+0.05}_{-0.04}$ \\
$\log N$(\FeII,j1)   & $14.76^{+0.06}_{-0.07}$ & $13.91^{+0.07}_{-0.05}$ & $13.21^{+0.04}_{-0.04}$ & $13.01^{+0.13}_{-0.15}$ & $12.53^{+0.17}_{-0.25}$ & $12.56^{+0.26}_{-0.55}$ & $12.01^{+0.12}_{-0.12}$ & $12.48^{+0.03}_{-0.06}$ \\
$\log N$(\FeII,j2)   & $14.50^{+0.06}_{-0.07}$ & $13.78^{+0.06}_{-0.09}$ & $13.10^{+0.04}_{-0.04}$ & $12.81^{+0.14}_{-0.13}$ & $12.63^{+0.12}_{-0.29}$ & $12.60^{+0.25}_{-0.58}$ & $11.68^{+0.27}_{-0.39}$ & $12.12^{+0.09}_{-0.09}$ \\
$\log N$(\FeII,j3)   & $14.64^{+0.05}_{-0.08}$ & $13.43^{+0.10}_{-0.05}$ & $12.97^{+0.04}_{-0.03}$ & $12.73^{+0.10}_{-0.13}$ & $12.40^{+0.21}_{-0.42}$ & $12.48^{+0.21}_{-0.96}$ & $11.55^{+0.26}_{-0.33}$ & $12.20^{+0.06}_{-0.09}$ \\
$\log N$(\FeII,j4)   & $14.25^{+0.06}_{-0.08}$ & $13.13^{+0.13}_{-0.14}$ & $12.67^{+0.07}_{-0.03}$ & $<12.5$ & $<12.2$ & $<12.1$ & $<11.7$ & $<11.5$ \\
$\log N$(\FeII,j5)   & $15.33^{+0.10}_{-0.08}$ & $14.26^{+0.10}_{-0.08}$ & $13.70^{+0.05}_{-0.04}$ & $13.22^{+0.32}_{-0.83}$ & $<13.5$ & $<13.7$ & $12.64^{+0.19}_{-0.30}$ & $<12.6$ \\
$\log N$(\FeII,j6)   & $14.92^{+0.09}_{-0.09}$ & $<13.5$ & $13.29^{+0.07}_{-0.09}$ & $<13.0$ & ... & ... & ... & ... \\
$\log N$(\FeII,j7)   & $14.63^{+0.20}_{-0.16}$ & $13.93^{+0.19}_{-0.31}$ & $<13.1$ & $<12.8$ & ... & ... & ... & ... \\
$\log N$(\FeII,j8)   & $14.88^{+0.08}_{-0.11}$ & $<12.80$ & $<13.0$ & $<12.2$ & ... & ... & ... & ... \\
$\log N$(\FeII,j9)   & $14.02^{+0.12}_{-0.08}$ & $<12.8$ & $12.55^{+0.05}_{-0.05}$ & $12.51^{+0.20}_{-0.30}$ & ... & ... & ... & ... \\
$\log N$(\FeII,j10)  & $14.20^{+0.08}_{-0.06}$ & $12.99^{+0.16}_{-0.12}$ & $12.16^{+0.09}_{-0.12}$ & $<11.8$ & ... & ... & ... & ... \\
$\log N$(\FeII,j11)  & $13.75^{+0.12}_{-0.15}$ & $<12.2$ & $12.00^{+0.09}_{-0.15}$ & $<11.3$ & ... & ... & ... & ... \\
$\log N$(\FeII,j12)  & $13.46^{+0.09}_{-0.11}$ & $<12.3$ & $11.50^{+0.24}_{-0.54}$ & $12.87^{+0.30}_{-0.44}$ & ... & ... & ... & ... \\
\hline
$\log N_{\rm tot}$ & $15.95^{+0.04}_{-0.04}$ & $14.86^{+0.05}_{-0.05}$ & $14.23^{+0.03}_{-0.04}$ & $13.80^{+0.12}_{-0.14}$ & $13.58^{+0.22}_{-0.15}$ & $13.83^{+0.12}_{-0.21}$ & $13.20^{+0.08}_{-0.10}$ & $13.39^{+0.05}_{-0.05}$ \\
\hline
$C_f$$^\ddagger$ & $0.050^{+0.002}_{-0.005}$ & $0.13^{+0.01}_{-0.01}$ & $0.36^{+0.01}_{-0.02}$ & $0.14^{+0.03}_{-0.03}$ & $0.11^{+0.01}_{-0.01}$ & $0.11^{+0.01}_{-0.01}$ & $0.32^{+0.04}_{-0.04}$ & $0.31^{+0.02}_{-0.02}$ \\
\hline
\end{tabular}
\begin{tablenotes}
\item $^\dagger$ relative to $z_{\rm sys}=0.868$ \citep[as reported by][]{Brotherton2001}. 
\item $^\ddagger$ covering factor 
\end{tablenotes}
\end{table*}

We used the measured population of the \FeII\ levels to constrain the physical conditions in the absorbing medium. We used the same model
as for \qso, where we considered the competition between collisions (with electrons) and radiative excitation (by UV pumping).
In Figs.~\ref{fig:2359_exc} and \ref{fig:2359_exc_2}, we show the excitation diagrams of the different \FeII\ levels together with the constrained
region of the parameter space of physical conditions, i.e., electron density, $n_{\rm e}$, and UV field strength. As in Sect.~\ref{sec:pumping}, the UV field  is expressed in
terms of distance to the central engine as estimated from the $r$-band \qsot\ magnitude of $\sim 17.0$ assuming a typical quasar spectral shape \citep{Selsing2016}.
The 2D posterior parameter distributions were obtained using the likelihood function assumed to be a product of individual likelihoods of the
comparison between the modelled and measured \FeII$^{i*}$/\FeII\ ratios\footnote{The latter ratios were obtained directly from the fitting procedure.
During the fit of the \FeII\ lines, the column densities of the different energy levels of \FeII\ were fitting parameters. However, since we used
the MCMC method to sample the posterior distributions, we could simply obtain the posterior distributions for the aforementioned ratios.}. We also assumed
flat priors on $\log n_e$ and $\log d$ emulating a wide prior distribution for these two parameters. In Figs.~\ref{fig:2359_exc} and \ref{fig:2359_exc_2},
one can see that in the components that have a large enough number of measured \FeII\ levels, the excitation is better reproduced by the model
of collisions with electrons only, and therefore these components provide robust constraints on the electron density. For the components at
$\Delta v=-1325$, $-1299$, and $-1298$~km\,s$^{-1}$, we found $n_e$ to be in the range between $5\times 10^3$ and $3\times 10^4$~cm$^{-3}$. For the other components,
the constrained posterior does not indicate a preferable source of excitation leaving the physical conditions poorly constrained over a wide range.
However, for the weaker and most-redshifted components at $\Delta v>-1200$~km\,s$^{-1}$, the excitation of the \FeII\ levels is less and if it were
dominated by collisions this would result in a significantly lower electron density, $\log n_e \sim 3.5$, than for the main components. In the three bluest components, where $n_e$ is robustly measured, we can get an upper limit on the ionization parameter which was found to be
$\log U \lesssim -3$, $-1.5$, and $-2.5$ for the components at $\Delta v=-1325$, $-1299$, and $-1298$~km\,s$^{-1}$, respectively. These values are reasonably
consistent with the constraint of $\log U\sim -2.4$ obtained from photo-ionization modelling of this system by \citet{Korista2008}. It is also in line
with the characteristic values we obtained in Sects.~\ref{sect:analysis} and \ref{sect:modeling} in the case of the FeLoBAL towards \qso.

\begin{figure*}
\begin{tabular}{c}
    \includegraphics[trim={0.0cm 0.0cm 0.0cm 0.0cm},clip,width=1.0\textwidth]{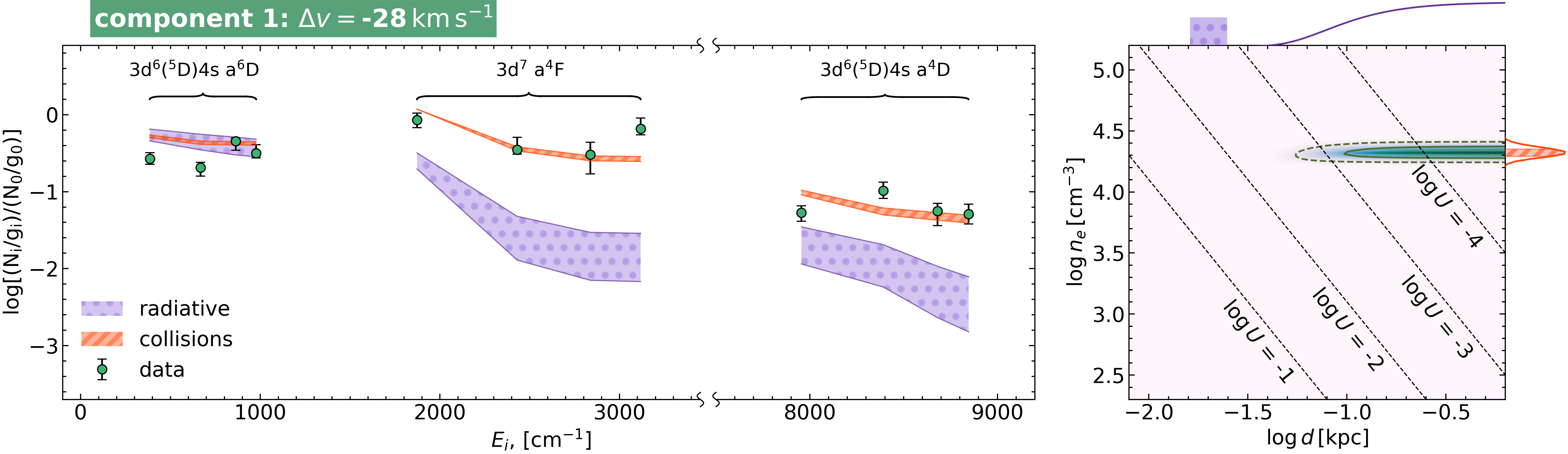} \\  
    \includegraphics[trim={0.0cm 0.0cm 0.0cm 0.0cm},clip,width=1.0\textwidth]{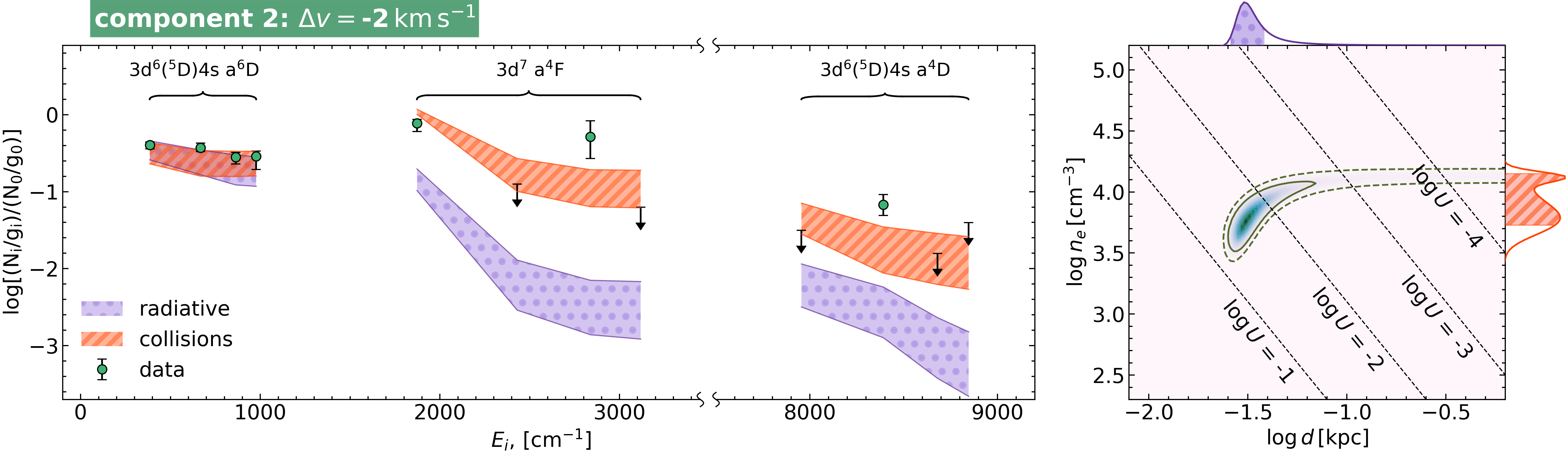} \\
    \includegraphics[trim={0.0cm 0.0cm 0.0cm 0.0cm},clip,width=1.0\textwidth]{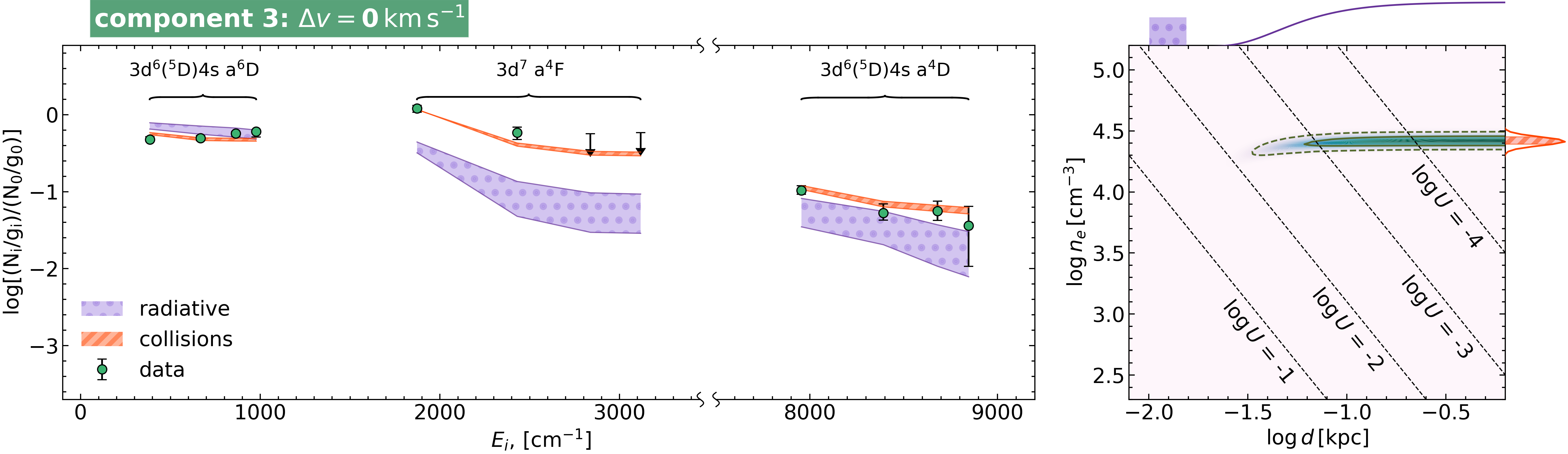} \\  
    \includegraphics[trim={0.0cm 0.0cm 0.0cm 0.0cm},clip,width=1.0\textwidth]{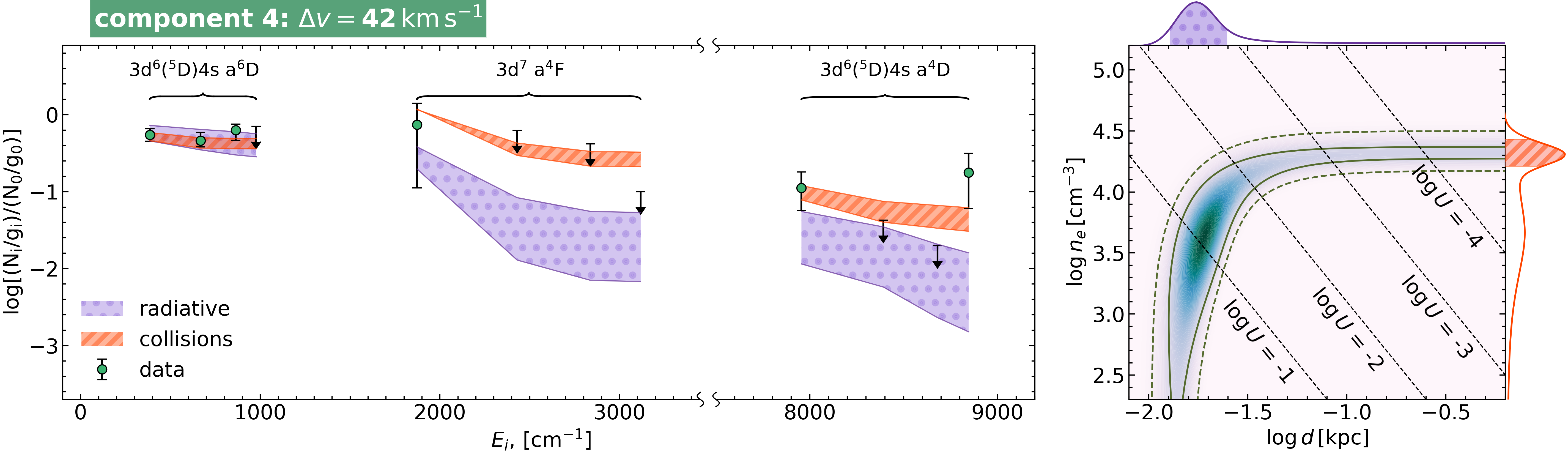} \\
\end{tabular}
\caption{{\sl Left panels:} Excitation diagrams of \FeII\ levels in the FeLoBAL towards \qsot. {The y-axes indicate the ratio of the measured column density divided by statistical weight of i-th level to the ground level, while the x-axes provide the energy of the levels.} The text in the upper part of each panel indicates
the level terms. Each panel corresponds to a given velocity component as indicated in a green box on top of the panel. {\sl Right panels:}
Constrained physical conditions using the excitation of the \FeII\ levels shown in the left panels. The solid and dashed lines correspond to the
$1\sigma$ and $2\sigma$ confidence intervals of the 2D posterior probability function, respectively. The violet and red curves on the top x-axis and
right y-axis, respectively, show the 1D marginalized probability functions. The red and violet hatched regions below them indicate the approximate
solutions where the population of the levels is dominated by collisions or radiative excitation, respectively. For illustrative purposes,
the corresponding regions of the excitation diagrams are shown in the left panels using the same colours and hatch code.
\label{fig:2359_exc}}
\end{figure*}

\begin{figure*}
\begin{tabular}{c}
    \includegraphics[trim={0.0cm 0.0cm 0.0cm 0.0cm},clip,width=1.0\textwidth]{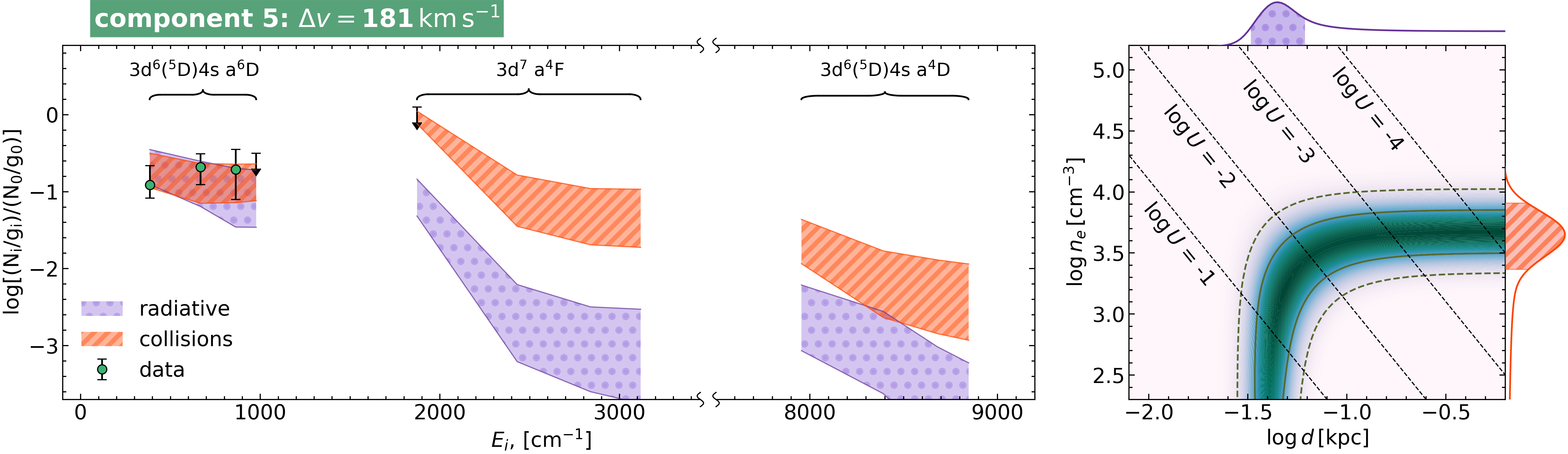} \\  
    \includegraphics[trim={0.0cm 0.0cm 0.0cm 0.0cm},clip,width=1.0\textwidth]{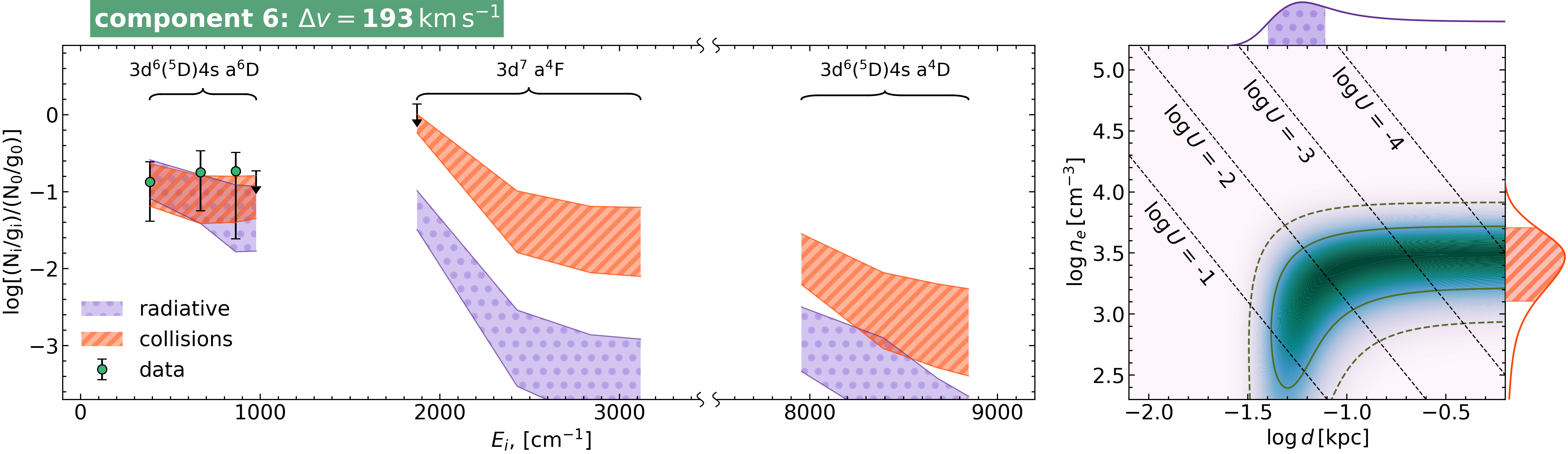} \\
    \includegraphics[trim={0.0cm 0.0cm 0.0cm 0.0cm},clip,width=1.0\textwidth]{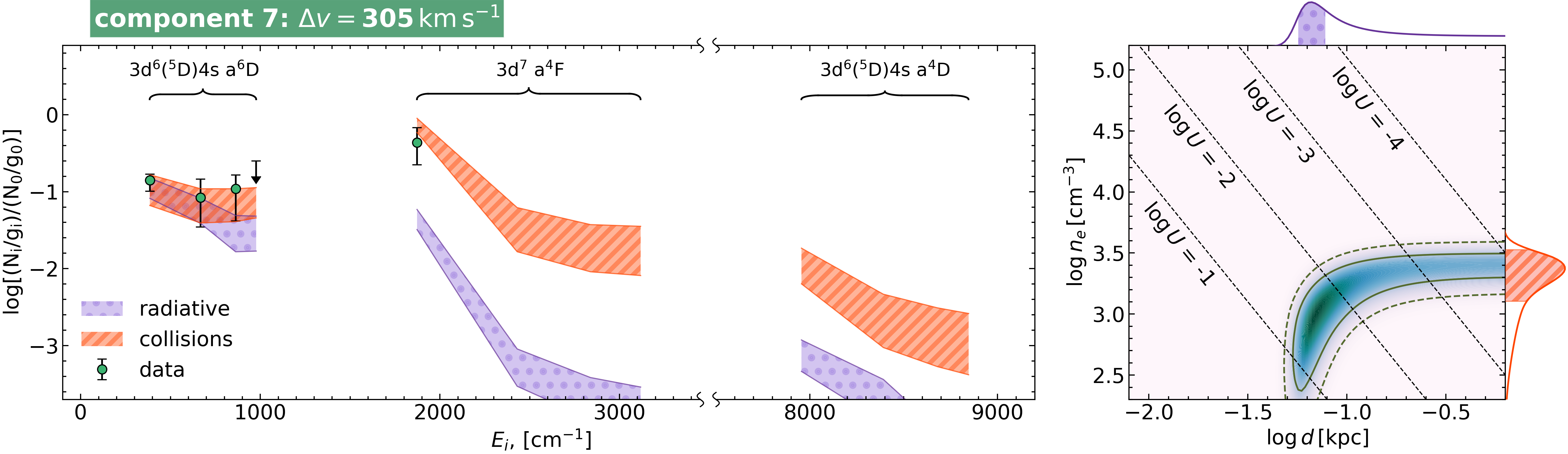} \\  
    \includegraphics[trim={0.0cm 0.0cm 0.0cm 0.0cm},clip,width=1.0\textwidth]{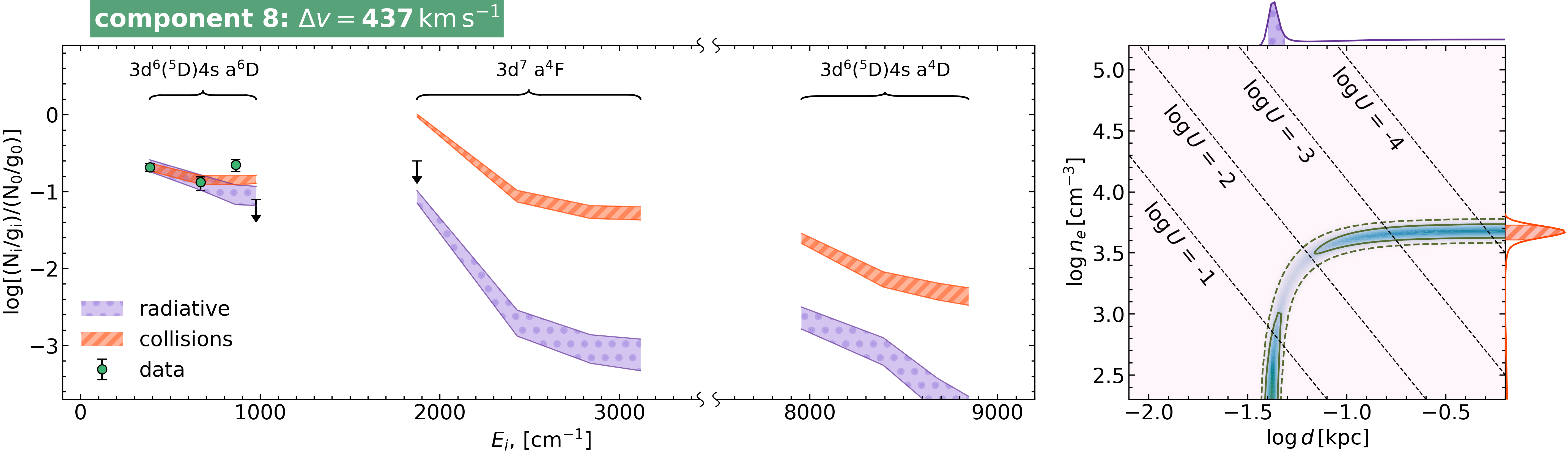} \\
\end{tabular}
\caption{Continuation of Fig.~\ref{fig:2359_exc}. 
\label{fig:2359_exc_2}}
\end{figure*}

In comparison with
\cite{Korista2008} and \cite{Bautista2010}, we derived significantly higher column densities for \FeII\ levels in all components (except at $-992$~km\,s$^{-1}$) and
total column density as well, which is dominated by the velocity components at $-1325$ and $-1299$~km\,s$^{-1}$. This discrepancy is explained by the small covering factors of these
two components, which allow the lines to be significantly saturated. However, the relative excitation of the \FeII\ levels even in these saturated components remains similar. Furthermore, the two central (in terms of apparent optical depth) components at $-1298$ and $-1256$~km\,s$^{-1}$ indicate an excitation structure
consistent with the results of \citet{Korista2008}. We note that determining the exact profile decomposition is not trivial in such systems. We attempted to increase the number of fitted velocity
components in \FeII\ lines but obtained more or less similar results, as additional components turned out to be weak if any. Moreover, the systematic uncertainty is most likely dominated by our choice of partial covering model, which is uniform with no mutual intersection (see Sect.~\ref{sect:fits}). In terms of number density, we obtained similar results
as \citet{Korista2008} who reported that $\log n_{\rm H}\sim 4.4\pm 0.1$ considering total \FeII\ column densities only and a smaller number of energy levels than we do. 

Overall, our approach (multi-component model with a uniform covering factor) provides well-consistent results with those from earlier works, especially for derived physical quantity. While FeLoBALs in most cases are quite complicated for spectral analysis, such systems as one towards \qsot\ provide an important example and testbed for the assumptions (e.g. collisionally-dominated excitation) that can ease the analysis of more complex objects. 

\subsection{Similarities and differences between FeLoBALs}

Previous studies indicate a wide range of physical conditions in FeLoBALs (and other intrinsic \FeII\ absorbers) with some kind of bimodal distribution, where part of the population is located at mild distances of $\sim 0.1-10$~kpc and has number densities of $<10^5$~cm$^{-3}$, while the second part shows more extreme properties with number densities $>10^8$~cm$^{-3}$ and distances to the nuclei down to $\sim 1$~pc. This may be in line with state-of-the-art models of AGN outflows formation \citep{Faucher-Giguere2012, Costa2020}, where the wind is a complex phenomenon that is driven by different mechanisms at different scales. 
It can be launched in the close vicinity of the accretion disk by radiative pressure and at a much larger distance by shocks, produced either by a wind from the accretion disk or the jet \cite[e.g.][]{Proga2000, Costa2020}. 

On the other hand, we note that ionization parameters around $-1..-3$ have been found for almost all detected FeLoBALs, while one would  expect a wider range of values. This can be a selection effect, where such values of the ionization parameter are favourable for FeLoBAL observation. However, since $U \propto n^{-1} d^{-2}$ the observed bimodality can be an artefact of improper constraints on the physical conditions from the modelling. 
Indeed, the modelling of FeLoBALs is quite complex and ambiguous, and there are many factors that cannot be resolved using line-of-sight observations. FeLoBALs always exhibit a multi-component structure, which makes the derivation of the physical conditions a difficult task, since several solutions are possible. A typical question arising from the photo-ionization modelling setup is what is the relative position in physical space of the "gaseous clouds" associated with each component? This impacts both the column density estimation, due to unknown mutual partial covering, and ionization properties, since the closest to AGN clouds will shield the farthest located ones. A relevant example of this situation is the FeLoBAL towards J\,104459.6$+$365605, where \FeII\ excitation and simplistic photo-ionization modelling suggest relatively low number densities, $\log n_{\rm e}<4$, and a distance of $\sim 700$~pc, while more sophisticated wind models \citep{Everett2002}, which take shielding effects into account, yield number densities $10^4$ times higher, and a distance of $\sim 4$~pc. We note however that in the latter model, the excitation of \FeII\ levels is expected to be much higher than what is observed. In that sense, the usage of the excited fine-structure levels and \HeI* may provide less degenerated constraints than the relative abundances of the ions, the latter is also suffering from complications due to the unknown metallicity and depletion pattern (see discussion in Sect.~\ref{sect:modeling}).

\subsection{Relation between FeLoBALs and other intrinsic absorbers}

It is worth mentioning that a recently-identified class of associated quasar absorbers bearing H$_2$ molecules exhibits distances to the AGN of $\sim 1-10$~kpc \citep{Noterdaeme2019,Noterdaeme2021,Noterdaeme2023}, slightly higher, but comparable to
what is derived for FeLoBALs. Interestingly, while the medium in such systems is neutral (and even at the \HI-H$_2$ transition), in contrast with FeLoBALs which arise in the ionized phase or at the boundary of the ionization
front \citep[e.g.,][]{Korista2008}, they exhibit number densities of $\gtrsim10^4$~cm$^{-3}$, similar to FeLoBALs. In the case of H$_2$-bearing systems, such number densities are required for H$_2$ to survive in the vicinity
of the AGN \citep{Noterdaeme2019} as the radiation fields are greatly enhanced. While in the case of \qso\, we were not able to get constraints on the H$_2$ column density since the lines are out of the range of the spectrum, there is no H$_2$ detection in other {FeLoBAL} so far. Additionally, presented \Cloudy\ modelling (given in Sect.~\ref{sect:modeling}) suggests that the column densities and ionization parameter are not enough in FeLoBAL towards \qso\ to expect the presence of the H$_2$ in such kind of medium.

\begin{figure}
\centering
\includegraphics[trim={0.0cm 0.0cm 0.0cm 0.0cm},clip,width=0.5\textwidth]{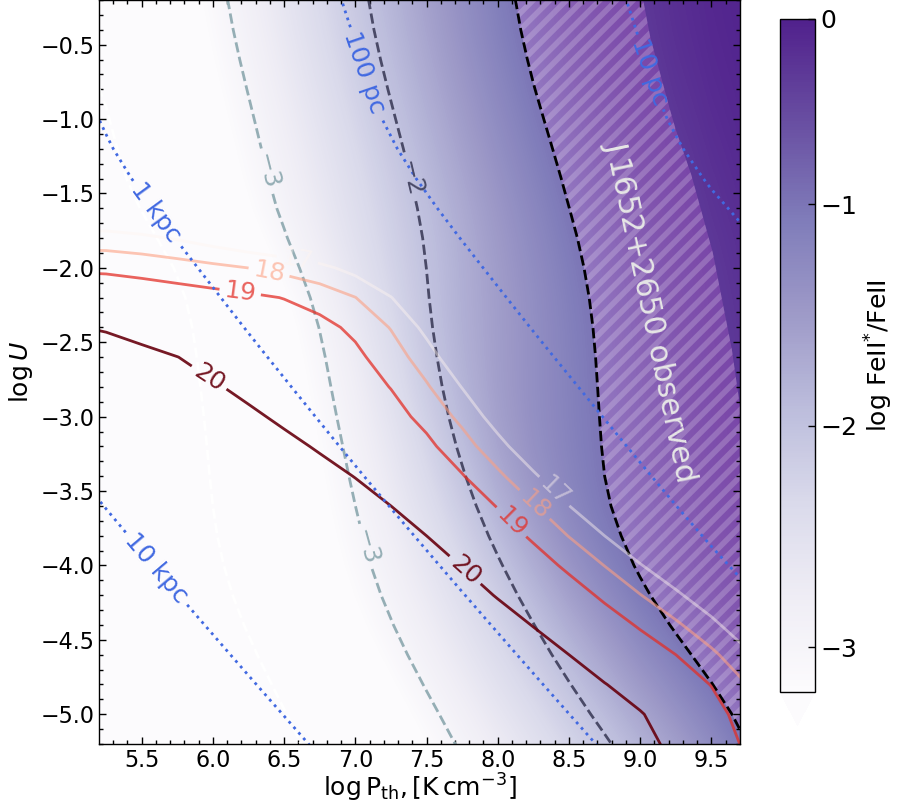}
\caption{Constraints on the ionization parameter and thermal pressure from a grid of \Cloudy\ photo-ionization models (for setup, see text). The violet colour gradient and dashed contours show the excitation of \FeII, defined as $\log N(\mbox{\FeII}^*)/ N(\mbox{\FeII})$. The characteristic range of \FeII\ excitation observed in \qso\ is shown by the hatched region. The red solid contours indicate the total H$_2$ column density, marked by values of $\log N(\rm H_2)$. The blue dotted contours show the distance to the central engine calculated using the parameters of \qso.
\label{fig:cloudy_H2}}
\end{figure}

Therefore, the difference between FeLoBALs and H$_2$-bearing systems may be related to the lower ionization parameters of the latter (akin to higher lower incident UV flux, or to their larger distances), which makes it possible for H$_2$ to survive or imply reasonable timescales to form H$_2$. To elaborate on this we ran a grid of \Cloudy\ models to see how the conditions for the presence of H$_2$ and excited \FeII\ levels compared in the physical parameter space. We considered an isobaric model of the medium with 0.3 metallicity relative to solar (we additionally scaled the Fe abundance by 0.3 to emulate typical depletion at such metallicity) exposed by the AGN-shaped radiation field and regular cosmic ray ionization rate, $2\times10^{-16}\,\rm s^{-1}$ (for atomic hydrogen). We varied the ionization parameter and thermal pressure in ranges $\log U = -6 .. 0$ (with 0.2 dex step) and $P_{\rm th} = 10^{5} .. 10^{10}\,\rm [K\,cm^{-3}]$ (with 0.5 dex step), respectively. We stopped the calculations either when the total H$_2$ or \FeII\ column densities reached characteristic values of $\log N(\rm H_2) = 20$ or $\log N (\mbox{\FeII}) = 15$, respectively. The obtained contours of \FeII\ excitation and the total H$_2$ column density are shown in Fig.~\ref{fig:cloudy_H2}. 

One can see that indeed, large H$_2$ column densities are found mostly outside the region of the parameter space where \FeII\ is highly excited, i.e. $\log \mbox{\FeII}^*/\mbox{\FeII} > -1$ (which is typical for the observed FeLoBAL systems), except only for the very high thermal pressures, $P_{\rm th} \gtrsim 10^9\,\rm K\,cm^{-3}$ and low ionization parameters $\log U\lesssim -4$. The presence of H$_2$ is mostly limited by the distances to the AGN, which in the case of \qso\ corresponds to the values of several hundreds of pc\footnote{Only a stationary cloud is considered here, i.e., we do not take into account time-dependent effects, which may be important for H$_2$ in harsh environments.}. 
In turn, the values measured \FeII$^*$/\FeII\ coupled with the constrained ionization parameter $\log U\sim -3$ points to a low molecular fraction. We note that this modelling is indicative only, and one needs to be careful when comparing the measured excitation of the \FeII\ levels with the modelled one in this simulation. Indeed, in this particular modelling case, we stopped at a relatively large total \FeII\ column density, at which we may include a significant part of the neutral medium, where \FeII\ excitation is not so high as in the ionized shell. 

{The region in $(\rm U, P_{th})$ parameter space where $\rm H_2$ is present in copious amount is adjacent to the region where \FeII\ is reasonably excited (e.g. $\log N(\mbox{\FeII}^*)/ N(\mbox{\FeII})>-1$). That indicate that we may witness the appearance of a global natural sequence among associated absorbers.}
The behaviour of this sequence is probably coupled with hydrodynamics processes in the outflowing gas, which set the thermal pressures and the sizes of the clumps and their dependence on the distance. Likely, a part of this sequence was observationally noticed by \citet{Fathivavsari2020}, regarding Coronographic \citep{Finley2013} and Ghostly \citep{Fathivavsari2017} DLAs, which also exhibit high excitation of fine-structure levels\footnote{There are examples of similar associated systems with fine-structure excitation observed in the past, e.g., \citep{Hamann2001}.}, but that do not show \FeII$^*$. The physical connection between these different classes of absorbers seems to be evident, with FeLoBALs representing predominantly gas at the ionization front, Coronographic and Ghostly DLAs being predominantly neutral, and associated H$_2$-bearing absorbers tracing the \HI-to-H$_2$ transition. Importantly, both the ionization and photo-dissociation fronts are controlled by the ratio of UV flux to number density, i.e., the ionization parameter, while the appearance of a certain class of absorber also depends on the ambient thermal pressure and the total column density of the medium.
In that sense, it will be important to observe and study systems of intermediate classes, displaying mixed properties. The search for such rare systems will be supported by the upcoming next-generation wide-field spectroscopic surveys such as 4MOST \citep{Krogager2023}, DESI \citep{DESI2023}, and WEAVE \citep{WEAVE2023}.

\section{Summary}
\label{sect:conclusions}

In this paper, we presented an analysis of the serendipitously-identified FeLoBAL system at z=0.3509 towards \qso\ performed using a high-resolution UVES spectrum. The main aim was to derive constraints on the physical conditions in the absorbing medium located near the AGN central engine. 

The absorption system consists of three kinematically-detached absorption complexes spanning -1700 to -5700 $\rm km\,s^{-1}$ relative to the QSO redshift. We detected lines profiles of \MgII, \MgI, \CaII, \HeIs, \MnII\ and \FeII. For, the latter species we detected lines from the various fine-structure levels of the ground and second excited electronic states, with energies up to $\approx8850\,\rm cm^{-1}$. The lines indicate a partial coverage of the continuum emission source, with a covering factor in the range from 0.98 to 0.2. A relatively simple kinematic structure (in comparison to the majority of known FeLoBALs) and an intermediate saturation allow us to perform joint multi-component Voigt profile fitting of the aforementioned species (except \MgII) to derive column densities in the simplistic homogeneous partial coverage assumption. Using an additional assumption during the fit, that excitation of \FeII\ levels dominated by the collisions with electrons, we obtained the constraints on the electron density in the medium to be $\sim 10^4\,\rm cm^{-3}$ with $\sim1$ dex dispersion. We also detected the lines from the first excited level of \MnII\ and constrained \MnIIs/\MnII\ column density ratios to be in the range from 0.1 to 0.5 across velocity components. However, the lack of collisional coefficients data for \MnII\ did not allow us to use \MnII\ excitation to infer the physical condition in the medium.  

Among the other elements detected in this FeLoBAL, \HeIs\ is most important since it allows obtaining constraints on the combination of ionization parameter and number density, even without measurement of the hydrogen column density, metallicity and depletion pattern, which is the case of \qso. We used \Cloudy\ code to model the characteristic column densities of \HeIs\ and obtained a value of the ionization parameter $\log U \sim -3$ assuming the number density derived from \FeII. Such values are typically measured in FeLoBAL systems, which likely represents the similarity among them, while line profiles in FeLoBALs can be drastically different. With the estimate of the UV flux from \qso, this translates to a constraint on the distance between the absorbing medium and the continuum source of $\sim 100$\,pc. 

We also discuss the connection of FeLoBAL systems with other types of intrinsic absorbers, including Coronographic and recently identified proximate H$_2$-bearing DLAs. The latter indicates a similar value of the number densities as measured in FeLoBAL, $\gtrsim10^4$\,cm$^{-3}$. Using \Cloudy\ modelling we showed that the FeLoBAL and H$_2$-bearing proximate systems located in the adjacent regions in the parameter space, representing the main global characteristics of the medium: thermal pressure and ionization parameter (or the number density and the distance to the AGN, which are mutually interconnected to them). This indicates a global natural sequence among associated absorbers, where FeLoBALs represent predominantly gas at the ionization front, Coronographic DLAs are predominantly neutral, and associated H$_2$-bearing absorbers trace the \HI-to-H$_2$ transition. This likely will be comprehensively explored with upcoming next-generation wide-field spectroscopic surveys. This shall greatly enhance our understanding of AGN feedback and cool-gas flows from the AGN central engine.

\section*{Data Availability}

The data published in this paper is available through Open Access via the ESO scientific archive, and in the SQUAD database \citep{Murphy2019}.
The reduced and co-added spectra can be shared upon request to the corresponding author.

\section*{Acknowledgements}

We thank the anonymous referee for constructive report, that significantly improved the paper. SAB acknowledges the hospitality and support from the Office for Science of the European Southern Observatory in Chile during a visit when
this project was initiated. SAB and KNT were supported by RSF grant 23-12-00166. S.L. acknowledges support by FONDECYT grant 1231187.




\bibliographystyle{mnras}
\bibliography{references}

\begin{thebibliography}{}
\makeatletter
\relax
\def\mn@urlcharsother{\let\do\@makeother \do\$\do\&\do\#\do\^\do\_\do\%\do\~}
\def\mn@doi{\begingroup\mn@urlcharsother \@ifnextchar [ {\mn@doi@}
  {\mn@doi@[]}}
\def\mn@doi@[#1]#2{\def\@tempa{#1}\ifx\@tempa\@empty \href
  {http://dx.doi.org/#2} {doi:#2}\else \href {http://dx.doi.org/#2} {#1}\fi
  \endgroup}
\def\mn@eprint#1#2{\mn@eprint@#1:#2::\@nil}
\def\mn@eprint@arXiv#1{\href {http://arxiv.org/abs/#1} {{\tt arXiv:#1}}}
\def\mn@eprint@dblp#1{\href {http://dblp.uni-trier.de/rec/bibtex/#1.xml}
  {dblp:#1}}
\def\mn@eprint@#1:#2:#3:#4\@nil{\def\@tempa {#1}\def\@tempb {#2}\def\@tempc
  {#3}\ifx \@tempc \@empty \let \@tempc \@tempb \let \@tempb \@tempa \fi \ifx
  \@tempb \@empty \def\@tempb {arXiv}\fi \@ifundefined
  {mn@eprint@\@tempb}{\@tempb:\@tempc}{\expandafter \expandafter \csname
  mn@eprint@\@tempb\endcsname \expandafter{\@tempc}}}

\bibitem[\protect\citeauthoryear{{Aoki}, {Oyabu}, {Dunn}, {Arav}, {Edmonds},
  {Korista}, {Matsuhara}  \& {Toba}}{{Aoki} et~al.}{2011}]{Aoki2011}
{Aoki} K.,  {Oyabu} S.,  {Dunn} J.~P.,  {Arav} N.,  {Edmonds} D.,  {Korista}
  K.~T.,  {Matsuhara} H.,   {Toba} Y.,  2011, \mn@doi [\pasj]
  {10.1093/pasj/63.sp2.S457}, \href
  {https://ui.adsabs.harvard.edu/abs/2011PASJ...63S.457A} {63, 457}

\bibitem[\protect\citeauthoryear{{Arav}, {Brotherton}, {Becker}, {Gregg},
  {White}, {Price}  \& {Hack}}{{Arav} et~al.}{2001}]{Arav2001}
{Arav} N.,  {Brotherton} M.~S.,  {Becker} R.~H.,  {Gregg} M.~D.,  {White}
  R.~L.,  {Price} T.,   {Hack} W.,  2001, \mn@doi [\apj] {10.1086/318244},
  \href {https://ui.adsabs.harvard.edu/abs/2001ApJ...546..140A} {546, 140}

\bibitem[\protect\citeauthoryear{{Arav}, {Moe}, {Costantini}, {Korista}, {Benn}
   \& {Ellison}}{{Arav} et~al.}{2008}]{Arav2008}
{Arav} N.,  {Moe} M.,  {Costantini} E.,  {Korista} K.~T.,  {Benn} C.,
  {Ellison} S.,  2008, \mn@doi [\apj] {10.1086/588651}, \href
  {https://ui.adsabs.harvard.edu/abs/2008ApJ...681..954A} {681, 954}

\bibitem[\protect\citeauthoryear{{Balashev}, {Petitjean}, {Ivanchik}, {Ledoux},
  {Srianand}, {Noterdaeme}  \& {Varshalovich}}{{Balashev}
  et~al.}{2011}]{Balashev2011}
{Balashev} S.~A.,  {Petitjean} P.,  {Ivanchik} A.~V.,  {Ledoux} C.,  {Srianand}
  R.,  {Noterdaeme} P.,   {Varshalovich} D.~A.,  2011, \mn@doi [\mnras]
  {10.1111/j.1365-2966.2011.19489.x}, \href
  {https://ui.adsabs.harvard.edu/abs/2011MNRAS.418..357B} {418, 357}

\bibitem[\protect\citeauthoryear{{Balashev} et~al.,}{{Balashev}
  et~al.}{2017}]{Balashev2017}
{Balashev} S.~A.,  et~al., 2017, \mn@doi [\mnras] {10.1093/mnras/stx1339},
  \href {https://ui.adsabs.harvard.edu/abs/2017MNRAS.470.2890B} {470, 2890}

\bibitem[\protect\citeauthoryear{{Balashev} et~al.,}{{Balashev}
  et~al.}{2019}]{Balashev2019}
{Balashev} S.~A.,  et~al., 2019, \mn@doi [\mnras] {10.1093/mnras/stz2707},
  \href {https://ui.adsabs.harvard.edu/abs/2019MNRAS.490.2668B} {490, 2668}

\bibitem[\protect\citeauthoryear{{Barlow} \& {Sargent}}{{Barlow} \&
  {Sargent}}{1997}]{Barlow1997}
{Barlow} T.~A.,  {Sargent} W.~L.~W.,  1997, \mn@doi [\aj] {10.1086/118239},
  \href {https://ui.adsabs.harvard.edu/abs/1997AJ....113..136B} {113, 136}

\bibitem[\protect\citeauthoryear{{Bautista}, {Dunn}, {Arav}, {Korista}, {Moe}
  \& {Benn}}{{Bautista} et~al.}{2010}]{Bautista2010}
{Bautista} M.~A.,  {Dunn} J.~P.,  {Arav} N.,  {Korista} K.~T.,  {Moe} M.,
  {Benn} C.,  2010, \mn@doi [\apj] {10.1088/0004-637X/713/1/25}, \href
  {https://ui.adsabs.harvard.edu/abs/2010ApJ...713...25B} {713, 25}

\bibitem[\protect\citeauthoryear{{Becker}, {Gregg}, {Hook}, {McMahon}, {White}
  \& {Helfand}}{{Becker} et~al.}{1997}]{Becker1997}
{Becker} R.~H.,  {Gregg} M.~D.,  {Hook} I.~M.,  {McMahon} R.~G.,  {White}
  R.~L.,   {Helfand} D.~J.,  1997, \mn@doi [\apjl] {10.1086/310594}, \href
  {https://ui.adsabs.harvard.edu/abs/1997ApJ...479L..93B} {479, L93}

\bibitem[\protect\citeauthoryear{{Becker}, {White}, {Gregg}, {Brotherton},
  {Laurent-Muehleisen}  \& {Arav}}{{Becker} et~al.}{2000}]{Becker2000}
{Becker} R.~H.,  {White} R.~L.,  {Gregg} M.~D.,  {Brotherton} M.~S.,
  {Laurent-Muehleisen} S.~A.,   {Arav} N.,  2000, \mn@doi [\apj]
  {10.1086/309099}, \href
  {https://ui.adsabs.harvard.edu/abs/2000ApJ...538...72B} {538, 72}

\bibitem[\protect\citeauthoryear{{Boroson}, {Meyers}, {Morris}  \&
  {Persson}}{{Boroson} et~al.}{1991}]{Boroson1991}
{Boroson} T.~A.,  {Meyers} K.~A.,  {Morris} S.~L.,   {Persson} S.~E.,  1991,
  \mn@doi [\apjl] {10.1086/185967}, \href
  {https://ui.adsabs.harvard.edu/abs/1991ApJ...370L..19B} {370, L19}

\bibitem[\protect\citeauthoryear{{Bowler}, {Hewett}, {Allen}  \&
  {Ferland}}{{Bowler} et~al.}{2014}]{Bowler2014}
{Bowler} R. A.~A.,  {Hewett} P.~C.,  {Allen} J.~T.,   {Ferland} G.~J.,  2014,
  \mn@doi [\mnras] {10.1093/mnras/stu1730}, \href
  {https://ui.adsabs.harvard.edu/abs/2014MNRAS.445..359B} {445, 359}

\bibitem[\protect\citeauthoryear{{Brotherton}, {Arav}, {Becker}, {Tran},
  {Gregg}, {White}, {Laurent-Muehleisen}  \& {Hack}}{{Brotherton}
  et~al.}{2001}]{Brotherton2001}
{Brotherton} M.~S.,  {Arav} N.,  {Becker} R.~H.,  {Tran} H.~D.,  {Gregg} M.~D.,
   {White} R.~L.,  {Laurent-Muehleisen} S.~A.,   {Hack} W.,  2001, \mn@doi
  [\apj] {10.1086/318243}, \href
  {https://ui.adsabs.harvard.edu/abs/2001ApJ...546..134B} {546, 134}

\bibitem[\protect\citeauthoryear{{Byun}, {Arav}  \& {Walker}}{{Byun}
  et~al.}{2022a}]{Byun2022b}
{Byun} D.,  {Arav} N.,   {Walker} A.,  2022a, \mn@doi [\mnras]
  {10.1093/mnras/stac2194}, \href
  {https://ui.adsabs.harvard.edu/abs/2022MNRAS.516..100B} {516, 100}

\bibitem[\protect\citeauthoryear{{Byun}, {Arav}  \& {Hall}}{{Byun}
  et~al.}{2022b}]{Byun2022a}
{Byun} D.,  {Arav} N.,   {Hall} P.~B.,  2022b, \mn@doi [\mnras]
  {10.1093/mnras/stac2638}, \href
  {https://ui.adsabs.harvard.edu/abs/2022MNRAS.517.1048B} {517, 1048}

\bibitem[\protect\citeauthoryear{{Byun}, {Arav}  \& {Hall}}{{Byun}
  et~al.}{2022c}]{Byun2022c}
{Byun} D.,  {Arav} N.,   {Hall} P.~B.,  2022c, \mn@doi [\apj]
  {10.3847/1538-4357/ac503d}, \href
  {https://ui.adsabs.harvard.edu/abs/2022ApJ...927..176B} {927, 176}

\bibitem[\protect\citeauthoryear{{Chaussidon} et~al.,}{{Chaussidon}
  et~al.}{2023}]{DESI2023}
{Chaussidon} E.,  et~al., 2023, \mn@doi [\apj] {10.3847/1538-4357/acb3c2},
  \href {https://ui.adsabs.harvard.edu/abs/2023ApJ...944..107C} {944, 107}

\bibitem[\protect\citeauthoryear{{Choi}, {Leighly}, {Terndrup}, {Dabbieri},
  {Gallagher}  \& {Richards}}{{Choi} et~al.}{2022}]{Choi2022}
{Choi} H.,  {Leighly} K.~M.,  {Terndrup} D.~M.,  {Dabbieri} C.,  {Gallagher}
  S.~C.,   {Richards} G.~T.,  2022, \mn@doi [\apj] {10.3847/1538-4357/ac61d9},
  \href {https://ui.adsabs.harvard.edu/abs/2022ApJ...937...74C} {937, 74}

\bibitem[\protect\citeauthoryear{{Costa}, {Pakmor}  \& {Springel}}{{Costa}
  et~al.}{2020}]{Costa2020}
{Costa} T.,  {Pakmor} R.,   {Springel} V.,  2020, \mn@doi [\mnras]
  {10.1093/mnras/staa2321}, \href
  {https://ui.adsabs.harvard.edu/abs/2020MNRAS.497.5229C} {497, 5229}

\bibitem[\protect\citeauthoryear{{Crenshaw}, {Kraemer}, {Hutchings}, {Danks},
  {Gull}, {Kaiser}, {Nelson}  \& {Weistrop}}{{Crenshaw}
  et~al.}{2000}]{Crenshaw2000}
{Crenshaw} D.~M.,  {Kraemer} S.~B.,  {Hutchings} J.~B.,  {Danks} A.~C.,  {Gull}
  T.~R.,  {Kaiser} M.~E.,  {Nelson} C.~H.,   {Weistrop} D.,  2000, \mn@doi
  [\apjl] {10.1086/317333}, \href
  {https://ui.adsabs.harvard.edu/abs/2000ApJ...545L..27C} {545, L27}

\bibitem[\protect\citeauthoryear{{Dekker}, {D'Odorico}, {Kaufer}, {Delabre}  \&
  {Kotzlowski}}{{Dekker} et~al.}{2000}]{Dekker2000}
{Dekker} H.,  {D'Odorico} S.,  {Kaufer} A.,  {Delabre} B.,   {Kotzlowski} H.,
  2000, in {Iye} M.,  {Moorwood} A.~F.,  eds,  Society of Photo-Optical
  Instrumentation Engineers (SPIE) Conference Series Vol. 4008, \procspie. pp
  534--545, \mn@doi{10.1117/12.395512}

\bibitem[\protect\citeauthoryear{{Dere}, {Del Zanna}, {Young}, {Landi}  \&
  {Sutherland}}{{Dere} et~al.}{2019}]{Dere2019}
{Dere} K.~P.,  {Del Zanna} G.,  {Young} P.~R.,  {Landi} E.,   {Sutherland}
  R.~S.,  2019, \mn@doi [\apjs] {10.3847/1538-4365/ab05cf}, \href
  {https://ui.adsabs.harvard.edu/abs/2019ApJS..241...22D} {241, 22}

\bibitem[\protect\citeauthoryear{{Drake} \& {Morton}}{{Drake} \&
  {Morton}}{2007}]{Drake07}
{Drake} G.~W.~F.,  {Morton} D.~C.,  2007, \mn@doi [\apjs] {10.1086/512239},
  \href {https://ui.adsabs.harvard.edu/abs/2007ApJS..170..251D} {170, 251}

\bibitem[\protect\citeauthoryear{{Dunn} et~al.,}{{Dunn}
  et~al.}{2010}]{Dunn2010}
{Dunn} J.~P.,  et~al., 2010, \mn@doi [\apj] {10.1088/0004-637X/709/2/611},
  \href {https://ui.adsabs.harvard.edu/abs/2010ApJ...709..611D} {709, 611}

\bibitem[\protect\citeauthoryear{{Elvis}}{{Elvis}}{2017}]{Elvis2017}
{Elvis} M.,  2017, \mn@doi [\apj] {10.3847/1538-4357/aa82b6}, \href
  {https://ui.adsabs.harvard.edu/abs/2017ApJ...847...56E} {847, 56}

\bibitem[\protect\citeauthoryear{{Everett}, {K{\"o}nigl}  \& {Arav}}{{Everett}
  et~al.}{2002}]{Everett2002}
{Everett} J.,  {K{\"o}nigl} A.,   {Arav} N.,  2002, \mn@doi [\apj]
  {10.1086/339346}, \href
  {https://ui.adsabs.harvard.edu/abs/2002ApJ...569..671E} {569, 671}

\bibitem[\protect\citeauthoryear{{Farrah} et~al.,}{{Farrah}
  et~al.}{2012}]{Farrah2012}
{Farrah} D.,  et~al., 2012, \mn@doi [\apj] {10.1088/0004-637X/745/2/178}, \href
  {https://ui.adsabs.harvard.edu/abs/2012ApJ...745..178F} {745, 178}

\bibitem[\protect\citeauthoryear{{Fathivavsari}}{{Fathivavsari}}{2020}]{Fathivavsari2020}
{Fathivavsari} H.,  2020, \mn@doi [\apj] {10.3847/1538-4357/ab59da}, \href
  {https://ui.adsabs.harvard.edu/abs/2020ApJ...888...85F} {888, 85}

\bibitem[\protect\citeauthoryear{{Fathivavsari}, {Petitjean}, {Zou},
  {Noterdaeme}, {Ledoux}, {Kr{\"u}hler}  \& {Srianand}}{{Fathivavsari}
  et~al.}{2017}]{Fathivavsari2017}
{Fathivavsari} H.,  {Petitjean} P.,  {Zou} S.,  {Noterdaeme} P.,  {Ledoux} C.,
  {Kr{\"u}hler} T.,   {Srianand} R.,  2017, \mn@doi [\mnras]
  {10.1093/mnrasl/slw233}, \href
  {https://ui.adsabs.harvard.edu/abs/2017MNRAS.466L..58F} {466, L58}

\bibitem[\protect\citeauthoryear{{Faucher-Gigu{\`e}re} \&
  {Quataert}}{{Faucher-Gigu{\`e}re} \& {Quataert}}{2012}]{Faucher-Giguere2012}
{Faucher-Gigu{\`e}re} C.-A.,  {Quataert} E.,  2012, \mn@doi [\mnras]
  {10.1111/j.1365-2966.2012.21512.x}, \href
  {https://ui.adsabs.harvard.edu/abs/2012MNRAS.425..605F} {425, 605}

\bibitem[\protect\citeauthoryear{{Ferland} et~al.,}{{Ferland}
  et~al.}{2017}]{Ferland2017}
{Ferland} G.~J.,  et~al., 2017, \rmxaa, \href
  {https://ui.adsabs.harvard.edu/abs/2017RMxAA..53..385F} {53, 385}

\bibitem[\protect\citeauthoryear{{Finley} et~al.,}{{Finley}
  et~al.}{2013}]{Finley2013}
{Finley} H.,  et~al., 2013, \mn@doi [\aap] {10.1051/0004-6361/201321745}, \href
  {https://ui.adsabs.harvard.edu/abs/2013A&A...558A.111F} {558, A111}

\bibitem[\protect\citeauthoryear{{Fitzpatrick} \& {Massa}}{{Fitzpatrick} \&
  {Massa}}{2007}]{Fitzpatrick2007}
{Fitzpatrick} E.~L.,  {Massa} D.,  2007, \mn@doi [\apj] {10.1086/518158}, \href
  {https://ui.adsabs.harvard.edu/abs/2007ApJ...663..320F} {663, 320}

\bibitem[\protect\citeauthoryear{{Fynbo} et~al.,}{{Fynbo}
  et~al.}{2014}]{Fynbo2014}
{Fynbo} J.~P.~U.,  et~al., 2014, \mn@doi [\aap] {10.1051/0004-6361/201424726},
  \href {https://ui.adsabs.harvard.edu/abs/2014A&A...572A..12F} {572, A12}

\bibitem[\protect\citeauthoryear{{Gibson} et~al.,}{{Gibson}
  et~al.}{2009}]{Gibson2009}
{Gibson} R.~R.,  et~al., 2009, \mn@doi [\apj] {10.1088/0004-637X/692/1/758},
  \href {https://ui.adsabs.harvard.edu/abs/2009ApJ...692..758G} {692, 758}

\bibitem[\protect\citeauthoryear{{Goodman} \& {Weare}}{{Goodman} \&
  {Weare}}{2010}]{Goodman2010}
{Goodman} J.,  {Weare} J.,  2010, \mn@doi [Communications in Applied
  Mathematics and Computational Science] {10.2140/camcos.2010.5.65}, \href
  {https://ui.adsabs.harvard.edu/abs/2010CAMCS...5...65G} {5, 65}

\bibitem[\protect\citeauthoryear{{Hall}, {Hutsem{\'e}kers}, {Anderson},
  {Brinkmann}, {Fan}, {Schneider}  \& {York}}{{Hall} et~al.}{2003}]{Hall2003}
{Hall} P.~B.,  {Hutsem{\'e}kers} D.,  {Anderson} S.~F.,  {Brinkmann} J.,  {Fan}
  X.,  {Schneider} D.~P.,   {York} D.~G.,  2003, \mn@doi [\apj]
  {10.1086/376409}, \href
  {https://ui.adsabs.harvard.edu/abs/2003ApJ...593..189H} {593, 189}

\bibitem[\protect\citeauthoryear{{Hamann}, {Barlow}, {Chaffee}, {Foltz}  \&
  {Weymann}}{{Hamann} et~al.}{2001}]{Hamann2001}
{Hamann} F.~W.,  {Barlow} T.~A.,  {Chaffee} F.~C.,  {Foltz} C.~B.,   {Weymann}
  R.~J.,  2001, \mn@doi [\apj] {10.1086/319733}, \href
  {https://ui.adsabs.harvard.edu/abs/2001ApJ...550..142H} {550, 142}

\bibitem[\protect\citeauthoryear{{Hamann}, {Tripp}, {Rupke}  \&
  {Veilleux}}{{Hamann} et~al.}{2019}]{Hamann2019}
{Hamann} F.,  {Tripp} T.~M.,  {Rupke} D.,   {Veilleux} S.,  2019, \mn@doi
  [\mnras] {10.1093/mnras/stz1408}, \href
  {https://ui.adsabs.harvard.edu/abs/2019MNRAS.487.5041H} {487, 5041}

\bibitem[\protect\citeauthoryear{{Hazard}, {McMahon}, {Webb}  \&
  {Morton}}{{Hazard} et~al.}{1987}]{Hazard1987}
{Hazard} C.,  {McMahon} R.~G.,  {Webb} J.~K.,   {Morton} D.~C.,  1987, \mn@doi
  [\apj] {10.1086/165823}, \href
  {https://ui.adsabs.harvard.edu/abs/1987ApJ...323..263H} {323, 263}

\bibitem[\protect\citeauthoryear{{Hewett} \& {Foltz}}{{Hewett} \&
  {Foltz}}{2003}]{Hewett2003}
{Hewett} P.~C.,  {Foltz} C.~B.,  2003, \mn@doi [\aj] {10.1086/368392}, \href
  {https://ui.adsabs.harvard.edu/abs/2003AJ....125.1784H} {125, 1784}

\bibitem[\protect\citeauthoryear{{Ishita}, {Misawa}, {Itoh}, {Charlton}  \&
  {Eracleous}}{{Ishita} et~al.}{2021}]{Ishita2021}
{Ishita} D.,  {Misawa} T.,  {Itoh} D.,  {Charlton} J.~C.,   {Eracleous} M.,
  2021, arXiv e-prints, \href
  {https://ui.adsabs.harvard.edu/abs/2021arXiv210706496I} {p. arXiv:2107.06496}

\bibitem[\protect\citeauthoryear{{Jin} et~al.,}{{Jin} et~al.}{2023}]{WEAVE2023}
{Jin} S.,  et~al., 2023, \mn@doi [\mnras] {10.1093/mnras/stad557}, \href
  {https://ui.adsabs.harvard.edu/abs/2023MNRAS.tmp..715J} {}

\bibitem[\protect\citeauthoryear{{Kling} \& {Griesmann}}{{Kling} \&
  {Griesmann}}{2000}]{Kling2000}
{Kling} R.,  {Griesmann} U.,  2000, \mn@doi [\apj] {10.1086/308490}, \href
  {https://ui.adsabs.harvard.edu/abs/2000ApJ...531.1173K} {531, 1173}

\bibitem[\protect\citeauthoryear{{Knigge}, {Scaringi}, {Goad}  \&
  {Cottis}}{{Knigge} et~al.}{2008}]{Knigge2008}
{Knigge} C.,  {Scaringi} S.,  {Goad} M.~R.,   {Cottis} C.~E.,  2008, \mn@doi
  [\mnras] {10.1111/j.1365-2966.2008.13081.x}, \href
  {https://ui.adsabs.harvard.edu/abs/2008MNRAS.386.1426K} {386, 1426}

\bibitem[\protect\citeauthoryear{{Korista}, {Bautista}, {Arav}, {Moe},
  {Costantini}  \& {Benn}}{{Korista} et~al.}{2008}]{Korista2008}
{Korista} K.~T.,  {Bautista} M.~A.,  {Arav} N.,  {Moe} M.,  {Costantini} E.,
  {Benn} C.,  2008, \mn@doi [\apj] {10.1086/592140}, \href
  {https://ui.adsabs.harvard.edu/abs/2008ApJ...688..108K} {688, 108}

\bibitem[\protect\citeauthoryear{{Kraemer} et~al.,}{{Kraemer}
  et~al.}{2001}]{Kraemer2001}
{Kraemer} S.~B.,  et~al., 2001, \mn@doi [\apj] {10.1086/320244}, \href
  {https://ui.adsabs.harvard.edu/abs/2001ApJ...551..671K} {551, 671}

\bibitem[\protect\citeauthoryear{{Krogager} et~al.,}{{Krogager}
  et~al.}{2023}]{Krogager2023}
{Krogager} J.~K.,  et~al., 2023, \mn@doi [The Messenger]
  {10.18727/0722-6691/5310}, \href
  {https://ui.adsabs.harvard.edu/abs/2023Msngr.190...38K} {190, 38}

\bibitem[\protect\citeauthoryear{{Leighly}, {Dietrich}  \& {Barber}}{{Leighly}
  et~al.}{2011}]{Leighly2011}
{Leighly} K.~M.,  {Dietrich} M.,   {Barber} S.,  2011, \mn@doi [\apj]
  {10.1088/0004-637X/728/2/94}, \href
  {https://ui.adsabs.harvard.edu/abs/2011ApJ...728...94L} {728, 94}

\bibitem[\protect\citeauthoryear{{Liu} et~al.,}{{Liu} et~al.}{2015}]{Liu2015}
{Liu} W.-J.,  et~al., 2015, \mn@doi [\apjs] {10.1088/0067-0049/217/1/11}, \href
  {https://ui.adsabs.harvard.edu/abs/2015ApJS..217...11L} {217, 11}

\bibitem[\protect\citeauthoryear{{Lucy}, {Leighly}, {Terndrup}, {Dietrich}  \&
  {Gallagher}}{{Lucy} et~al.}{2014}]{Lucy2014}
{Lucy} A.~B.,  {Leighly} K.~M.,  {Terndrup} D.~M.,  {Dietrich} M.,
  {Gallagher} S.~C.,  2014, \mn@doi [\apj] {10.1088/0004-637X/783/1/58}, \href
  {https://ui.adsabs.harvard.edu/abs/2014ApJ...783...58L} {783, 58}

\bibitem[\protect\citeauthoryear{{Morton}}{{Morton}}{2003}]{Morton2003}
{Morton} D.~C.,  2003, \mn@doi [\apjs] {10.1086/377639}, \href
  {https://ui.adsabs.harvard.edu/abs/2003ApJS..149..205M} {149, 205}

\bibitem[\protect\citeauthoryear{{Murphy} \& {Bernet}}{{Murphy} \&
  {Bernet}}{2016}]{Murphy2016}
{Murphy} M.~T.,  {Bernet} M.~L.,  2016, \mn@doi [\mnras]
  {10.1093/mnras/stv2420}, \href
  {https://ui.adsabs.harvard.edu/abs/2016MNRAS.455.1043M} {455, 1043}

\bibitem[\protect\citeauthoryear{{Murphy}, {Kacprzak}, {Savorgnan}  \&
  {Carswell}}{{Murphy} et~al.}{2019}]{Murphy2019}
{Murphy} M.~T.,  {Kacprzak} G.~G.,  {Savorgnan} G. A.~D.,   {Carswell} R.~F.,
  2019, \mn@doi [\mnras] {10.1093/mnras/sty2834}, \href
  {https://ui.adsabs.harvard.edu/abs/2019MNRAS.482.3458M} {482, 3458}

\bibitem[\protect\citeauthoryear{{Nave} \& {Johansson}}{{Nave} \&
  {Johansson}}{2013}]{Nave2013}
{Nave} G.,  {Johansson} S.,  2013, \mn@doi [\apjs] {10.1088/0067-0049/204/1/1},
  \href {https://ui.adsabs.harvard.edu/abs/2013ApJS..204....1N} {204, 1}

\bibitem[\protect\citeauthoryear{{Negrete} et~al.,}{{Negrete}
  et~al.}{2018}]{Negrete2018}
{Negrete} C.~A.,  et~al., 2018, \mn@doi [\aap] {10.1051/0004-6361/201833285},
  \href {https://ui.adsabs.harvard.edu/abs/2018A&A...620A.118N} {620, A118}

\bibitem[\protect\citeauthoryear{{Noterdaeme}, {Balashev}, {Krogager},
  {Srianand}, {Fathivavsari}, {Petitjean}  \& {Ledoux}}{{Noterdaeme}
  et~al.}{2019}]{Noterdaeme2019}
{Noterdaeme} P.,  {Balashev} S.,  {Krogager} J.~K.,  {Srianand} R.,
  {Fathivavsari} H.,  {Petitjean} P.,   {Ledoux} C.,  2019, \mn@doi [\aap]
  {10.1051/0004-6361/201935371}, \href
  {https://ui.adsabs.harvard.edu/abs/2019A&A...627A..32N} {627, A32}

\bibitem[\protect\citeauthoryear{{Noterdaeme}, {Balashev}, {Krogager},
  {Laursen}, {Srianand}, {Gupta}, {Petitjean}  \& {Fynbo}}{{Noterdaeme}
  et~al.}{2021}]{Noterdaeme2021}
{Noterdaeme} P.,  {Balashev} S.,  {Krogager} J.~K.,  {Laursen} P.,  {Srianand}
  R.,  {Gupta} N.,  {Petitjean} P.,   {Fynbo} J.~P.~U.,  2021, \mn@doi [\aap]
  {10.1051/0004-6361/202038877}, \href
  {https://ui.adsabs.harvard.edu/abs/2021A&A...646A.108N} {646, A108}

\bibitem[\protect\citeauthoryear{{Noterdaeme} et~al.,}{{Noterdaeme}
  et~al.}{2023}]{Noterdaeme2023}
{Noterdaeme} P.,  et~al., 2023, \mn@doi [\aap] {10.1051/0004-6361/202245554},
  \href {https://ui.adsabs.harvard.edu/abs/2023A&A...673A..89N} {673, A89}

\bibitem[\protect\citeauthoryear{{P{\^a}ris} et~al.,}{{P{\^a}ris}
  et~al.}{2018}]{Paris2018}
{P{\^a}ris} I.,  et~al., 2018, \mn@doi [\aap] {10.1051/0004-6361/201732445},
  \href {https://ui.adsabs.harvard.edu/abs/2018A&A...613A..51P} {613, A51}

\bibitem[\protect\citeauthoryear{{Proga}, {Stone}  \& {Kallman}}{{Proga}
  et~al.}{2000}]{Proga2000}
{Proga} D.,  {Stone} J.~M.,   {Kallman} T.~R.,  2000, \mn@doi [\apj]
  {10.1086/317154}, \href
  {https://ui.adsabs.harvard.edu/abs/2000ApJ...543..686P} {543, 686}

\bibitem[\protect\citeauthoryear{{Reichard} et~al.,}{{Reichard}
  et~al.}{2003}]{Reichard2003}
{Reichard} T.~A.,  et~al., 2003, \mn@doi [\aj] {10.1086/368244}, \href
  {https://ui.adsabs.harvard.edu/abs/2003AJ....125.1711R} {125, 1711}

\bibitem[\protect\citeauthoryear{{Safronova} \& {Safronova}}{{Safronova} \&
  {Safronova}}{2011}]{Safronova2011}
{Safronova} M.~S.,  {Safronova} U.~I.,  2011, \mn@doi [\pra]
  {10.1103/PhysRevA.83.012503}, \href
  {https://ui.adsabs.harvard.edu/abs/2011PhRvA..83a2503S} {83, 012503}

\bibitem[\protect\citeauthoryear{{Sarazin} \& {Roddier}}{{Sarazin} \&
  {Roddier}}{1990}]{Sarazin1990}
{Sarazin} M.,  {Roddier} F.,  1990, \aap, \href
  {https://ui.adsabs.harvard.edu/abs/1990A&A...227..294S} {227, 294}

\bibitem[\protect\citeauthoryear{{Schnabel}, {Schultz-Johanning}  \&
  {Kock}}{{Schnabel} et~al.}{2004}]{Schnabel2004}
{Schnabel} R.,  {Schultz-Johanning} M.,   {Kock} M.,  2004, \mn@doi [\aap]
  {10.1051/0004-6361:20031685}, \href
  {https://ui.adsabs.harvard.edu/abs/2004A&A...414.1169S} {414, 1169}

\bibitem[\protect\citeauthoryear{{Selsing}, {Fynbo}, {Christensen}  \&
  {Krogager}}{{Selsing} et~al.}{2016}]{Selsing2016}
{Selsing} J.,  {Fynbo} J.~P.~U.,  {Christensen} L.,   {Krogager} J.~K.,  2016,
  \mn@doi [\aap] {10.1051/0004-6361/201527096}, \href
  {https://ui.adsabs.harvard.edu/abs/2016A&A...585A..87S} {585, A87}

\bibitem[\protect\citeauthoryear{{Shen} et~al.,}{{Shen}
  et~al.}{2016}]{Shen2016}
{Shen} Y.,  et~al., 2016, \mn@doi [\apj] {10.3847/0004-637X/831/1/7}, \href
  {https://ui.adsabs.harvard.edu/abs/2016ApJ...831....7S} {831, 7}

\bibitem[\protect\citeauthoryear{{Smith}, {Schmidt}, {Allen}  \&
  {Angel}}{{Smith} et~al.}{1995}]{Smith1995}
{Smith} P.~S.,  {Schmidt} G.~D.,  {Allen} R.~G.,   {Angel} J.~R.~P.,  1995,
  \mn@doi [\apj] {10.1086/175589}, \href
  {https://ui.adsabs.harvard.edu/abs/1995ApJ...444..146S} {444, 146}

\bibitem[\protect\citeauthoryear{{Tolea}, {Krolik}  \& {Tsvetanov}}{{Tolea}
  et~al.}{2002}]{Tolea2002}
{Tolea} A.,  {Krolik} J.~H.,   {Tsvetanov} Z.,  2002, \mn@doi [\apjl]
  {10.1086/344563}, \href
  {https://ui.adsabs.harvard.edu/abs/2002ApJ...578L..31T} {578, L31}

\bibitem[\protect\citeauthoryear{{Trump} et~al.,}{{Trump}
  et~al.}{2006}]{Trump2006}
{Trump} J.~R.,  et~al., 2006, \mn@doi [\apjs] {10.1086/503834}, \href
  {https://ui.adsabs.harvard.edu/abs/2006ApJS..165....1T} {165, 1}

\bibitem[\protect\citeauthoryear{{Veilleux}, {Mel{\'e}ndez}, {Tripp}, {Hamann}
  \& {Rupke}}{{Veilleux} et~al.}{2016}]{Veilleux2016}
{Veilleux} S.,  {Mel{\'e}ndez} M.,  {Tripp} T.~M.,  {Hamann} F.,   {Rupke}
  D.~S.~N.,  2016, \mn@doi [\apj] {10.3847/0004-637X/825/1/42}, \href
  {https://ui.adsabs.harvard.edu/abs/2016ApJ...825...42V} {825, 42}

\bibitem[\protect\citeauthoryear{{V{\'e}ron-Cetty} \&
  {V{\'e}ron}}{{V{\'e}ron-Cetty} \& {V{\'e}ron}}{2010}]{Veron2010}
{V{\'e}ron-Cetty} M.~P.,  {V{\'e}ron} P.,  2010, \mn@doi [\aap]
  {10.1051/0004-6361/201014188}, \href
  {https://ui.adsabs.harvard.edu/abs/2010A&A...518A..10V} {518, A10}

\bibitem[\protect\citeauthoryear{{Walker}, {Arav}  \& {Byun}}{{Walker}
  et~al.}{2022}]{Walker2022}
{Walker} A.,  {Arav} N.,   {Byun} D.,  2022, \mn@doi [\mnras]
  {10.1093/mnras/stac2349}, \href
  {https://ui.adsabs.harvard.edu/abs/2022MNRAS.516.3778W} {516, 3778}

\bibitem[\protect\citeauthoryear{{Wampler}, {Chugai}  \& {Petitjean}}{{Wampler}
  et~al.}{1995}]{Wampler1995}
{Wampler} E.~J.,  {Chugai} N.~N.,   {Petitjean} P.,  1995, \mn@doi [\apj]
  {10.1086/175551}, \href
  {https://ui.adsabs.harvard.edu/abs/1995ApJ...443..586W} {443, 586}

\bibitem[\protect\citeauthoryear{{Weselak}, {Galazutdinov}, {Musaev},
  {Beletsky}  \& {Kre{\l}owski}}{{Weselak} et~al.}{2009}]{Weselak2009}
{Weselak} T.,  {Galazutdinov} G.~A.,  {Musaev} F.~A.,  {Beletsky} Y.,
  {Kre{\l}owski} J.,  2009, \mn@doi [\aap] {10.1051/0004-6361:200810348}, \href
  {https://ui.adsabs.harvard.edu/abs/2009A&A...495..189W} {495, 189}

\bibitem[\protect\citeauthoryear{{Weymann}, {Carswell}  \& {Smith}}{{Weymann}
  et~al.}{1981}]{Weymann1981}
{Weymann} R.~J.,  {Carswell} R.~F.,   {Smith} M.~G.,  1981, \mn@doi [\araa]
  {10.1146/annurev.aa.19.090181.000353}, \href
  {https://ui.adsabs.harvard.edu/abs/1981ARA&A..19...41W} {19, 41}

\bibitem[\protect\citeauthoryear{{Xu}, {Arav}, {Miller}, {Korista}  \&
  {Benn}}{{Xu} et~al.}{2021}]{Xu2021}
{Xu} X.,  {Arav} N.,  {Miller} T.,  {Korista} K.~T.,   {Benn} C.,  2021,
  \mn@doi [\mnras] {10.1093/mnras/stab1866}, \href
  {https://ui.adsabs.harvard.edu/abs/2021MNRAS.506.2725X} {506, 2725}

\bibitem[\protect\citeauthoryear{{de Kool}, {Arav}, {Becker}, {Gregg}, {White},
  {Laurent-Muehleisen}, {Price}  \& {Korista}}{{de Kool}
  et~al.}{2001}]{deKool2001}
{de Kool} M.,  {Arav} N.,  {Becker} R.~H.,  {Gregg} M.~D.,  {White} R.~L.,
  {Laurent-Muehleisen} S.~A.,  {Price} T.,   {Korista} K.~T.,  2001, \mn@doi
  [\apj] {10.1086/318996}, \href
  {https://ui.adsabs.harvard.edu/abs/2001ApJ...548..609D} {548, 609}

\bibitem[\protect\citeauthoryear{{de Kool}, {Becker}, {Gregg}, {White}  \&
  {Arav}}{{de Kool} et~al.}{2002a}]{deKool2002b}
{de Kool} M.,  {Becker} R.~H.,  {Gregg} M.~D.,  {White} R.~L.,   {Arav} N.,
  2002a, \mn@doi [\apj] {10.1086/338490}, \href
  {https://ui.adsabs.harvard.edu/abs/2002ApJ...567...58D} {567, 58}

\bibitem[\protect\citeauthoryear{{de Kool}, {Becker}, {Arav}, {Gregg}  \&
  {White}}{{de Kool} et~al.}{2002b}]{deKool2002a}
{de Kool} M.,  {Becker} R.~H.,  {Arav} N.,  {Gregg} M.~D.,   {White} R.~L.,
  2002b, \mn@doi [\apj] {10.1086/339793}, \href
  {https://ui.adsabs.harvard.edu/abs/2002ApJ...570..514D} {570, 514}

\makeatother
\end{thebibliography}



\appendix

\section{\CaII, \MgI, and \HeIs\ fits}

Here we present the figures of all the fitted \CaII, \MgI, \HeIs, \FeII\ and \MnII\ lines.

\begin{figure*}
\centering
\includegraphics[trim={0.0cm 0.0cm 0.0cm 0.0cm},clip,width=0.32\textwidth]{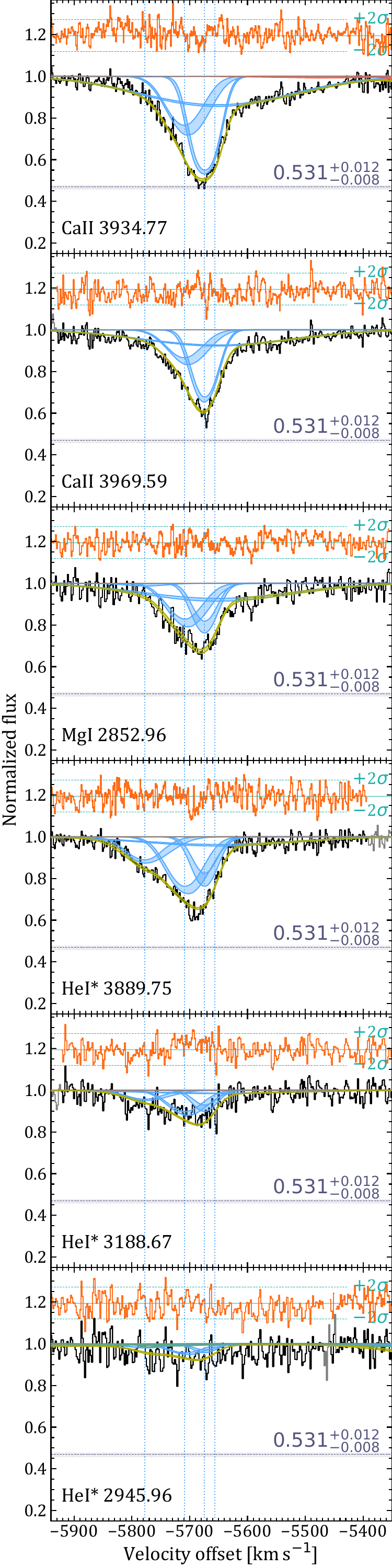}
\includegraphics[trim={0.0cm 0.0cm 0.0cm 0.0cm},clip,width=0.32\textwidth]{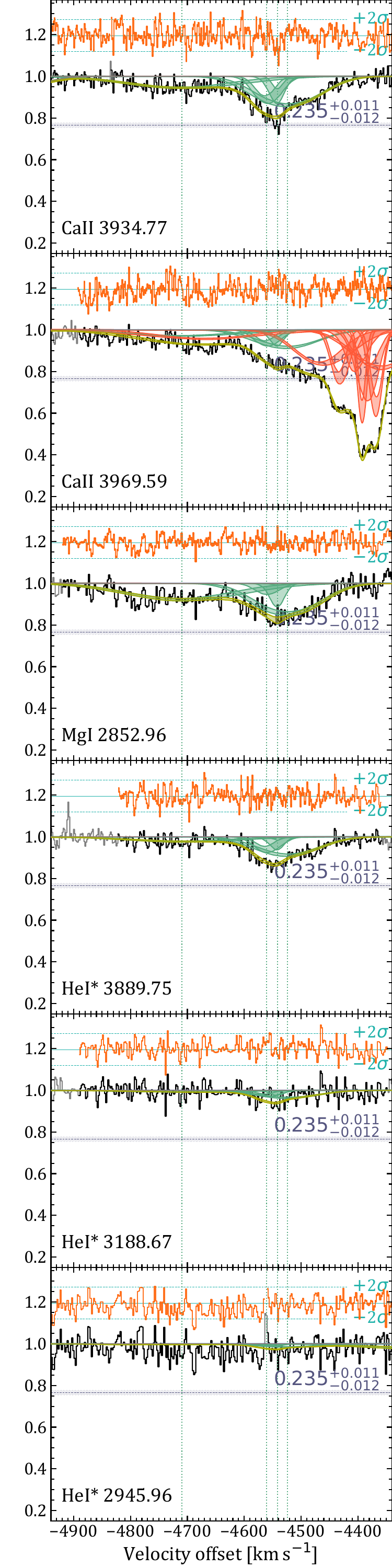}
\includegraphics[trim={0.0cm 0.0cm 0.0cm 0.0cm},clip,width=0.32\textwidth]{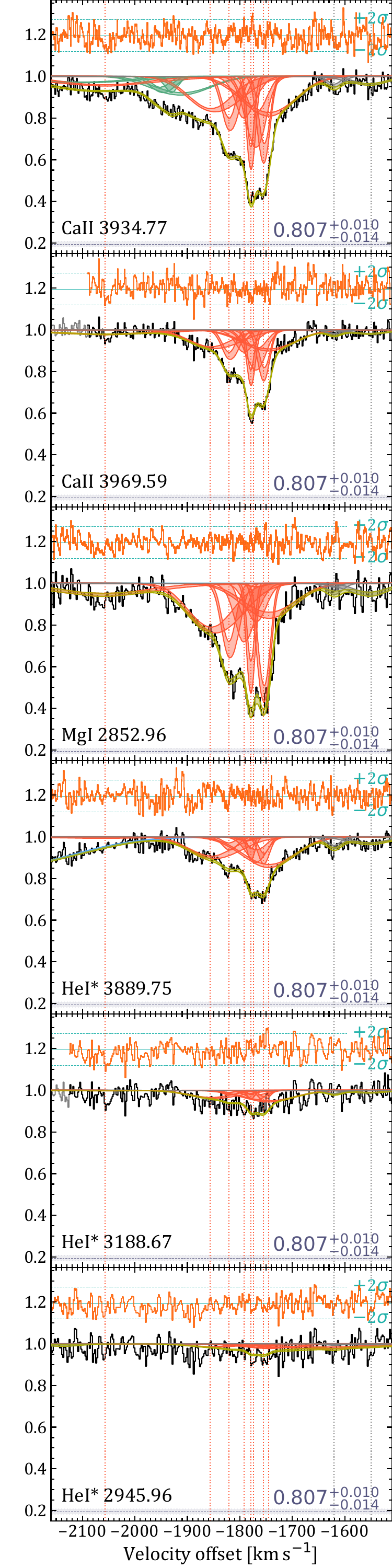}
\caption{{\sl Left to Right}: Voigt-profile fits to \CaII, \MgI, and \HeIs\ absorption lines in complexes $A$ (Left), $B$ (Middle), and $C$ (Right),
at $z_{\rm abs}=0.3253$, 0.3304, and 0.3429, respectively, towards \qso. The graphical information is the same as in Fig.~\ref{fig:fit_overview}. The red lines at the top of each panel show the residuals between the spectrum and fit, with green dashed horizontal lines corresponding to 2$\sigma$ deviation. 
\label{fig:CaII_MgI_HeI}}
\end{figure*}

\begin{figure*}
\centering
\includegraphics[trim={0.0cm 0.0cm 0.0cm 0.0cm},clip,width=0.32\textwidth]{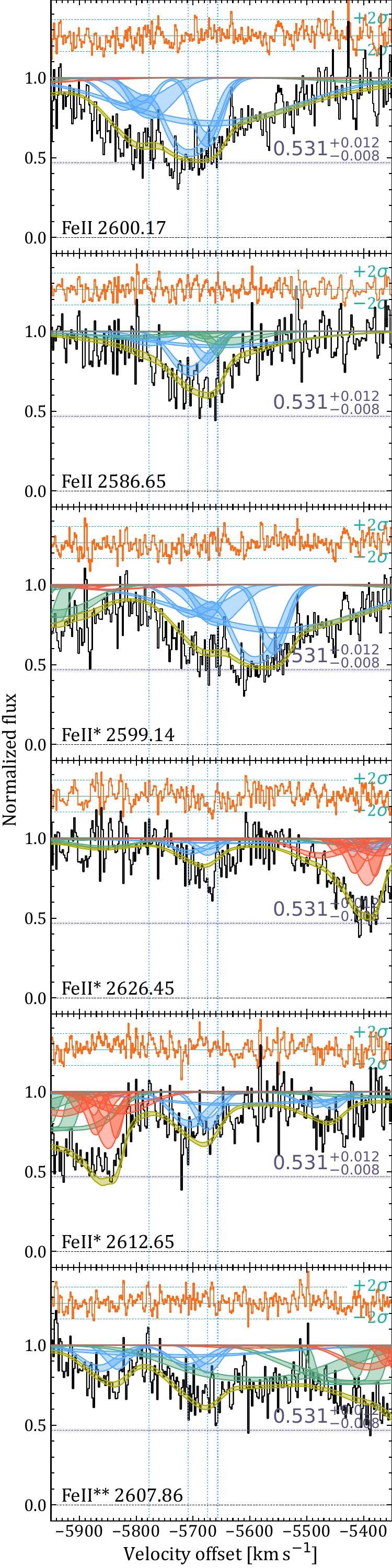}
\includegraphics[trim={0.0cm 0.0cm 0.0cm 0.0cm},clip,width=0.32\textwidth]{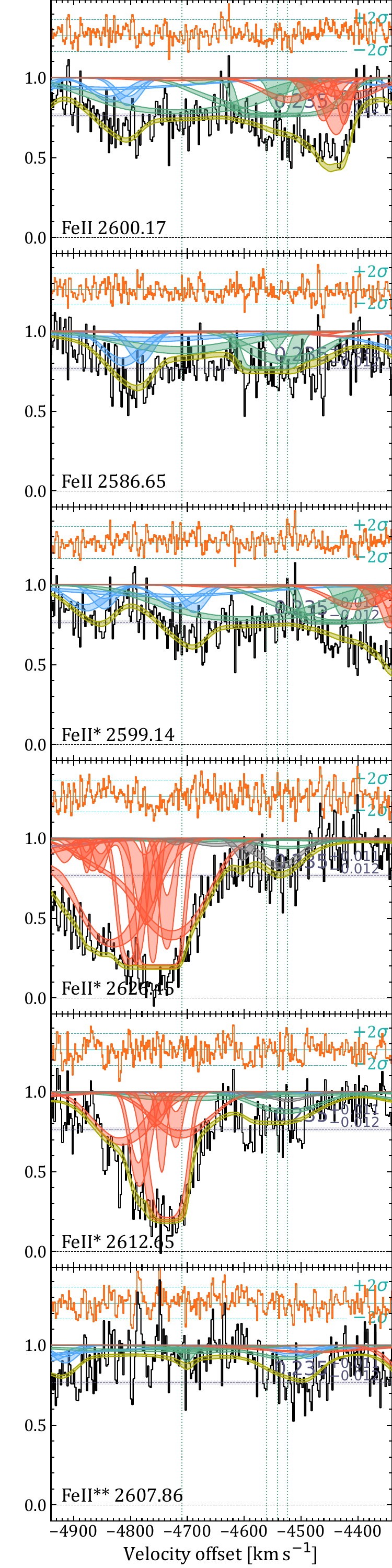}
\includegraphics[trim={0.0cm 0.0cm 0.0cm 0.0cm},clip,width=0.32\textwidth]{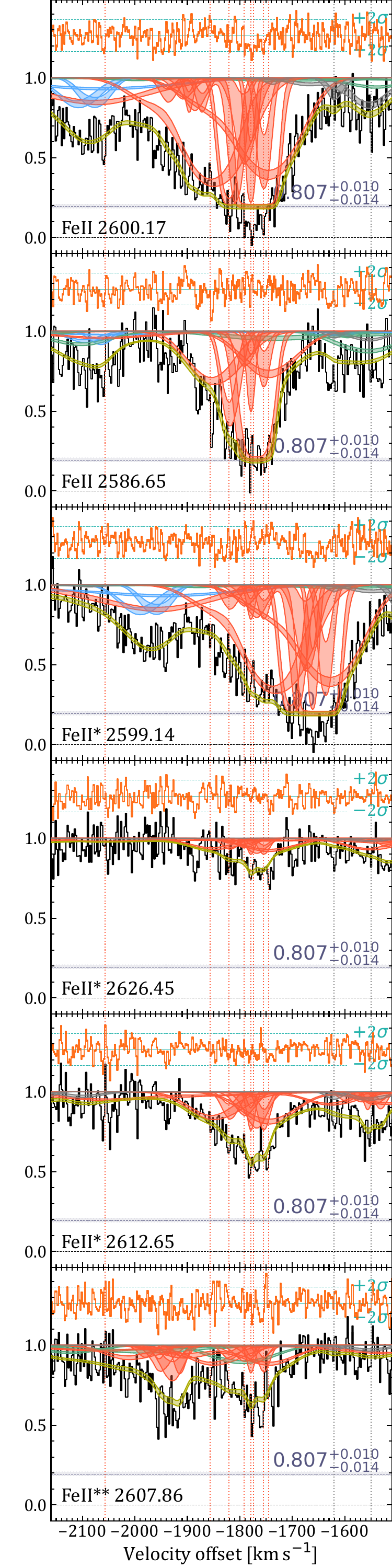}
\caption{{\sl Left to Right}: Voigt-profile fits to \FeII\ absorption lines in complexes $A$ (Left), $B$ (Middle), and $C$ (Right),
at $z_{\rm abs}=0.3253$, 0.3304, and 0.3429, respectively, towards \qso. Horizontal dashed lines and surrounding grey areas indicate
the extent of partial covering determined by fitting each clump independently with its own covering factor.
\label{fig:FeII_low_1}}
\end{figure*}

\begin{figure*}
\centering
\includegraphics[trim={0.0cm 0.0cm 0.0cm 0.0cm},clip,width=0.32\textwidth]{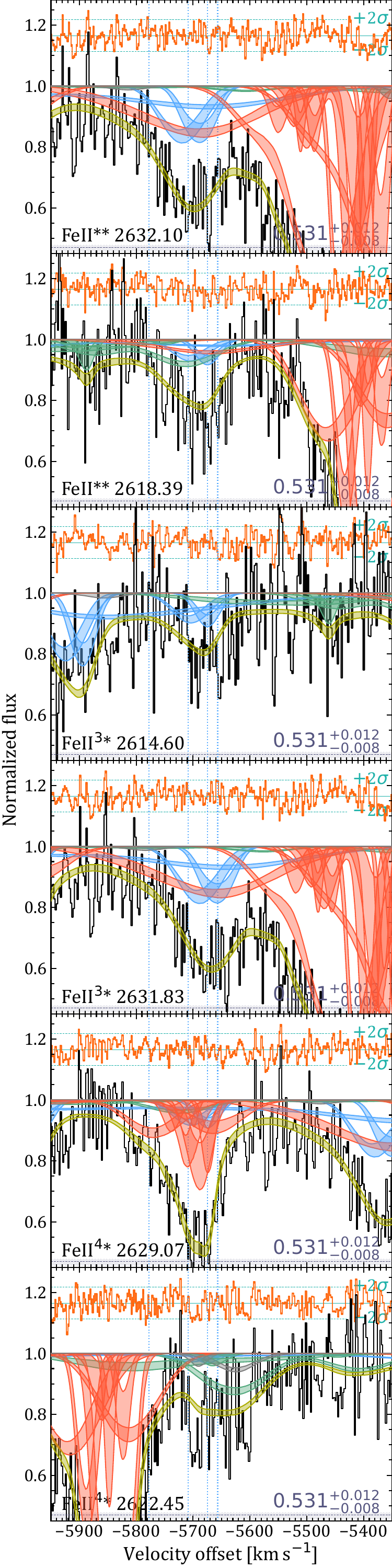}
\includegraphics[trim={0.0cm 0.0cm 0.0cm 0.0cm},clip,width=0.32\textwidth]{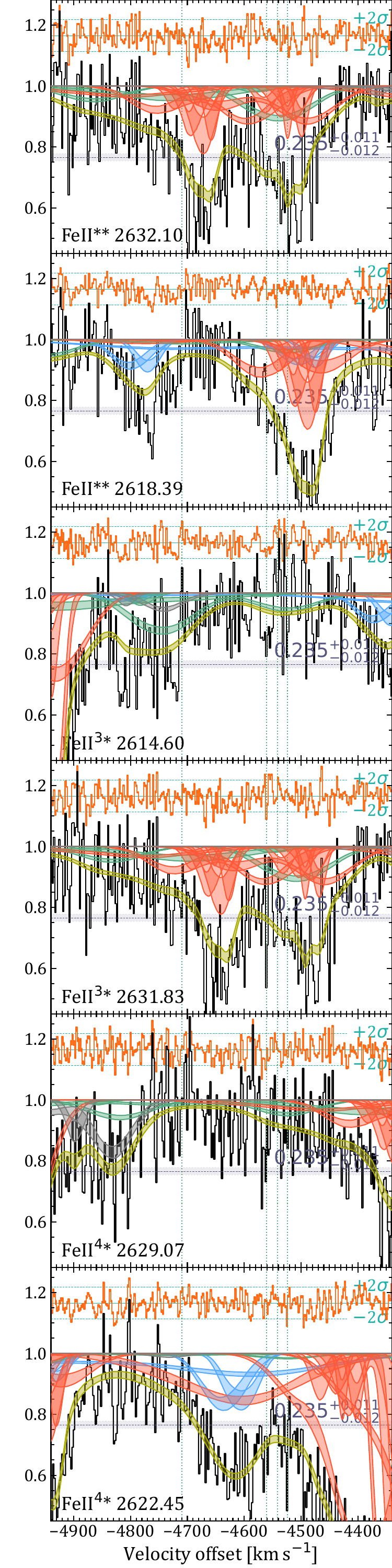}
\includegraphics[trim={0.0cm 0.0cm 0.0cm 0.0cm},clip,width=0.32\textwidth]{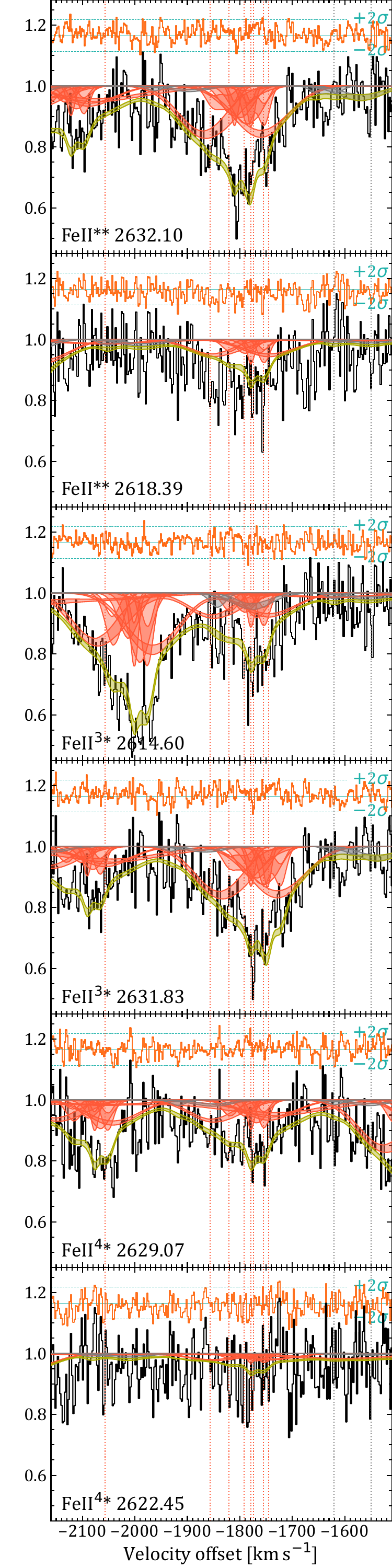}
\caption{Same as Fig.~\ref{fig:FeII_low_1} for higher-excitation \FeII\ lines.
\label{fig:FeII_low_2}}
\end{figure*}

\begin{figure*}
\centering
\includegraphics[trim={0.0cm 0.0cm 0.0cm 0.0cm},clip,width=0.32\textwidth]{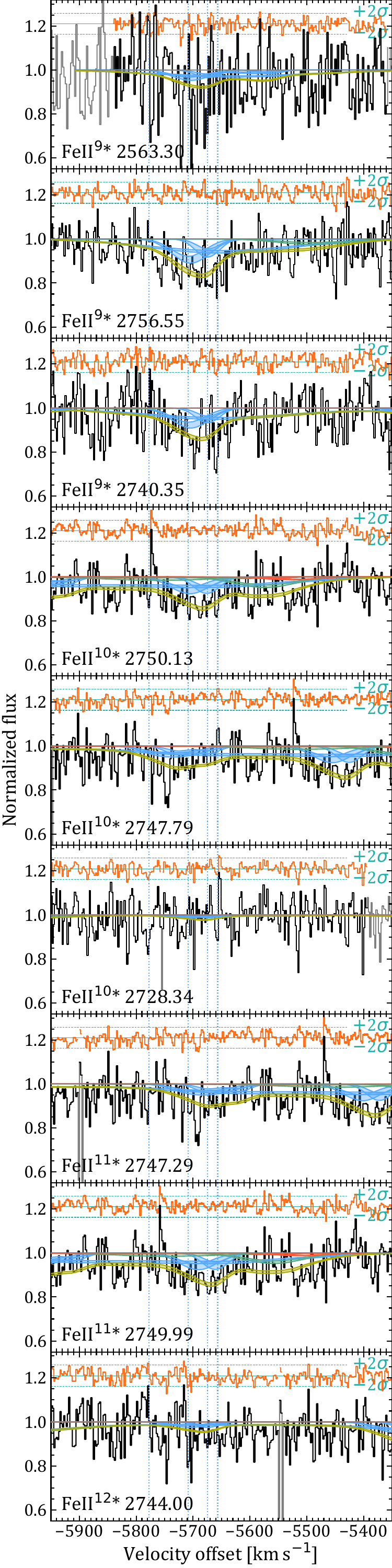}
\includegraphics[trim={0.0cm 0.0cm 0.0cm 0.0cm},clip,width=0.32\textwidth]{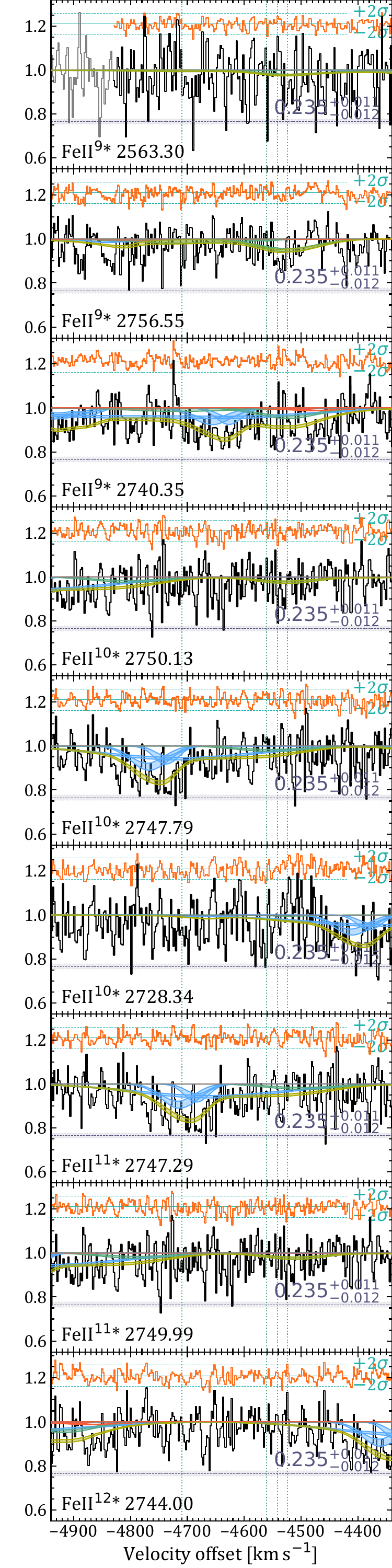}
\includegraphics[trim={0.0cm 0.0cm 0.0cm 0.0cm},clip,width=0.32\textwidth]{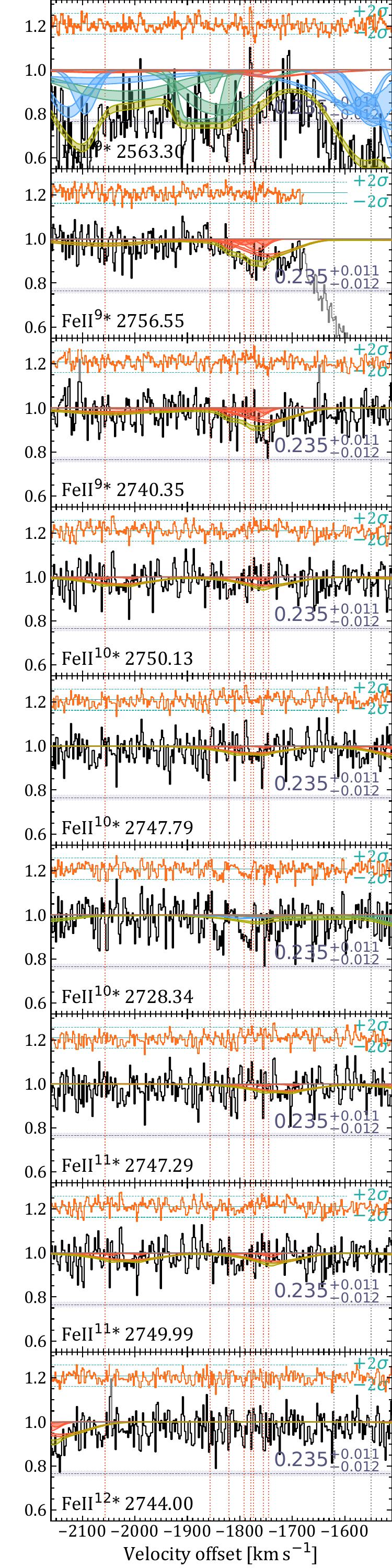}
\caption{Same as Fig.~\ref{fig:FeII_low_1} for higher-excitation \FeII\ lines (continued).
\label{fig:FeII_high}}
\end{figure*}

\begin{figure*}
\centering
\includegraphics[trim={0.0cm 0.0cm 0.0cm 0.0cm},clip,width=0.32\textwidth]{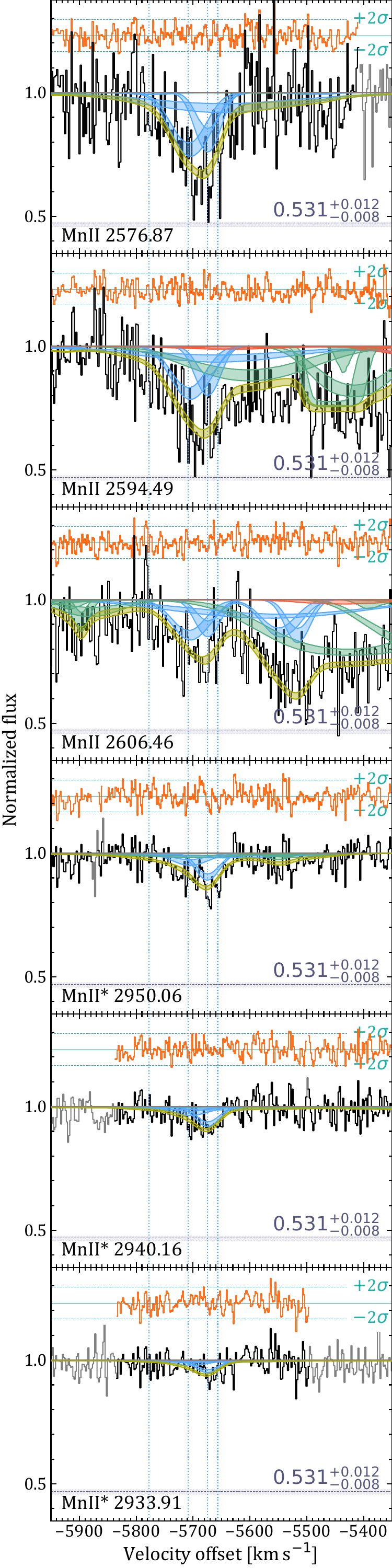}
\includegraphics[trim={0.0cm 0.0cm 0.0cm 0.0cm},clip,width=0.32\textwidth]{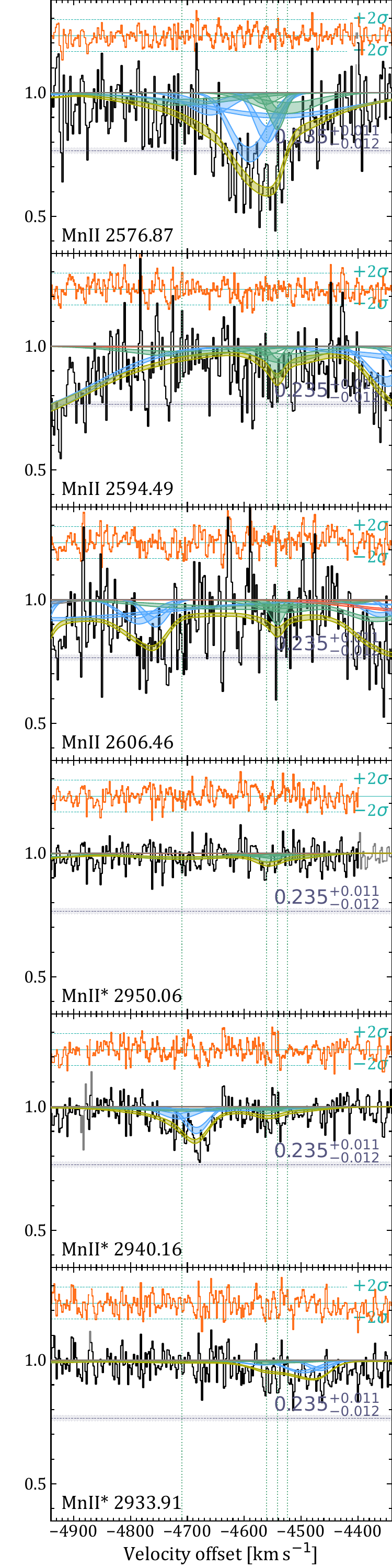}
\includegraphics[trim={0.0cm 0.0cm 0.0cm 0.0cm},clip,width=0.32\textwidth]{MnII_B.pdf}
\caption{Same as Fig.~\ref{fig:FeII_low_1} for \MnII\ lines.
\label{fig:MnII}}
\end{figure*}

\begin{figure*}
\centering
\includegraphics[trim={0.0cm 0.0cm 0.0cm 0.0cm},clip,width=0.98\textwidth]{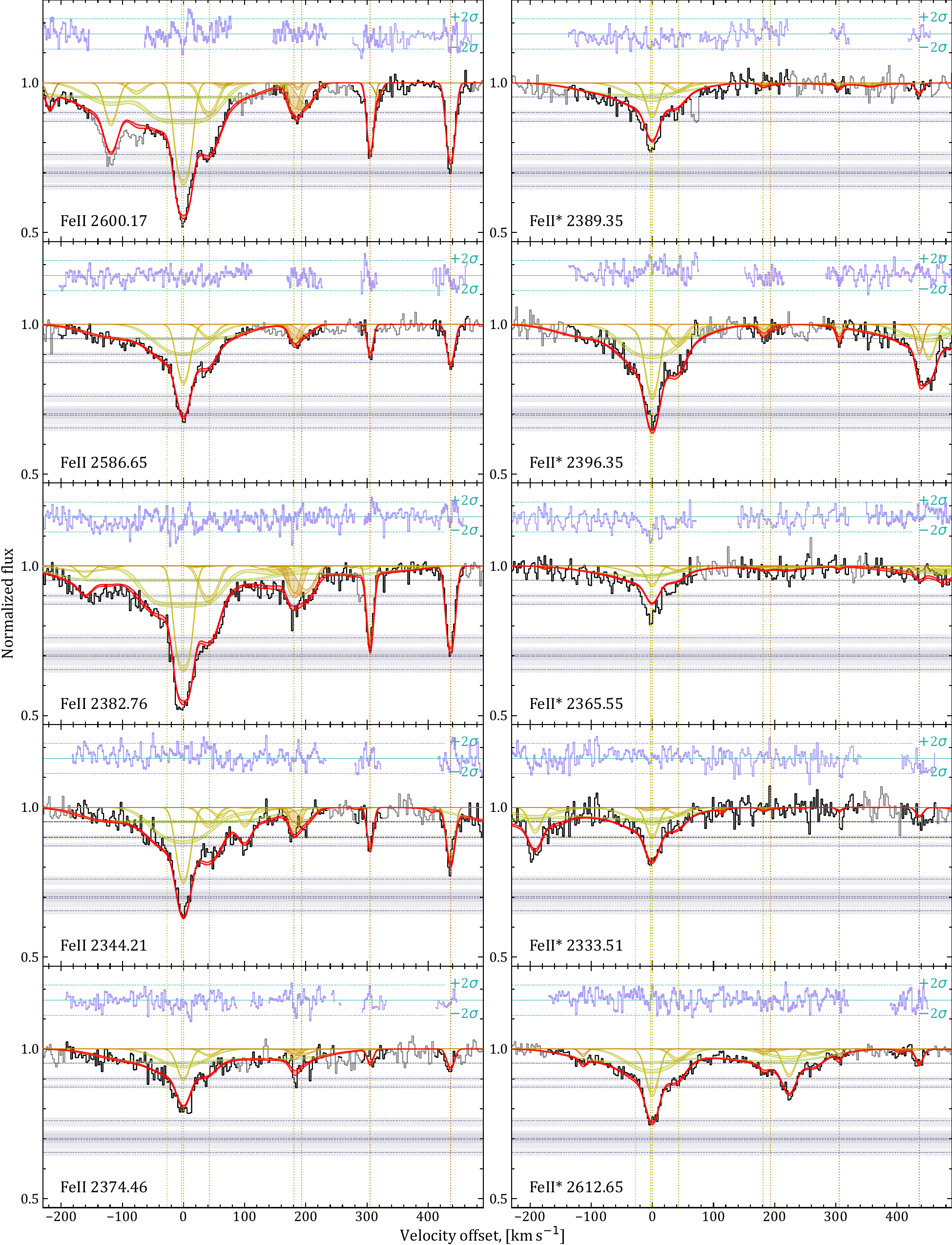}
\caption{Voigt-profile fits to \FeII\ absorption lines at $z_{\rm abs}=0.85993$ towards \qsot. The graphical information is similar to that in Fig.~\ref{fig:CaII_MgI_HeI}. 
\label{fig:J2359_FeII_low_1}}
\end{figure*}

\begin{figure*}
\centering
\includegraphics[trim={0.0cm 0.0cm 0.0cm 0.0cm},clip,width=0.98\textwidth]{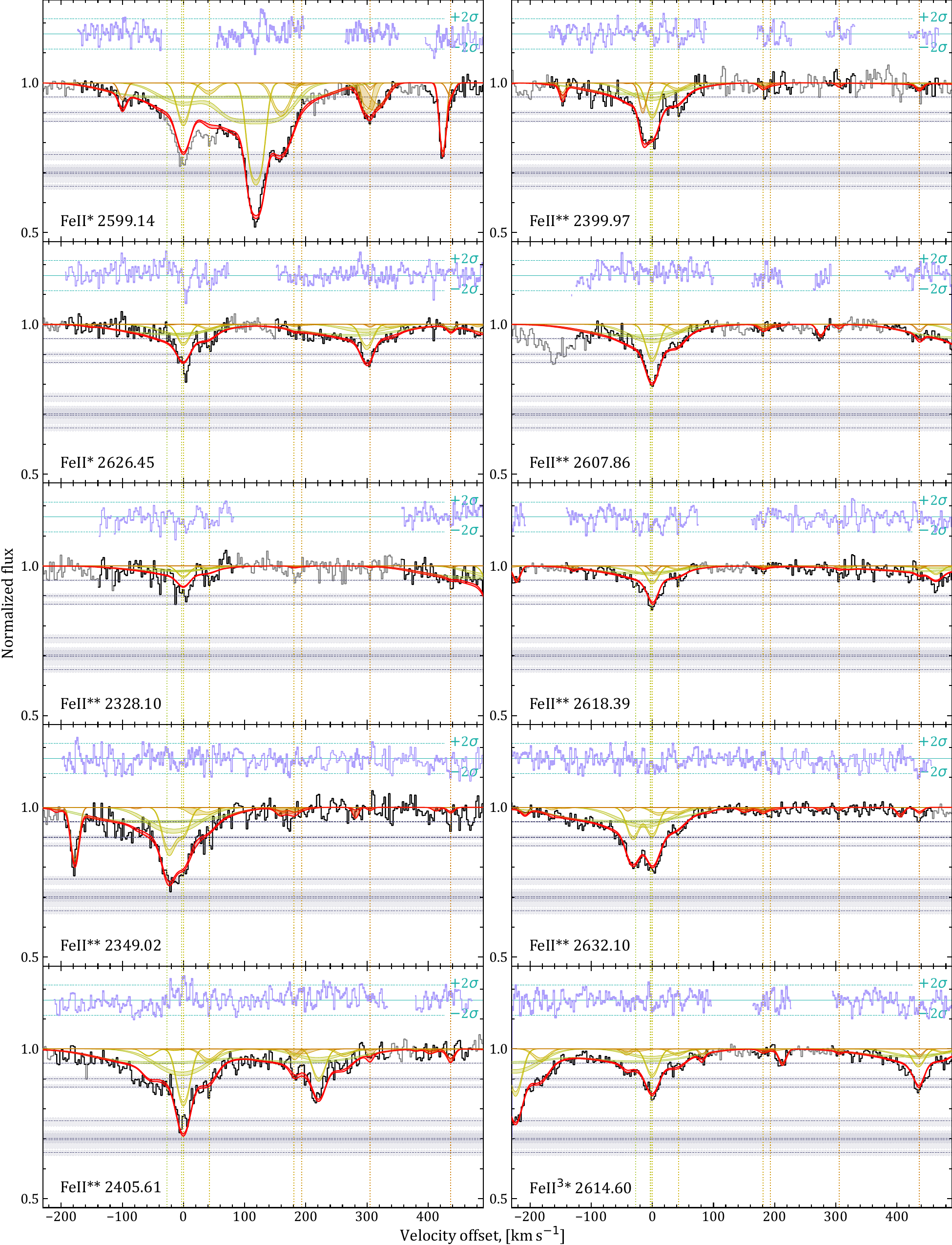}
\caption{Same as Fig.~\ref{fig:J2359_FeII_low_1} for higher-excitation \FeII\ lines.
\label{fig:J2359_FeII_low_2}}
\end{figure*}

\begin{figure*}
\centering
\includegraphics[trim={0.0cm 0.0cm 0.0cm 0.0cm},clip,width=0.98\textwidth]{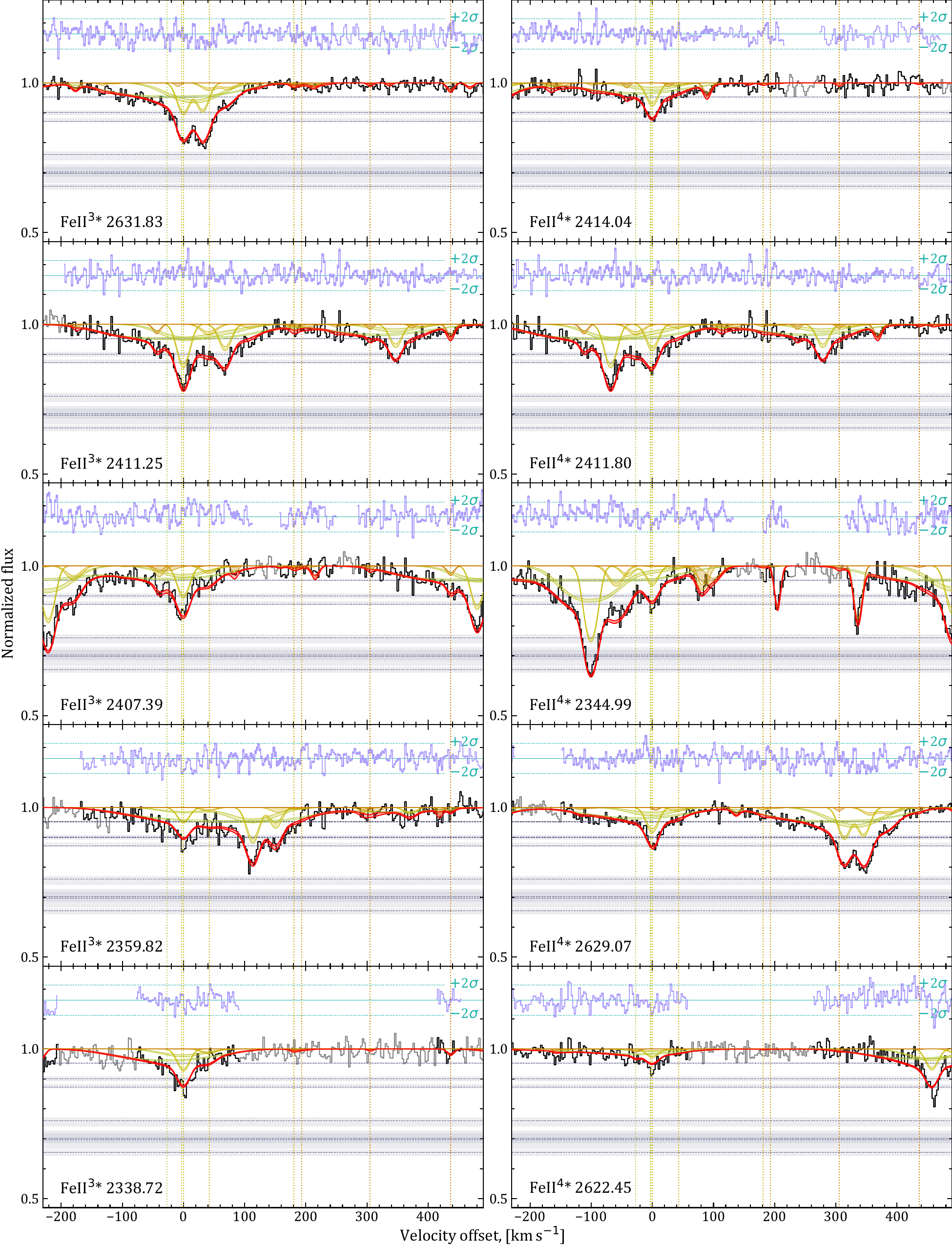}
\caption{Same as Fig.~\ref{fig:J2359_FeII_low_1} for higher-excitation \FeII\ lines (continued).
\label{fig:J2359_FeII_low_3}}
\end{figure*}

\begin{figure*}
\centering
\includegraphics[trim={0.0cm 0.0cm 0.0cm 0.0cm},clip,width=0.98\textwidth]{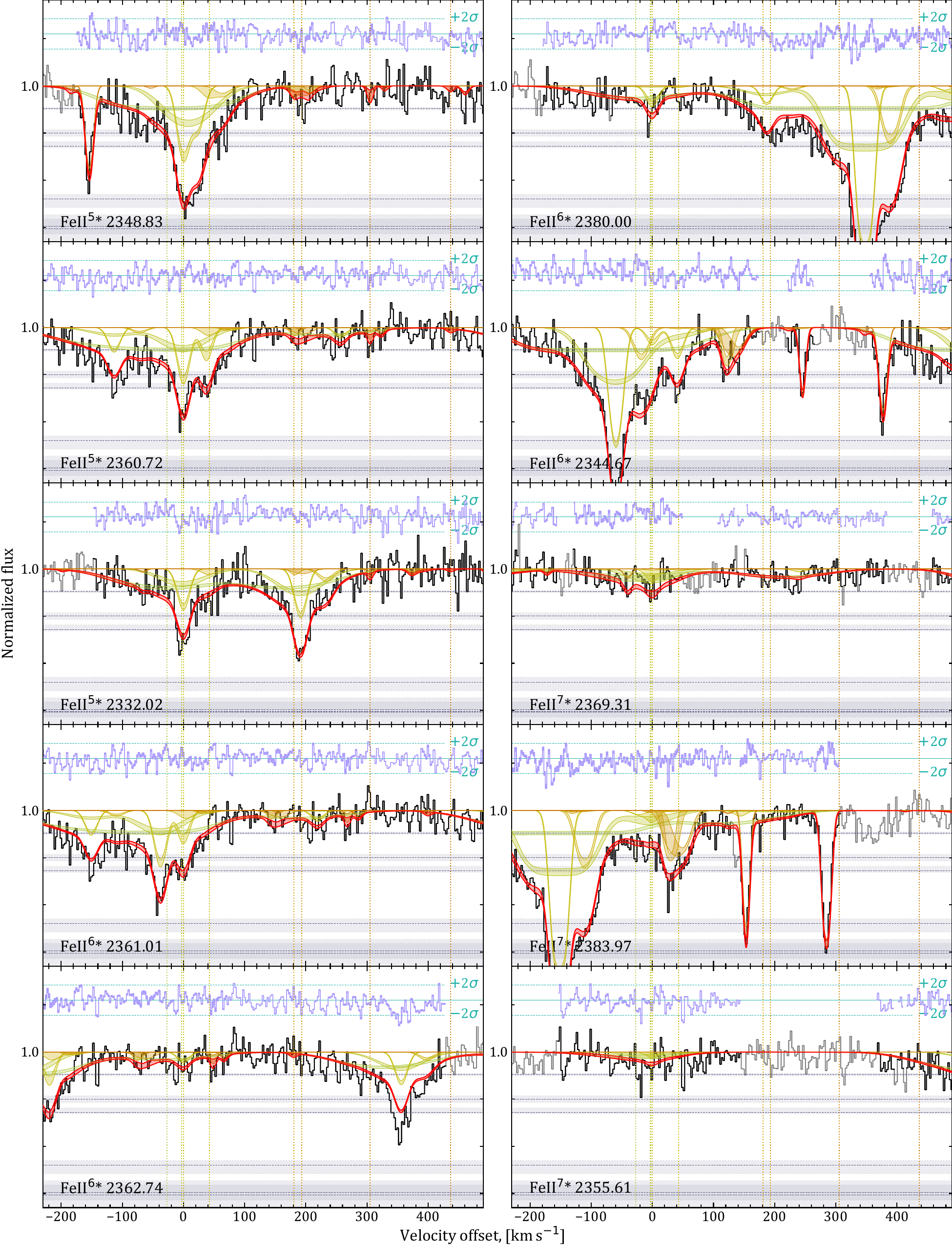}
\caption{Same as Fig.~\ref{fig:J2359_FeII_low_1} for higher-excitation \FeII\ lines (continued).
\label{fig:J2359_FeII_high_1}}
\end{figure*}

\begin{figure*}
\centering
\includegraphics[trim={0.0cm 0.0cm 0.0cm 0.0cm},clip,width=0.98\textwidth]{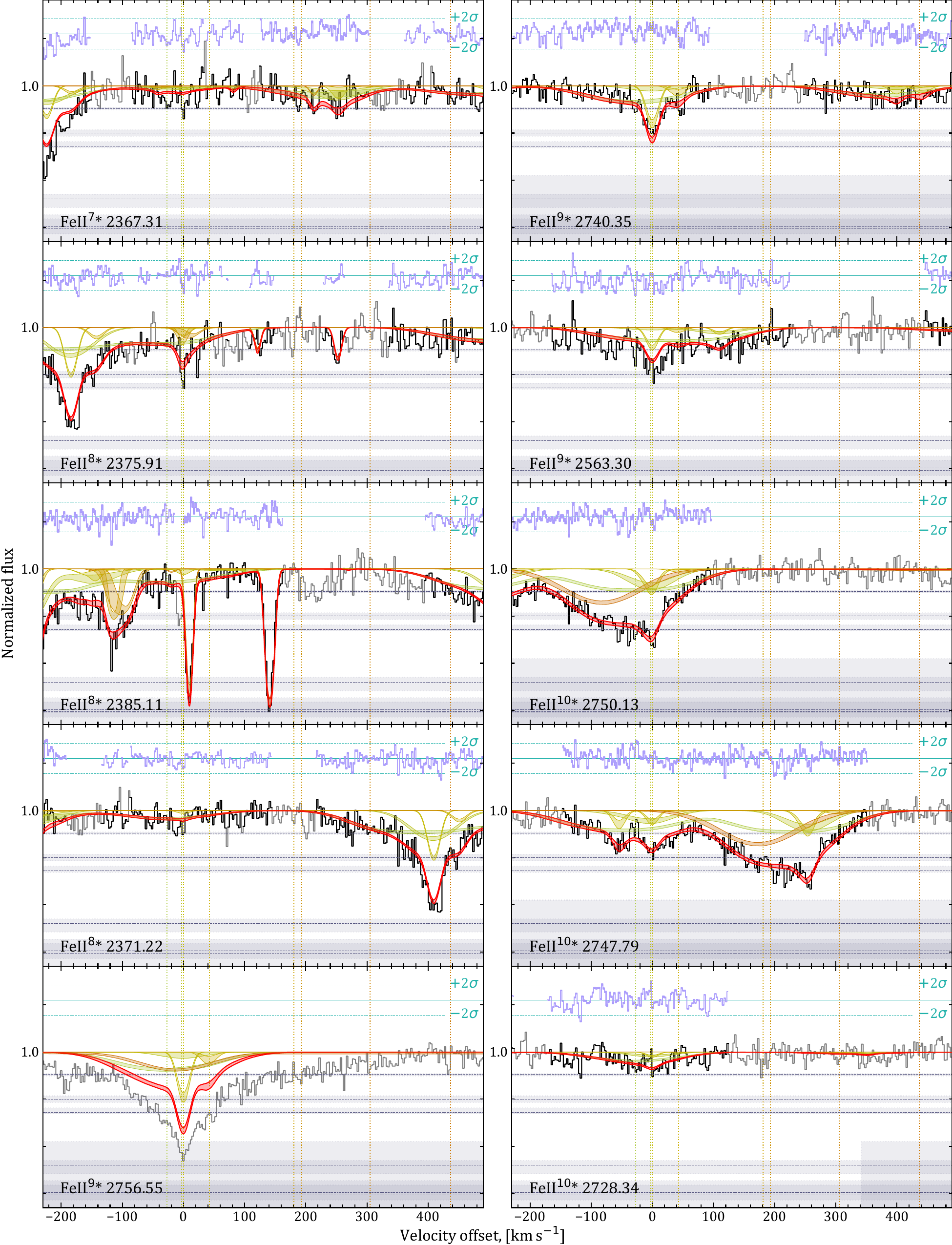}
\caption{Same as Fig.~\ref{fig:J2359_FeII_low_1} for higher-excitation \FeII\ lines (continued).
\label{fig:J2359_FeII_high_2}}
\end{figure*}

\begin{figure*}
\centering
\includegraphics[trim={0.0cm 0.0cm 0.0cm 0.0cm},clip,width=0.98\textwidth]{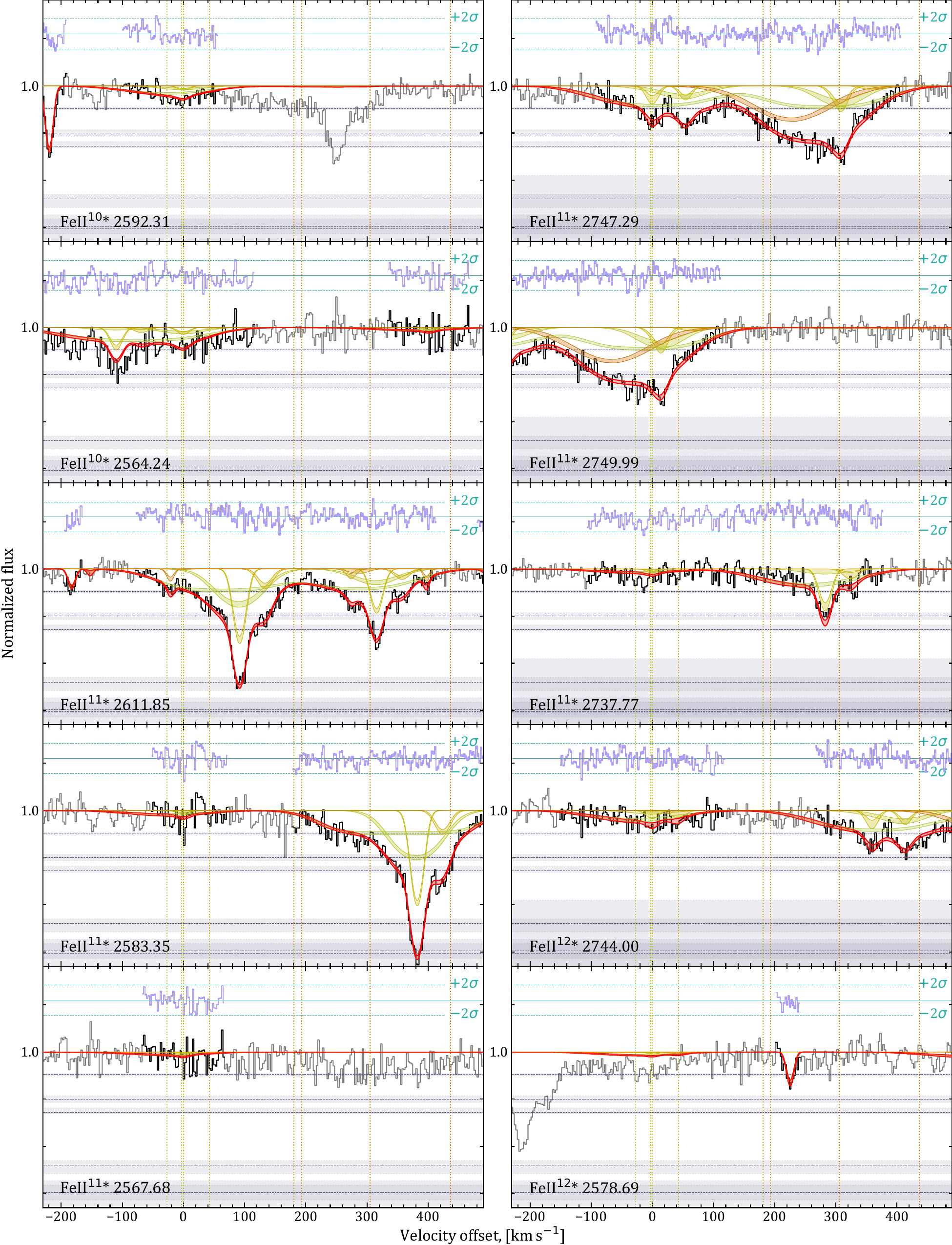}
\caption{Same as Fig.~\ref{fig:J2359_FeII_low_1} for higher-excitation \FeII\ lines (continued).
\label{fig:J2359_FeII_high_3}}
\end{figure*}


\bsp	
\label{lastpage}
\end{document}